\documentstyle[12pt,aaspp4]{article}
\begin{document}




\newcommand\approxgt{\mbox{$^{>}\hspace{-0.24cm}_{\sim}$}}
\newcommand\approxlt{\mbox{$^{<}\hspace{-0.24cm}_{\sim}$}}


\def\capdelvtrans
{A line of sight through a $\sigma_8 = 0.7$
CDM simulation (produced using the truncated Zel'dovich approximation)
at $\bar z=3$, with $h = 0.5$. The transfer function is taken from Ma
(1996). Box size is $12.8 \, {\rm Mpc}$
with a grid spacing of 
$0.032 \, {\rm Mpc}$. The parameters are $\Omega_b h^2 = 0.0125$, $J_{\rm HI} =
0.5$, $T_0 = 10^4 \, {\rm K}$ 
,$\gamma = 1.5$ and $k_{\rm S} = 2.3 \, {\rm Mpc^{-1}}$. All distances are
comoving. See Sec.~\protect{\ref{ingr}} for definitions of the
symbols. The abscissas for the lower two panels are the comoving
distances along the line of sight in units of ${\rm Mpc}$. The lower
of the two panels is the profile of overdensity $\delta_{\rm b}$
(eq. [\protect{\ref{deltadef}}]) and the upper 
one is the profile of velocity $u$ (eq. [\protect{\ref{ux}}]). 
The top two panels are both transmission profiles where $\tau$ is the
Ly$\alpha$ optical depth and the abscissas represent $u_\circ$, which is
related to the observed frequency through equation
(\protect{\ref{uuO2}}).
The profile with solid line is obtained using the full density and
peculiar velocity fields. 
The profile with dashed line is obtained using the same density field
but setting the peculiar velocity to zero everywhere (in which case,
$u$ becomes linear in $x$). 
}

\def\capdelvtransix
{The same as in Fig.~\protect{\ref{delvtrans7_21_50PCL1_52}} but a
different line of sight.}

\def\capcda
{The column density distributions for the
same model and parameters as those in 
Fig.~\protect{\ref{delvtrans7_21_50PCL1_52}}.  
The quantity $d^2 N_{{\rm Ly}\alpha}/dN_{\rm HI}/dz$ has units of ${\rm
cm^2}$. Crosses represent the distribution obtained by applying the
Threshold-Algorithm (threshold at $0.89$, which is the mean
transmission) to spectra generated using the density and
peculiar velocity fields predicted by the truncated Zel'dovich approximation.
Open triangles represent the same except that peculiar velocities are
set to zero. Open squares are obtained by applying the
Threshold-Deblending-Algorithm at the threshold of $0.89$.} 




\def\capmodcd
{Same parameters as in
Fig.~\protect{\ref{delvtrans7_21_50PCL1_52}}. 
The column densities computed using two different methods are plotted
against each other. 
First, we identify absorption
lines by the Threshold-Deblending-Algorithm using a transmission threshold of
$0.89$ and assign  
column densities according to equation (\protect\ref{tcd}), which are
plotted as the abscissas. 
We then take each absorption
line identified using the Threshold-Deblending-Algorithm and 
search for the corresponding maximum in $\delta_{\rm b}$ and apply
equation (\protect{\ref{npk3}}) to assign a second set of column
densities, which are plotted as the ordinates. 
}

\def\capzhang
{The model CDM1b (Table~\protect{\ref{cdmmodels}}) at $\bar z=3$, $J_{\rm HI} =
0.325$, $T_0 = 10^4 K$ and $\gamma = 1.45$. 
Solid triangles  
represent the distribution obtained by 
applying the Voigt-profile-fitting-technique to synthetic spectra
from a full hydrodynamic simulation (Zhang et al. 1996) with box size
of $9.6 \, \,{\rm Mpc}$ comoving and grid spacing of $0.075 \, {\rm Mpc}$. Open
triangles and open squares are the predictions of the Density-Peak-Ans\"{a}tz
(DPA) coupled with the truncated ($k_{\rm S} = 2.3 \,{\rm
Mpc^{-1}}$) Zel'dovich approximation, the former using the same box size
and grid-spacing as the hydrodynamic simulation and the latter using a
box size of $12.8$ Mpc and grid spacing of $0.05$ Mpc.
Crosses are the results of applying the
Density-Peak-Ans\"{a}tz and the Threshold-Algorithm (TA; transmission
threshold at $0.83$, the mean transmission) to the same density field
as for the open squares.
The short-dashed and long-dashed curves are the predictions of the
Density-Peak-Ans\"{a}tz 
coupled with the lognormal approximation, the former with $k_{\rm S} =
2.3 \,{\rm Mpc^{-1}}$ and the latter with the smoothing scale chosen
so that the final
rms density fluctuation matches that of the Zel'dovich approximation
($k_{\rm S} = 3.6 \,{\rm
Mpc^{-1}}$).
}



\def\capresolution
{The solid triangles are same as those in Fig.~\protect{\ref{zhang}},
representing the column density distribution 
obtained by applying the Voigt-profile-fitting-technique to synthetic
spectra from a full hydrodynamic simulation.
The rest of the points represent the column density distributions
obtained using the 
Density-Peak-An\"{a}tz in conjunction with the truncated Zel'dovich 
approximation for two simulations of different resolutions but 
the same cosmological, thermal and ionization parameters as
in Fig.~\protect{\ref{zhang}}.
The open squares and open triangles represent the distributions of a
simulation with box size of $12.8 \, {\rm Mpc}$ and grid spacing of
$0.05 \, {\rm Mpc}$ (all distances quoted are comoving). The open
squares here are the same as those in Fig.~\protect{\ref{zhang}} where
each maximum over three cells is identified as a peak. The open
triangles are the result of a different definition of peaks:
a local maximum over five cells with the density slope on each side
of the maximum not changing signs. 
Similarly, the crosses and open hexagons are the distributions for a
simulation of box size $12.8 \, {\rm Mpc}$ and grid spacing $0.0284
\, \,{\rm Mpc}$, using the three-cell and five-cell definition of peaks
respectively. 
}

\def\capFcd
{The column density distribution of the $\sigma_8 = 0.7$ CDM model with
no tilt (CDM1a in Table~\protect{\ref{cdmmodels}}), obtained using the
Density-Peak-Ans\"{a}tz and the truncated Zel'dovich approximation
($k_{\rm S} = 
2.3 \,{\rm Mpc^{-1}}$). The redshift is $\bar z=3$. Box size 
is $12.8 \,{\rm Mpc}$ with grid spacing of $0.05 \,{\rm Mpc}$. Open
squares: $F=0.25$ (eq. [\protect{\ref{Fdef}}]). Crosses:  
$F=1$. Open triangles: $F=5$. All of them have the equation of state described
by $\gamma = 1.5$ (eq. [\protect{\ref{tgamma}}]). The points with
error bars are the observational data at
about $\bar z = 3$ which have been corrected 
for incompleteness, taken from Hu et al. (1995).}  

\def\capeqs
{Column density distributions of the CDM1a model
(Table~\protect{\ref{cdmmodels}}) for three different equations of
state (eq. [\protect{\ref{tgamma}}]).
Open squares: $\gamma = 1.2$; crosses (same as crosses in
Fig. ~\protect{\ref{Fcd}}): $\gamma = 1.5$; open
triangles: $\gamma = 1.7$. $F = 1$ (defined  
in eq. [\protect{\ref{Fdef}}]) for all three. Points with error-bars  are
the same observational data as in Fig.~\protect{\ref{Fcd}}.
Long-dashed and short-dashed lines have the approximate slopes ($\beta
= 1.62$
and $\beta = 1.48$)
(normalization is chosen by hand) as
given in equation (\protect{\ref{slope}}) for the open triangles
and open squares
respectively.
}


\def\capnorm
{The effect of the amplitude of power spectrum on the slope of the column
density distribution. Open trianges represent the CDM2 model with $\sigma_0 =
1.13$ and $n_{\rm eff} = -1.94$ and open squares represent the CDM3 model with
$\sigma_0 = 0.81$ and the same $n_{\rm eff}$ (see
Table~\protect{\ref{cdmmodels}}). The normalizations of the column density
distributions are chosen by hand. The solid and dashed lines have
the approximate slopes ($\beta = 1.53$
and $\beta = 1.72$) as given in equation (\protect{\ref{slope}}) for the open
triangles and squares respectively.}


\def\capan
{The effect of the shape of the power spectrum on the slope of the column
density distribution. Open trianges represent the CDM4 model with $\sigma_0 =
1.01$ and $n_{\rm eff} = -2.17$ and open squares represent the CDM5 model with
the same $\sigma_0$ and $n_{\rm eff} = -2.59$ (see
Table~\protect{\ref{cdmmodels}}). The normalizations of the column density
distributions are chosen by hand. The solid and dashed lines have
the approximate slopes ($\beta = 1.67$
and $\beta = 1.79$) as given in equation (\protect{\ref{slope}}) for the open
triangles and squares respectively.}

\def\capsmooth
{Column density distributions of the CDM1a model
(Table~\protect{\ref{cdmmodels}}) for three different initial
smoothing scales. Crosses (same as crosses in 
Fig.~\protect{\ref{Fcd}}): $k_{\rm S} = 2.3 \, \,{\rm Mpc^{-1}}$, which is
the smoothing scale according to the standard prescription
(eq. [\protect{\ref{ks}}]). 
Open triangles: $k_{\rm S} = 1.15 \, \,{\rm Mpc^{-1}}$. Open squares:
$k_{\rm S} = 8.4 
\, \,{\rm Mpc^{-1}}$, which is the Jeans scale for $T_0 = 1$ and
$\gamma = 1.5$. Significant amount of orbit-crossing has probably taken
place for the last case. We adopt
$F = 1$ (eq. [\protect{\ref{Fdef}}]) and $\gamma = 1.5$ (eq.
[\protect{\ref{tgamma}}]) for all three cases.} 


\def\cappsmMDM
{$\sigma_0$ versus $k_{\rm S}$ (eq. [\protect{\ref{sigma0}}]) at $\bar z
= 3$. Going from the top, the solid line is the same $\sigma_8 = 0.7$, $h
= 0.5$ CDM model with no tilt as in Fig.~\protect{\ref{Fcd}}; the next
four sets of 
points/lines close together are all 
$\Omega_\nu = 0.1$ CHDM models, A1, B1, C1 and D1 in
Table~\protect{\ref{mdmmodels}} ; the next four sets are all $\Omega_\nu
= 0.2$ CHDM models, A2, B2, C2 and D2; the last set, solid triangles, is
an $\Omega_\nu = 
0.3$ CHDM model, A3. Note that $\sigma_0 (k_{\rm S}) \propto k_{\rm S}^{(n_{\rm
eff} + 3)/2}$ (see eq. [\protect{\ref{neff}}]).}


\def\capsmoothMDM
{Column density distributions of the B2 CHDM model (see
Table~\protect{\ref{mdmmodels}}) for three 
initial smoothing 
scales. Crosses: $k_{\rm S} = 19.2 \,{\rm Mpc^{-1}}$ (standard truncation
prescription, $k_{\rm S} = 1.5 \, k_{\rm NL}$ according to eq.
[\protect{\ref{ks}}]). Open
triangles: $k_{\rm S} = 8.4 
\,{\rm Mpc^{-1}}$ (Jeans scale for $T_0 = 10^4$ and $\gamma = 1.5$). Open
squares: no smoothing at all. Points with error-bars are the observational
data as in Fig.~\protect{\ref{Fcd}}. For all models, $F =1$ (eq.
[\protect{\ref{Fdef}}]) and $\gamma = 1.5$ (eq. [\protect{\ref{tgamma}}])
are used.}

\def\capmdmcdnuone
{Column density distributions for four $\Omega_\nu = 0.1$ CHDM
models. Points with error-bars  are the observational data as in 
Fig.~\protect{\ref{Fcd}}. 
All models have $\Omega_\nu = 0.1$. We use $\gamma = 1.5$ in the
equation of state for all of them (eq.
[\protect{\ref{tgamma}}]). Table~\protect{\ref{mdmmodels}} contains
descriptions of each of the following models.
Open hexagons: B1, $F = 3.33$. Open triangles: A1, $F = 3.33$. 
Crosses: D1, $F = 5.7$. Open squares: C1, $F = 5.7$. $F$ is defined in
equation (\protect{\ref{Fdef}}).}


\def\capFnutwofourtwo
{The column density distribution for the $A2$ CHDM model
(Table~\protect{\ref{mdmmodels}}). Three values of $F$ (eq.
[\protect{\ref{Fdef}}]) are shown: $F=1$
(open squares), $F=2.5$ (crosses) and $F=5$ (open triangles). We
choose $\gamma =
1.5$ for all three (eq. [\protect{\ref{tgamma}}]). Points with
error-bars  are the
observational data as in Fig.~\protect{\ref{Fcd}}.
The dashed line has the slope of $\beta = 1.86$, as 
given in equation (\protect{\ref{slope}}). The normalization of the
line is chosen
by hand.}

\def\capFnutwothreethreeb
{The column density distribution for the 
$B2$ CHDM model (Table~\protect{\ref{mdmmodels}}). Three values of $F$
are shown (eq.
[\protect{\ref{Fdef}}]): $F=1$ 
(open squares), $F=2.5$ (crosses) and $F=5$ (open triangles). $\gamma =
1.5$ for all three  (eq. [\protect{\ref{tgamma}}]). Points with
error-bars  are the 
observational data as in Fig.~\protect{\ref{Fcd}}. The dashed line has
a slope of $\beta = 1.86$, as 
given in equation (\protect{\ref{slope}}).} 

\def\capFnutwofourfour
{The column density distribution for the $D2$
(Table~\protect{\ref{mdmmodels}}) CHDM
model. Three values of $F$ are shown (eq.
[\protect{\ref{Fdef}}]): $F=7.14$
(open triangles), $F=3.57$ (crosses) and $F=1.79$ (open squares). $\gamma
= 1.5$ for all three (eq. [\protect{\ref{tgamma}}]). Points with
error-bars are the  same
observational data as in Fig.~\protect{\ref{Fcd}}. The dashed line has
a slope of $\beta = 1.78$, as 
given in equation (\protect{\ref{slope}}).}


\def\capeqsnutwofourtwo
{The column density distribution of the $A2$ CHDM model
(Table~\protect{\ref{mdmmodels}}) for three different values of
$\gamma$ (eq. [\protect{\ref{tgamma}}]). $F = 2.5$ (eq.
[\protect{\ref{Fdef}}])
is adopted. Three values of $\gamma$ are 
shown: $\gamma=1.2$ 
(open squares), $\gamma=1.5$ (crosses) and $\gamma=1.7$ (open
triangles). Points with error-bars are the
observational data as in Fig.~\protect{\ref{Fcd}}. The dashed line has
a slope of $\beta = 1.75$, which is the value given in equation
(\protect{\ref{slope}}) for $\gamma=1.2$, $\sigma_0$ and $n_{\rm eff}$ as given
in 
Table~\protect{\ref{mdmmodels}}.}

\title{The Statistics of Density Peaks and the Column Density Distribution of
the Lyman-Alpha Forest}

\author{Lam Hui\altaffilmark{1,2,5}, Nickolay
Y. Gnedin\altaffilmark{1,3,6} and Yu Zhang\altaffilmark{4}}
\altaffiltext{1}{Department of Physics, Massachusetts Institute of
Technology, Cambridge, MA 02139}
\altaffiltext{2}{Center for Theoretical Physics, Laboratory for
Nuclear Science, MIT}
\altaffiltext{3}{Princeton University Observatory, Peyton Hall,
Princeton, NJ 08544}
\altaffiltext{4}{Laboratory for Computational Astrophysics, National
Center for Supercomputing Applications, University of Illinois at
Urbana-Champaign, Urbana, IL 61801}
\altaffiltext{5}{Present Address: NASA/Fermilab Astrophysics Center, Fermi
National Accelerator Laboratory, Batavia, IL 60510; e-mail: \it
lhui@hydra.fnal.gov} 
\altaffiltext{6}{Present Address: Department of Astronomy, University
of California, Berkeley, CA 94720} 

\begin{abstract}
We develop a method to calculate the column density distribution of
the Ly$\alpha$ forest for column densities in the range $10^{12.5} -
10^{14.5} \, {\rm cm^{-2}}$. The Zel'dovich approximation, with
appropriate smoothing, is used to compute the density and peculiar
velocity fields. The effect of the latter on absorption profiles
is discussed and it is shown to have little effect on the column
density distribution. An approximation is introduced in which the
column density distribution is related to a statistic of density peaks
(involving its height and first and second derivatives along the line
of sight) in real space. We show that the slope of the column density
distribution is determined by the temperature-density relation as
well as the power spectrum on scales $2 \, h {\rm Mpc^{-1}} \,
\approxlt  
\, k \, \approxlt \, 20 \, h {\rm Mpc^{-1}}$. An expression relating the
three is given. We find very good agreement between the column density
distribution obtained by applying the Voigt-profile-fitting technique
to the output of a full hydrodynamic simulation and that obtained
using our approximate method for a test model. 
This formalism then is applied to study a group of CDM as well as
CHDM models. We show that the amplitude of the column density
distribution depends on the combination of parameters $(\Omega_b h^2)^2
T_0^{-0.7} 
J_{\rm HI}^{-1}$, which is not well-constrained by independent
observations. The slope of 
the distribution, on the other hand, can be used to
distinguish between different models: those with a smaller amplitude and
a steeper slope of the power spectrum on small scales give
rise to steeper distributions, for the range of column densities we
study. Comparison with high resolution Keck data is made.  

\end{abstract}


\section{Introduction}

There is a long history of theoretical efforts to place the study of the
Ly$\alpha$ forest within the framework of cosmological structure
formation theories (Doroshkevich \& Shandarin 1977; Rees 1986; Bond, Szalay \&
Silk 1988; McGill 1990; 
Bi, B\"{o}rner \& Chu 1992). Recent
work making use of numerical simulations has greatly advanced our
understanding in this direction
(\cite{cmor94}; Zhang, Anninos \& Norman 1995;
\cite{hkwm95}; \cite{pmk95}; \cite{mcor96}). (See also Bi, Ge \& Fang 1995.  
for a linear theory calculation).
The emerging picture is that it is possible to account for all the
observed properties of the Ly$\alpha$ forest (with column densities
less than about $10^{17} \, \,{\rm cm^{-2}}$) by assuming it originates from the
small scale structure, including the network of filaments, pancakes
and mild density fluctuations, which arises naturally in hierarchical
clustering cosmological models (\cite{whkm96}). 

A commonly used statistic to characterize the forest is its column
density distribution, the number
of absorption lines per unit neutral hydrogen column density per unit
redshift as a 
function of column density. Other useful statistics include line-line
correlations and the distributions of $b$-values and equivalent widths
(\cite{mhpb86}; \cite{clpw91}; \cite{prs93}; \cite{cdfgs95}). There
have also been 
proposals of new statistical tools 
(\cite{mb95}; \cite{mcor96}; \cite{pf96}). 
(See  
\cite{tytler92} for a general overview of the statistical issues
concerning quasar absorption systems.) We focus our attention on the
column density distribution in the present work.

One of the  most striking features of the observed column density distribution
of quasar absorption systems is that it can be approximated by a 
single power law that extends over many orders of magnitude. This was
emphasized by Tytler (1987), among others, who found that in the range
$10^{13} < N_{\rm HI} < 10^{22} \,{\rm cm^{-2}}$, the distribution was
reasonably well represented by a power law, $\propto N_{\rm HI}^{-\beta}$
with $\beta = 1.51 \pm 0.02$. However, there exists evidence of at least
one break. It has been demonstrated that there is a deficit of
absorption systems somewhere in the column 
density range $10^{14}$ to $10^{17} \,{\rm cm^{-2}}$ compared to a
power-law extrapolation of the distribution from lower column
densities (\cite{cwba87}; \cite{pwrcl93}; \cite{hkcsr95};
\cite{gcdfs96}). For reasons that have 
to do with the nature of the approximations that we make
(Sec.~\ref{rov}), we focus our attention on
absorption systems with column densities in the range $10^{12.5} <
N_{\rm HI} < 10^{14.5} \,{\rm cm^{-2}}$. Hu et al. (1995) obtained $\beta =
1.46$ with a $95 \%$ 
confidence range of $(1.37,1.51)$ in the column density range
$10^{12.3} < N_{\rm HI} < 10^{14.5} \,{\rm cm^{-2}}$ . Lu et al. (1996)
found the same best-fit $\beta$ for the same range of column densities.
 
An obvious ultimate goal of recent theoretical work on the Ly$\alpha$
forest is to constrain theories of structure formation. 
The natural question is: what determines the normalization and slope
of the column density distribution? What are the major determining factors, in
addition to the usual parameters specified by a given cosmological
model? To answer these questions, another question has to be
addressed: what are the analytical and/or computational tools
necessary to make accurate predictions 
for the column density distribution, given all the required parameters?

Accordingly, the present work can be divided into three parts, where
the tools are developed, the factors that influence the column density
distribution are analyzed and one application to a class of
cosmological models is discussed.

Numerical hydrodynamic simulations (\cite{cmor94}; \cite{zan95};
\cite{hkwm95}; \cite{mcor96})
provide the most obvious tools to study the Ly$\alpha$ forest.
Computational costs, however, prevent one from testing extensively several
cosmological models. We show in this paper that the Zel'dovich approximation
(\cite{zel70}), with appropriate smoothing, is an efficient
and accurate alternative (see also Doroshkevich \& Shandarin 1977).   
Our basic assumption is that the part of the Ly$\alpha$ forest with
column densities less than about $10^{14.5} \,{\rm cm^{-2}}$ 
arises mostly from regions which are slightly overdense (overdensity
$\approxlt \, 5$) or even underdense and  which have not undergone
orbit-crossing.  
The Zel'dovich approximation can then be coupled with the
equations governing the thermal and ionization states of the gas to
yield accurate predictions for the density of neutral hydrogen and the peculiar
velocity as a function of position. Absorption spectra 
are then generated and analyzed.  Basic expressions for the absorption
optical depth are presented in Sec.~\ref{fgp} and the approximations
that go into its computation are discussed in Sec.~\ref{ingr}.

Given an absorption spectrum, the column density distribution depends
on the method of identifying lines and assigning column
densities. This is discussed in Sec.~\ref{licdd}. We investigate the effects of
peculiar velocities on the column density distribution, using a method
described by Miralda-Escud\'{e} (1996).
We find that although peculiar velocities can strongly influence the
shapes of absorption profiles, they play a relatively minor role
in determining the column density distribution. 
The various interesting effects of peculiar velocities are discussed
in Sec.~\ref{v}. Motivated by this finding,
a very different way of assigning column densities is introduced
in Sec.~\ref{dpa},
in which no absorption spectrum needs to be generated. In the absence
of peculiar velocities, there is a one-to-one correspondence between
density peaks in real space (if they are separated by a distance
larger than a minimum corresponding to the thermal broadening width) and minima
of transmission (maxima in absorption) in the observed spectrum.
Under such conditions, we can simply associate each density peak in
real space with an absorption line and assign a
column density to each based on the height and curvature of the
peak. The column density distribution is then a statistic of density
peaks in real space. We apply this procedure (we
call it the Density-Peak-Ans\"{a}tz) to the density field
predicted by the truncated Zel'dovich approximation and test the
result against that of a full
hydrodynamic simulation. The column density distribution obtained in
this way is compared to that obtained from the hydrodynamic simulation
using the Voigt-profile-fitting-technique, which is the line-identification
method most widely used. The level of agreement is found to be
excellent. In Sec.~\ref{rov}, we discuss the range of parameters
in which our computed column density distribution is expected to be reliable.

Armed with the right tools, we turn our attention to the second
question: what factors determine the column density distribution? 
They can be divided into two categories. One has to do with 
properties of the intergalactic medium, including its temperature, the
equation of state (or temperature-density relation, which we
will use interchangeably; see Hui \& Gnedin 1996), the ionizing
radiation intensity and the baryon 
density. Uncertainties in all of them have to be taken into account
before one can meaningfully confront theories with observations. We
distinguish between the factors that mostly affect the normalization
of the column density distribution and those that mostly affect its
slope. It is found that the temperature-density relation (weakly) affects
the slope while the rest of the above factors influences the normalization. 
It is also 
emphasized that the temperature and the equation of 
state depend on the reionization
history of the universe (a fuller 
discussion of this point and related topics will be given in a
separate paper). The second set of factors 
affecting the column 
density distribution has to do with the specific cosmological model,
namely the normalization and shape of the corresponding power spectrum. 
We study a few variants of the Cold Dark Matter (CDM) model in
Sec.\ref{slopecd} for this purpose. 
It is found that the amplitude and slope of the linear power spectrum on
comoving scales of 
around $2 \, h \, \,{\rm Mpc^{-1}}$ to $20 \, h\, \,{\rm
Mpc^{-1}}$ are the 
most important factors in determining the slope of the column
density distribution (the equation of state also has a weak
effect on it). Decreasing the amplitude and/or steepening the slope of the
power 
spectrum tends to steepen the 
distribution in the column density range about $10^{12.5}$ to
$10^{14.5} {\rm cm^{-2}}$. We introduce an 
expression relating the slope of the column density distribution to
the equation of state and properties of the power spectrum on small scales.

We then study a class of Cold plus Hot Dark Matter
(CHDM) models in Sec.~\ref{mdm}, making use of the insights gained in
Sec.~\ref{cdm} and Sec.~\ref{slopecd}.  The $\Omega_\nu = 0.2$ (density
parameter in 
neutrino) models have steeper column density distributions compared to
those with $\Omega_\nu = 0.1$ because they have less power on the
relevant scales.  
In particular, the low Hubble constant ($H_0 = 50 \, \,{\rm km \, s^{-1} \,
Mpc^{-1}}$) $\Omega_\nu = 0.2$ models predict slopes that are
steeper than the observed one for most of the
parameter-space specifying the properties of the intergalactic
medium. Only for equations of state that are close to isothermal can the two be
made consistent with each other. We emphasize however that a more
detailed comparison between the models and observations, taking fully into
account instrumental noise and biases of the line-identification
method(s), is necessary before one can firmly reject any model.
We conclude in Sec.~\ref{gpconc} .

It is appropriate that we mention here two recent pieces of work along similar
lines as described above, but using a different dynamical approximation, namely
the lognormal approximation: Bi \& Davidsen (1997) and Gnedin \& Hui (1996).
The former, in particular, contains a very comprehensive and careful analysis 
of the many different observational properties of the Ly$\alpha$ forest.
One strong point of their analysis is that they tested their method using
VPFIT, a spectral analysis routine that is commonly used by observers.
We will discuss the predictions for the column density distribution by the
lognormal and the Zel'dovich approximations in Sec.~\ref{dpa}.

In our notation, bold faced letters are reserved for
three-dimensional vectors. The symbols ${\bf v_{\rm pec}}$ and ${\bf x}$
denote the three-dimensional peculiar velocity and comoving 
position while $v_{\rm pec}$ and $x$ are their counterparts along the
line of sight of interest. Standard symbols are used for cosmological
parameters: $H$ for the Hubble constant as a function of $z$, $H_0$
for the Hubble constant today, $h$ for $H_0 / 100 \, \,{\rm km s^{-1}
Mpc^{-1}}$, $\Omega_0$ for the density parameter today, with the
subscript $b$ to denote its baryon portion and $\nu$ its neutrino
content. We use the symbol ${\rm h}$ (as distinct from $h$) to denote
the Planck constant in a few places where it arises.  The term
multiple-streaming is reserved for the situation 
where a single observed redshift corresponds to more than one 
position in real space. We distinguish it from the term
orbit-crossing, which is commonly used interchangeably with
multiple-streaming in other contexts. Orbit-crossing refers to 
the case where a single position has more than one velocity.

\section{Cosmological Lyman-Alpha Absorption in a Fluctuating Medium:
Basic Results}
\label{fgp}

A photon emitted with energy higher than $10.196 \, {\rm eV}$ (wavelength
of $1216 \, \AA$) by a distant quasar is continuously redshifted as it
travels through the intergalactic medium until it reaches the
observer. At some intermediate point, the photon is redshifted to around 
$1216 \, \AA$ in the rest frame of the intervening medium, which may
contain neutral hydrogen. It can then excite the Ly$\alpha$ transition and be
absorbed. Let us consider a particular line of sight from the observer
to the quasar. The optical depth $\tau$ 
(the probability of transmission is given by $e^{-\tau}$) of a
photon at a given observed frequency $\nu_\circ$ is given
by:
\begin{equation}
\tau(\nu_\circ) = \int_{x_A}^{x_B} {n_{\rm HI}} \sigma_\alpha {dx\over{1+z}}
\, ,
\label{taux}
\end{equation}
where $x$ is the comoving radial coordinate of some intermediate point along
the line of sight, $z$ is the redshift and $n_{\rm HI}$ is the proper
number density of neutral hydrogen at that point. The limits of
integration, $x_A$ and $x_B$, are the comoving positions of the
observer and the quasar. The Ly$\alpha$ absorption 
cross-section is denoted by $\sigma_\alpha$. It is a function of the
frequency of the photon with 
respect to the rest frame of the intervening hydrogen at position
$x$. Let us call this frequency $\nu$. The 
cross-section is peaked when $\nu$ is equal to the Ly$\alpha$
frequency, $\nu_\alpha$. 

The frequency $\nu$ is related to the observed frequency $\nu_\circ$ by:
\begin{equation}
\nu = \nu_\circ (1+z) \left(1 + {v_{\rm pec}\over c}\right) \, ,
\label{nunu0}
\end{equation}
where $v_{\rm pec}$ is the peculiar velocity along the line of sight
at position $x$ and  
$1+z$ is the redshift factor {\it due to the uniform Hubble
expansion alone} at the same position.  The peculiar velocity of the
observer, which merely displaces the whole spectrum by a constant
amount (independent of $x$), is ignored. The quantity $v_{\rm pec}/ c$,
where $c$ is the speed 
of light, is much smaller than $1$. 

It proves convenient for later discussion to expand $z$ around some
mean redshift of interest $\bar z$, which 
could be the redshift of a simulation output or the average redshift
of an observed spectrum with limited redshift range.
Using $dx = c dt/ a$, where $a$ is the Hubble scale factor and $t$ is
the proper time, it can be shown that
\begin{eqnarray}
\label{ux}
\nu = \nu_\circ (1+ \bar z) \left(1 + {u\over c}\right) \, \, , \, \, u \equiv {\bar H
\over {1+\bar z}} (x-\bar x) + v_{\rm pec} (x) \, ,
\end{eqnarray}
where $\bar x$ is the position at which the redshift due to Hubble expansion
coincides exactly with $\bar z$. The Hubble constant at
the same redshift is denoted by $\bar H$. We assume the range of $x$
is small enough so that $u/c \ll 1$. The convention $a = 1$ today is
adopted. 

The velocity coordinate $u$ defined above contains contributions from
both the Hubble 
expansion and the peculiar motion. Without peculiar motion, $u$
increases monotonically with $x$ and is in fact linear in
$x$. Peculiar velocities destroy the linear relation and could give rise
to situations where a given $u$ corresponds to more than one position
$x$. It implies that a photon of a given observed frequency $\nu_\circ$
can have the same rest-frame 
frequency $\nu$ at more than one place in its trajectory from the
quasar to the observer. We reserve the term
multiple-streaming to this situation and distinguish it
from orbit-crossing where a given $x$ carries more than one
$v_{\rm pec}$ or $u$. We will return to the subject of
multiple-streaming in Sec.~\ref{v} and that of orbit-crossing in
Sec.~\ref{zasection} .

We define one more velocity coordinate $u_\circ$, which is
related to the observed frequency $\nu_\circ$ by:
\begin{equation}
\nu_\circ = {\nu_\alpha \over {1+\bar z}} \left(1 - {u_\circ\over c}\right) \,
\label{uuO2}
\end{equation}
where $\nu_\alpha$ is the Ly$\alpha$ frequency. The velocity
coordinate $u_\circ$ is simply equal to $u$ when $\nu$ coincides exactly with
$\nu_\alpha$ (this can be seen by comparing eq. [\ref{ux}] and
[\ref{uuO2}], bearing 
in mind that $u/c$ and $u_\circ/c$ are both assumed to be much less than $1$). 

With the definitions in place, we change
the variable from $x$ to $u$ in equation (\ref{taux}), which results
in the following expression for $\tau$, now a function of $u_\circ$:
\begin{equation}
\tau(u_\circ) = \sum \int_{u_A}^{u_B} {n_{\rm HI} \over {1+z}} \left\vert {du\over dx}\right\vert^{-1}
\sigma_\alpha du \,\,  , \,\, \sigma_\alpha = \sigma_{\alpha 0}
{c\over {b\sqrt{\pi}}} 
e^{-{(u-u_\circ)^2/b^2}} \, . 
\label{tau}
\end{equation}
The summation refers to a sum over multiple-streams (all the
$x$'s within the range $x_A - x_B$ that corresponds to a given $u$), and
$n_{\rm HI}$, $z$ and $|{du/dx}|^{-1}$ are now functions of $u$.
The limits of integration $u_A$ and $u_B$ are the velocity coordinates
corresponding to the positions $x_A$ and $x_B$ (assuming no
orbit-crossing so that each $x$ carries one $u$). 
Note that in practice, only a limited range of
$u$ contributes to $\tau$ for a limited range of $u_\circ$ so that one
can replace the redshift $z$ with $\bar z$. The same is also true for
equation (\ref{taux}). 
 
The Ly$\alpha$ cross-section is expressed as a function of $u-u_\circ$.
The constant $\sigma_{\alpha 0}$ is equal to the combination of
fundamental physical constants $0.416 
\pi q^2 / (m_e c \nu_\alpha)$, where $q$ is the charge of an electron
and $m_e$ is its 
mass. It is about $4.5 \times 10^{-18} \,{\rm cm^2}$. 
The parameter $b$ is equal to $\sqrt{2 k_B T / m_p}$ where
$k_B$ is the Boltzmann constant, $m_p$ is the mass of a proton and $T$
is the temperature of the gas at the velocity coordinate $u$.  

The form of the line profile function above takes into account thermal
broadening but ignores the natural line width. A more general profile
function involves a convolution of the two, resulting in the Voigt profile
(\cite{rl79}). However, the Voigt profiles are accurately thermal profiles
for column density less than about $10^{17} \,{\rm cm^{-2}}$. The reader is
referred to Spitzer (1978) and Press and Rybicki (1993) for
discussions of curve of growth analysis. 

Note also that it is sometimes assumed $b$ contains a component due to
turbulent motion. We do not include it explicitly in our formalism.
Bulk motion, on the other hand, is accounted for by $v_{\rm
pec}$ or $u$. 

Let us consider two different limits of equation (\ref{tau}).

Suppose there is a high local maximum in $n_{\rm HI} |du/dx|^{-1}$ at some
$u=u_{max}$ with width in velocity space much smaller than the
thermal width $b$. Then one can take the line profile function
associated with 
$\sigma_\alpha$ out of the integral in equation (\ref{tau}) because
$n_{\rm HI} |du/dx|^{-1}$ varies much more rapidly than the thermal
profile:
\begin{equation}
\tau(u_\circ) = \biggl(\int_{max} n_{\rm HI}(x) {dx\over {1+z}}\biggr)
\sigma_{\alpha 0} {c\over
{b\sqrt{\pi}}} e^{-(u_\circ-u_{max})^2/b^2} \, ,
\label{cloud}
\end{equation}
where the variable of integration has been changed back from $u$ to $x$.
The equation holds when $u_\circ$ is close enough to $u_{max}$. 
The integral is over the local maximum, assuming that the amount of
neutral hydrogen away from the maximum does not cause significant 
absorption (until another maximum is encountered).
One then sees an absorption line with a Gaussian
profile in optical depth. While the width of the line
tells us about $b$, which is proportional to the square
root of the temperature, the depth of the line provides information about
both $b$ and the column density, which is the integral inside the
first pair of 
brackets on the right hand side. Let us call this the {\rm narrow-maximum-limit}. 

Consider another limit of the integral (eq. [\ref{tau}]) in which
$n_{\rm HI} |du/dx|^{-1}$ 
varies slowly with $u$. Suppose the scale of variation is
much larger than the thermal width. In this case, one can leave the
line profile function 
inside the integral but take the rest outside:\footnote{The expression
is not valid at velocity caustics, where $du/dx$ vanishes. Further
discussion on velocity caustics can be found in Sec.~\ref{v}.}
\begin{equation}
\tau(u_\circ) = \sum {n_{\rm HI}\over{1+\bar z}} \left\vert{du\over
dx}\right\vert^{-1} c\sigma_{\alpha 0} 
\, .
\label{noncloud}
\end{equation}
The velocity dependent terms on the right hand side are evaluated at
$u_\circ$. The profile function has been integrated out. 

In the above limit, $\tau$ does not necessarily have the thermal
profile around its maxima. We will call this the 
broad-maximum-limit. An extreme example is that of a
homogeneous medium, which gives rise to featureless and uniform absorption
(\cite{gp65}). 

Conventional analysis of quasar spectra involves identifying those
parts of the spectra that are due to the Ly$\alpha$ absorption and
fitting them with superpositions of the Voigt profiles (of which the
thermal profiles are a subset) until the residual signal is consistent
with noise. This technique was motivated by the picture of the
intergalactic medium as consisting of a smooth component which causes
relatively little absorption and a set of clouds that satisfy the
{\rm narrow-maximum-limit}. For each cloud, the best-fit Voigt profile then
yields its temperature and column density according to equation
(\ref{cloud}). 

However, it is clear that for a general fluctuating medium, not all
maxima in $\tau$ necessarily 
satisfy the conditions leading to equation (\ref{cloud}). In fact,
according to most structure formation 
theories, there invariably exist fluctuations in the intergalactic
medium on scales larger than the thermal width. In the
{\rm broad-maximum-limit}, the shape of a local maximum in optical depth is
determined by the distributions of $n_{\rm HI}$ and $|du/dx|$ around it.
Each maximum in $\tau$ can still be identified as an absorption line
and one can even apply standard techniques and try to
fit its shape with superposition of the Voigt profiles. 
Given the best-fit Voigt profiles, one can assign a $b$-value (width of
the profile) and a column density to each profile but it is no longer true, for
instance, that the $b$-value thus obtained is equal to $\sqrt{2k_B T/
m_p}$ (eq. [\ref{tau}]). 
A reasonable question to ask is whether there are other practical
methods of identifying absorption lines and assigning column densities
without assuming every absorption line consists of a
superposition of the Voigt 
profiles. This will be discussed in Sec.~\ref{licdd}. 
It should be borne in mind, however, that all existing observational
data on the column density distribution are obtained using the
Voigt-profile-fitting-technique, so for the purpose of comparing
theory with these observations, it is important the line-identification
algorithm one uses gives results consistent with the
profile-fitting-technique.

In general, there are regions of high density and limited extent,
galaxies for instance, which give rise to absorption profiles well
approximated by the {\rm narrow-maximum-limit}, but there are also
regions in the intergalactic medium with gentle fluctuations where the
{\rm broad-maximum-limit} holds.  Then there are those places where neither
limit applies, in which cases, a full integration of equation
(\ref{tau}) has to be carried out to compute the optical depth.
To do so, one needs to know how the neutral hydrogen density,
peculiar velocity and temperature vary with spatial position. This is the
subject of the next section. In any case, the above discussion should
make it clear that the quasar absorption spectrum contains a wealth of
information on the intergalactic medium.

\section{Ingredients for Generating Quasar Absorption Spectra}
\label{ingr}

There are four quantities that go into the computation of the optical
depth: temperature, peculiar velocity, overdensity and neutral
fraction. That the temperature and peculiar velocity are important
should be obvious from the expression for the absorption cross-section
in equation (\ref{tau}). The temperature determines the extent of
thermal broadening ($b$) and the peculiar velocity changes the frequency of the
photon in the fluid rest-frame (eq. [\ref{ux}]). 
Let us define carefully what we mean by the other two quantities, the
overdensity and the neutral fraction.

Suppose $n_{\rm H} ({\bf x})$ is the total proper number density of
all hydrogen species at position ${\bf x}$ and $\bar n_{\rm H}$ is its
spatial average.  The overdensity $\delta_{\rm b}$, which describes the
variation in space of 
$n_{\rm H} ({\bf x})$, satisfies: 
\begin{equation}
n_{\rm H} ({\bf x}) = \bar n_{\rm H} \left[1+\delta_{\rm b}({\bf x})\right]
\, \, , \, \, \rho_b ({\bf x}) = \bar \rho_b \left[1+\delta_{\rm b}({\bf
x})\right] \, .
\label{deltadef}
\end{equation} 
In the first expression, $\delta_{\rm b}$ is defined as the number overdensity of
hydrogen. In the second expression , we equate $\delta_{\rm b}$ with
the mass overdensity of baryons ($\rho_b$ is the proper mass density of
baryons and $\bar \rho_b$ is its mean), which is an excellent approximation
for our application because there is no significant conversion of
hydrogen into other elements, nor is there any interaction
that could cause the spatial distribution of hydrogen to deviate
significantly from that of other types of baryons.

What the Ly$\alpha$ absorption directly probes is not the total
hydrogen density but its neutral component. The neutral fraction
$X_{\rm HI}$ is
defined by the following relation:
\begin{equation}
n_{\rm HI} ({\bf x}) = n_{\rm H} ({\bf x}) X_{\rm HI} ({\bf x})\, ,
\label{Xdef}
\end{equation}
where $n_{\rm HI}$ is the proper number density of neutral hydrogen as
a function of space. The neutral fraction is determined by the balance
between recombination and ionization, the rates of which are dictated by
the temperature and radiation intensity respectively.

In general, all four quantities, overdensity $\delta_{\rm b}$, peculiar
velocity $v_{\rm pec}$, temperature $T$ and neutral fraction $X_{\rm
HI}$, are functions of position. In the next two sub-sections, we
discuss first how to determine the spatial distributions of $\delta_{\rm b}$
and $v_{\rm pec}$, and second how $T$ and $X_{\rm HI}$ vary with
position through their dependence on $\delta_{\rm b}$.
All quantities are evaluated at $\bar z = 3$. Although most of the
material in this section is standard textbook fare, it consists of a
somewhat unusual combination of methods, so it is worth going through
the basic equations and stating our assumptions carefully.

\subsection{The Zel'dovich Approximation}
\label{zasection}

In cosmological models where dark matter (a term we use
interchangeably with non-interacting matter) dominates the mass density of
the universe, $\delta_{\rm b}$ as defined in 
equation (\ref{deltadef}) coincides with the dark matter overdensity
$\delta_{\rm DM}$ on large scales. We define $\delta_{\rm DM}$ in an
analogous manner as before (eq. [\ref{deltadef}]):
\begin{equation}
\rho_{\rm DM} ({\bf x}) = \bar\rho_{\rm DM} [1+\delta_{\rm DM}({\bf x})] \, ,
\label{deltaDM}
\end{equation}
where $\rho_{\rm DM}$ is the mass density of dark matter at position ${\bf
x}$ and $\bar\rho_{\rm DM}$ is its mean. The equality $\delta_{\rm b} =
\delta_{\rm DM}$ is equivalent to the statement that the hydrogen density
(which we assume is simply proportional to the baryon density) varies with 
position in the same manner as the dark matter density does. This is
true on large scales where gas pressure is insignificant compared to
the gravitational attraction of the dark matter, provided the baryons and dark
matter start out having the same spatial distribution, which is approximately
true 
for most popular cosmological models. Moreover, with no significant
interaction that distinguishes between the two on large scales, the
baryons and dark matter share the same peculiar velocity field.
On small scales, however, gas pressure can cause the spatial distributions of
baryons and dark matter and their velocities to differ. We will return
to this point below.  

Hence, on sufficiently large scales (how large is large, an obviously
important question, will be addressed later), it is adequate to know
the overdensity and peculiar velocity of the dark matter.
The Zel'dovich approximation (\cite{zel70}) is a well-tested
approximation to compute the density and velocity distributions of
dark matter in the mildly nonlinear regime
($\delta_{\rm DM}\,\approxlt\, 5$) before orbit-crossing. 
The reader is referred to the article by Shandarin and Zel'dovich
(1989) for a comprehensive review (see also Hui and Bertschinger 1996
for an alternative formulation of the approximation). 

The starting point of the Zel'dovich approximation is the following
equation for the displacement of a given mass element or particle:
\begin{equation}
{\bf x}({\bf q}, t) = {\bf q} + D_+(t) {\bf\nabla_q} \psi({\bf q}) \, ,
\label{zad}
\end{equation}
The coordinate ${\bf q}$ is the initial position of the mass element
and ${\bf x}$ is its comoving position as a function of time. The
displacement is then $D_+(t) {\bf\nabla_q} \psi({\bf q})$. Its time
dependent part $D_+(t)$ is the linear growth factor (Peebles 1980), which,
for a universe with critical matter density, can be equated with $a$,
the Hubble scale factor. The time independent function ${\bf\nabla_q}
\psi({\bf q})$ is determined by initial conditions. Growing mode
initial conditions dictate that it is curl-free, hence its form as
the gradient of the potential $\psi$ (${\bf\nabla_q}$ is the spatial gradient
in $q$ space). 

Expressions for the peculiar velocity and overdensity follow
immediately from equation (\ref{zad}):
\begin{equation}
{\bf v}_{\rm pec} = a \dot D_+ {\bf\nabla_q} \psi\, \, , \, \, 1 +
\delta_{\rm DM} = \,{\rm det}^{-1} \left[\delta_{ij} + D_+(t)
{\partial^2\psi\over {\partial q_i\partial q_j}} \right] \, .
\label{zav}
\end{equation}
The dot in the first expression denotes differentiation with respect
to proper time $t$. The peculiar velocity is defined by ${\bf v}_{\rm pec} =
a d{\bf x}/dt$. The second expression follows from mass conservation i.e. 
$(1+\delta_{\rm DM}) d^3 x = d^3 q$. The right hand side of the second
expression is simply the Jacobian of the $q$-$x$ transformation matrix.

The function $\psi({\bf q})$ contains all the information on the specific
cosmological model one chooses to study. For the cosmological models
we study in this paper, it is a Gaussian random field in ${\bf q}$
space. Suppose $\tilde 
\psi ({\bf k})$ is its Fourier counterpart defined by $\psi
= \int d^3 k \, \tilde\psi \, e^{i{\bf k}\cdot{\bf q}}$. The two-point correlation
of $\tilde\psi$ is related to the commonly used power spectrum $P$ by
\begin{equation}
\langle \tilde \psi ({\bf k}) \tilde \psi^{*}({\bf k'}) \rangle =
k^{-4} P(k) \delta^3 ({\bf k} - {\bf k'})  \, ,
\label{Pdef}
\end{equation}
where $P$ is related to the root-mean-squared (rms) linear
overdensity fluctuation by
\begin{equation}
\langle \delta^2 \rangle = D_+^2(t)\int_0^\infty 4\pi P(k) k^2 dk \, .
\label{rmsdelta}
\end{equation}
Note that $D_+^2 = (1+\bar z)^{-2}$ for a universe with critical
matter density, choosing $D_+ = 1$ today.

To produce a realization of the density and velocity fields for a
given cosmological model, we employ the following procedure:
first, we use the corresponding power spectrum to generate the Gaussian
random field $\psi (\bf q)$ on a grid; second, we displace particles from
their initial grid positions (${\bf q}$) according
to equation (\ref{zad}) for $D_+(t)$ corresponding to ${\bar z} = 3$; 
third, a peculiar velocity is assigned to each 
particle according to the first expression in equation (\ref{zav});
finally, we use the TSC (Triangular-Shaped density Cloud) scheme
(\cite{he88}) to interpolate the 
particle positions and velocities to become momentum and mass
densities on the grid and divide one by the other to obtain the velocity
itself. The interpolation to obtain mass density is our way of
enforcing mass conservation, as is expressed in the second expression
of equation (\ref{zav}). In the last procedure, we smooth the
momentum and mass density fields over a small number of grid cells (in fact, we
use one and have checked that the precise number is not important as
long as it is small) before performing the division to obtain the
velocity field so that we have well-defined velocities even in places
with zero density after the TSC interpolation (\cite{kbgnd94}). 
Any line of sight can then be chosen
through the simulation box and the above set of steps gives the
overdensity and peculiar velocity (in fact only the component parallel
to the line of sight is needed) at each grid point on it.

The procedure just outlined is very efficient because there is no need to
integrate any equation of motion. One simply multiplies the displacement
of each particle by an appropriate factor of $D_+(t)$.
However, the first step of the procedure has to be slightly modified
to address two 
problems.

The first one is orbit-crossing. The Zel'dovich approximation is
known to predict too rapid growth of the thickness of the post-collapse
pancake (Shandarin and Zel'dovich 1989). A number of cures have been
proposed (\cite{kps90}; \cite{mlms92}; \cite{bsv93}; \cite{bagpad94})
but the one that consistently gives good agreement with 
N-body simulations is the truncated Zel'dovich approximation
(Kofman 1991; \cite{cms93}; \cite{mbw95}). The basic idea is
to smooth the initial power spectrum on small scales so that the amount of
orbit-crossing that might have occurred by the time of interest is not
significant enough to destroy the accuracy of the Zel'dovich
approximation on large scales, where the fluctuations are still mildly
nonlinear. The initial power spectrum $P(k)$ is multiplied by a 
Gaussian smoothing kernel of the form 
$e^{-k^2/k_{\rm S}^2}$, before it is used to generate the
Zel'dovich displacement field (eq. [\ref{Pdef}]). (This is equivalent to
smoothing the initial density field $\delta({\bf x})$ through the following 
convolution: $(2\pi k_{\rm S}^{-2})^{-0.5} \int \delta({\bf x'})  e^{-(k_{\rm
S}^{2}|{\bf x}-{\bf x'}|^2/2)} d^3 x'$.)  
The smoothing
wavenumber $k_{\rm S}$ is chosen according to the following
prescription
\begin{equation}
k_{\rm S} = 1.5 \, k_{\rm NL} \, , \quad {\rm where} \, \, 1 =
D_+^2(t) \int_0^{k_{\rm NL}} 4\pi P(k) k^2 dk \, . 
\label{ks}
\end{equation}
Note that $P(k)$ above is the initial power spectrum before any smoothing.
The proportionality constant between $k_{\rm S}$ and $k_{\rm NL}$
actually depends
somewhat on the power spectrum, with more smoothing (smaller $k_{\rm S}$)
required for models  that have relatively more power on small scales
(\cite{melott94}). The choice above has been shown to give good
agreement with N-body simulations 
for CDM models (Melott et al. 1995).
We will see that for those CHDM models with relatively little power on
small scales, the precise value of $k_{\rm S}$ is not important. The
procedure described above is commonly called the truncated Zel'dovich
approximation.

The second problem is one we have pointed out before, namely that
$\delta_{\rm DM}$ is not necessarily equal to $\delta_{\rm b}$ (which
is what we are interested in ultimately) on small scales.  
In linear theory, it is possible to show that for a dark matter
dominated universe, the Fourier components of the two quantities obey
$\tilde\delta_{\rm b} (k) = \tilde\delta_{\rm DM} (k)$ when $k \ll
k_{\rm J}$ and $\tilde\delta_{\rm b} (k) = k^2_{\rm J}
\tilde\delta_{\rm DM} (k) / 
k^2$ when $k \gg k_{\rm J}$. Under some restrictive assumptions (see
Appendix A), it can be shown that
$\tilde\delta_{\rm b} (k) = \tilde\delta_{\rm DM} (k) (1+k^2/k_{\rm 
J}^2)^{-1}$. The quantity $k_{\rm J}^{-1}$ is 
known as the Jeans scale and is defined by:
\begin{equation}
k_{\rm J}^{-1} = \sqrt{\gamma k_B \bar T \over {4\pi a^2 G
\mu\bar\rho_{\rm DM}}}
\, ,
\label{jk}
\end{equation}
where $k_B$ is the Boltzmann constant, $\bar T$ is the spatially averaged
temperature of the gas, $\mu$ is the mean mass per particle (for fully
ionized gas with primordial abundances, it is about $0.6 m_p$
where $m_p$ is the mass of the proton) and $\gamma$ describes the
relation between the temperature $T$ (the actual, not average, value) and
$1+\delta_{\rm b}$: $T \propto (1+\delta_{\rm b})^{\gamma - 1}$. Note
that $\gamma$ does not necessarily equal $5/3$ unless the gas behaves
adiabatically. The proofs of the above 
assertions can be found in the Appendix A (see also Bi et al.  1992;
\cite{peebles1}). It is sufficient to note here that in the linear
regime, the baryon density field is smoother than that of the dark
matter on small scales due to the effect of gas pressure. 

Now, the above relations between $\tilde\delta_{\rm b}$ and
$\tilde\delta_{\rm DM}$ hold only in the linear regime when both
quantities are small. To take into account the effect of gas pressure
in the mildly nonlinear regime, one possibility is to smooth the
initial power spectrum by a factor of $(1+ k^2/k^2_{\rm J})^{-2}$ before
generating the displacement field, similar to
what is done in the case of the truncated Zel'dovich approximation.
This method was used by Reisenegger \& Miralda-Escud\'{e} (1995) to
study the fluctuating Gunn-Peterson effect. In practice, we smooth the
initial power spectrum by a Gaussian kernel $e^{-k^2/k_{\rm J}^2}$ and
find that the two ways of smoothing give very similar column
density distributions. 

To give an idea of scale, for $\gamma = 1.5$, $\bar T = 10^4 \, {\rm K}$
and a universe at critical density, $k_{\rm J}$ is equal to $16.8 \, h\, {\rm
Mpc^{-1}}$.
It turns out that for all models considered in this paper except the
$\Omega_\nu = 0.2$ CHDM models, the truncation scales $k_{\rm S}^{-1}$
according 
to equation (\ref{ks}) are larger than $k_{\rm J}^{-1}$
(eq. [\ref{jk}]), for reasonable 
ranges of temperature and $\gamma$. 
 For these models, it is unnecessary to smooth the 
initial power spectrum over the Jeans scale because the truncated
Zel'dovich approximation already prescribes more smoothing.  The
opposite is true for the
$\Omega_\nu = 0.2$ CHDM models. In fact, the amount of small
scale power is so insignificant for these models that the precise
scale of smoothing does not affect the column density distribution for
column densities 
of interest (Sec.~\ref{mdm}). Orbit-crossing is probably not very
severe for this class of models. Hence, uncertainty in the Jeans
smoothing scale due to uncertainties in the temperature and equation
of state of the intergalactic medium is not a concern.

To sum up, we smooth the initial power spectrum on the scale of
$k_{\rm S}^{-1}$ (eq. [\ref{ks}]) or $k_{\rm J}^{-1}$
(eq. [\ref{jk}]), depending on which is bigger, before it is used to
generate the displacement field (eq. [\ref{Pdef}]) (except for a few test-cases
discussed in Sec.~\ref{slopecd}). The rest of the 
procedure to obtain the overdensity $\delta_b$ and peculiar velocity
$v_{\rm pec}$ on a grid follows as before. 
The implicit assumption underlying 
the whole procedure is that fluctuations on scales smaller than the
smoothing scale do not contribute significantly to the number of
absorption lines at our column densities of interest, about $10^{12.5} -
10^{14.5} \, \,{\rm cm^{-2}}$. The upper limit is related to the
maximum overdensity ($\delta_b$ around $5$) beyond which the
Zel'dovich approximation is not expected to be reliable and the lower
limit is set by our resolution (see Sec.~\ref{dpa} for more
details). 
Note that while the Jeans scale smoothing is meant to capture
the actual smoothing of the density field by gas pressure, the
nonlinear scale smoothing is an approximation technique to avoid the problem
of orbit-crossing. As such, the validity of the latter in the present
application has to be checked. We show in Sec.~\ref{dpa} a comparison between
the column density distribution computed using the approximate method described
here and that using a full hydrodynamic simulation. The level of
agreement lends support to our assumption. Another consistency check
is to see if shock-heating is important for regions with overdensities
(or underdensities) associated with the above range of column
densities.  A plot of density versus temperature like Fig. 2 in
Weinberg et al. (1996) shows that shock-heating, and by extension
orbit-crossing, is not important for regions of underdensity or low
overdensity.

\subsection{The Thermal and Ionization State}
\label{tXrho}

Given the evolution of $\delta_{\rm b}$ predicted by the Zel'dovich
approximation, it is possible to integrate evolution equations for $T$
and for $X_{\rm HI}$ as well as the abundance of other species to
obtain their relations with $\delta_{\rm b}$. \footnote{Since these equations
are local, in the sense that each mass element evolves independently of
the others, there is actually no need to generate a full three-dimensional
realization for the purpose of studying the thermal and
ionization evolution. A simpler approach is to generate a set of
eigenvalues of the deformation matrix $\partial^2\psi/\partial q_i\partial q_j$
according to the prescription of Doroshkevich (1970) and determine
the density evolution through the second part of equation (\ref{zav}).}

Details of the computation will be given in a separate paper (Hui and
Gnedin 1996). A brief discussion can be found in Appendix B of this paper. 
We summarize the main relevant conclusions here.

First, ionization equilibrium is maintained at high
accuracy except during the period of initial reionization. Ionization
equilibrium implies that the neutral hydrogen fraction
(eq. [\ref{Xdef}]) satisfies 
\begin{equation}
X_{\rm HI} \sim 1.6 \times 10^{-6} \left({T\over 10^4 K}\right)^{-0.7}
\left({\Omega_b 
h^2\over 0.0125}\right)  
\left({J_{\rm HI}\over 0.5}\right)^{-1} (1+\delta_{\rm b}) \left({{1+\bar
z}\over 4}\right)^3 \, ,
\label{nh1}
\end{equation}
where we have adopted the approximate form of the 
recombination coefficient of hydrogen: $4.29 \times 10^{-13} (T/10^4 {\rm
K})^{-0.7} {\rm cm^3 s^{-1}}$, which is sufficient for our purpose (see Hui \&
Gnedin 1996 for a more accurate analytical fit).
The quantity $J_{\rm HI}$ is a measure of the radiation intensity defined
as follows (analogous to the
definition in Miralda-Escud\'{e} et al. 1996 but differs by a factor of
$10^{-21} \,{\rm ergs\, Hz^{-1}\, s^{-1}\, cm^{-2}\, ster^{-1}} $):
\begin{equation}
J_{\rm HI} = {{\int_{\nu_{\rm HI}}^{\infty} 4 \pi J_\nu \sigma_{\rm
HI} {d\nu\over {\rm h}\nu}}
\over {\int_{\nu_{\rm HI}}^\infty 4 \pi \sigma_{\rm HI} {d\nu\over
{\rm h}\nu}}}
(10^{-21} \,{\rm ergs\, Hz^{-1}\, s^{-1}\, cm^{-2}\, ster^{-1}})^{-1}\, ,
\label{JHI}
\end{equation}
where $J_\nu$ is the specific intensity as a function of frequency
$\nu$ in the units given above, ${\rm h}$ is the Planck constant,
${\rm h}\nu_{\rm HI}$ is $13.6 \, {\rm eV}$ and $\sigma_{\rm HI}$ is the
ionization cross-section. 
The photoionization rate is simply equal to $4 \times 10^{-12} J_{\rm
HI} \, {\rm s^{-1}}$. 

Observations indicate that $J_{\rm HI}$ is between about $0.1$ and
$2.0$ for $\bar z = 2-4$ (\cite{bdo88}; \cite{lwt91}; \cite{bech94};
\cite{gcdfs96}; \cite{cec97}).  
A perhaps more common way of characterizing the radiation intensity
is to quote its value, often referred to as $J_{912}$, at $\nu =
\nu_{\rm HI}$ or at wavelength $912 
\AA$, in units of $\,{\rm ergs\, Hz^{-1}\, s^{-1}\, cm^{-2}\, ster^{-1}}$. The
relation between $J_{912}$ and $J_{\rm HI}$ depends on the spectrum. A
good approximation for reasonable slopes of the spectrum right above
$\nu_{\rm HI}$ ($J_\nu
\propto \nu^{-m}$ for $m$ between $1$ and $1.5$) is $J_{\rm HI} = 0.7
J_{912}/10^{-21}$. 

Second, we find that
\begin{equation}
T = T_0 (1+\delta_{\rm b})^{\gamma-1} \, ,
\label{tgamma}
\end{equation}
where $T_0$ is not dependent on
position, is a good  approximation for overdensities of interest,
$\delta_{\rm b} \, \approxlt \, 5$, with a little flattening at the low end
($\delta_{\rm b}$ close to $0.1$) for some reionization scenarios.
We will call this our equation of state. Note that this implies that
the spatial dependence of $T$ (and by extension $X_{\rm HI}$) is
completely determined by that of $\delta_b$, which is true for
unshocked gas. Similar power law relations
between the overdensity and the temperature can be seen in Fig. 2 of Weinberg
et al. (1996) for low overdensity. 

Third, both $T_0$ and $\gamma$ depend on the reionization history, the
reasonable ranges being $1.2 < \gamma < 1.7$ and $3000 \, K
< T_0 < 30000 \, {\rm K}$ at $\bar z=3$. It is shown in Hui \& Gnedin (1996)
that $1.3 < \gamma < 1.62$ at $\bar z = 3$ if the universe reionizes before
$\bar z = 5$, assuming uniform radiation field. We allow for a larger range
here. 

Combining equations (\ref{tgamma}) and (\ref{nh1}), it can be seen
that the neutral hydrogen fraction is proportional to
$(1+\delta_b)^{1-0.7(\gamma-1)}$.

In conclusion to Sec.~\ref{ingr}, we have outlined a procedure to use the
Zel'dovich approximation, with appropriate initial smoothing, to
produce a realization of $\delta_{\rm b}$ and $v_{\rm pec}$ as a function of
position, and we have also shown how the relations between $T$, $X_{\rm
HI}$ and $\delta_b$ can be obtained (eq. [\ref{tgamma}] and [\ref{nh1}]).
All of them enter into the calculation of the optical depth $\tau$ (eq.
[\ref{taux}] or [\ref{tau}]). We can
compute $e^{-\tau}$, called the transmission, which is the ratio of
the observed to the emitted intensities. Observationally, its measurement
requires knowledge of the quasar emission spectrum. Moreover, one must
carefully choose the range of frequencies to consider if one is to
limit the source of absorption to that due to the Ly$\alpha$
transition. For a discussion of these issues, 
the reader is referred to Press, Rybicki and Schneider (1993).
To produce a realistic spectrum, one should also add noise and
convolve the transmission with a window function to mimic instrumental
resolution. This is important for a detailed comparison between theories
and observations, which we will defer to latter work. 
Our $x$ space grid cells, depending on the particular simulation,
have sizes ranging from $0.028\,{\rm Mpc}$ to $0.075 \,{\rm Mpc}$
comoving. Note 
that the true resolution in velocity space 
is not uniform because peculiar velocity varies from one place to
another. Without peculiar velocity, the above grid cell sizes
correspond to velocity cells  of $2.8 \, - 7.5 \, \,{\rm km
s^{-1}}$, for $h = 0.5$ at $\bar z = 3$ (eq. [\ref{ux}]). The
true velocity resolution is probably a little worse than that.
As a comparison, high quality Keck Telescope
data have a Full-Width-Half-Maximum of about $7 \,{\rm km\, s^{-1}}$ and 
signal to noise per pixel of the order of $30$ or higher
(\cite{hkcsr95}; \cite{lswt96}).

\section{The Peculiar Velocity: its Effects on Line Shapes and the
Column Density Distribution} 
\label{pvels}

We show in this section that while the peculiar velocity plays an
important role in determining the absorption profiles, its effect on
the column density distribution is minor. The procedures to obtain
the column density distribution are discussed first.

\subsection{Line Identification and the Column Density Distribution}
\label{licdd}



Fig.~\ref{delvtrans7_21_50PCL1_52} and
Fig.~\ref{delvtrans7_21_50PCL1_6} show the velocity, density and transmission
($e^{-\tau}$) along two lines of sight for a $\sigma_8 = 0.7$ CDM
simulation, with $h = 0.5$ (see Table \ref{cdmmodels}). 
The significance of the dashed 
transmission profile will be explained in the next sub-section.
The thermal and ionization
parameters are described in the caption of
Fig.~\ref{delvtrans7_21_50PCL1_52}. The truncation scale
$k_{\rm S}$ (eq. [\ref{ks}]) is $2.3 \, \,{\rm Mpc^{-1}}$. 
The transfer function is taken from Ma
(1996). We find that using instead the transfer function of Bardeen et
al. (1986) makes almost no difference to the resulting column
density distribution, for the range of column densities considered.

The first thing to note is that for the given parameters, 
\begin{equation}
b \equiv \sqrt{2 k_B T_0\over {m_p}} = 13 \, \,{\rm km s^{-1}}
(1+\delta_{\rm b})^{1/4} \, .
\label{b13}
\end{equation}
This might seem to be too small because the observed lower limit of the
$b$-value is about $15 - 20\, \,{\rm km s^{-1}}$ (\cite{hkcsr95};
\cite{lswt96}). There are two points to be made. First, $T_0$ is the
temperature of the gas at $\delta_b = 0$. The observed $b$-values might
originate from regions at higher $\delta_b$. It is true though that
the $1/4$ power of  
$1+\delta_b$ does not help very much. Second, a
distinction should be made between the observed $b$-value and the
$b$ defined above. The observed $b$-value is obtained by fitting
the quasar spectrum with superpositions of the Voigt profiles. Each
Voigt profile yields a column density and a $b$-value. All the
density peaks that give rise to absorption troughs in 
Fig.~\ref{delvtrans7_21_50PCL1_52} and
Fig.~\ref{delvtrans7_21_50PCL1_6} have velocity widths larger than or
comparable to the small thermal width defined in equation
(\ref{b13}). Therefore the 
{\rm narrow-maximum-limit} (eq. [\ref{cloud}]) does not apply and 
the absorption troughs do not exactly have the Voigt profile shapes. 
The $b$-value obtained from the best-fit Voigt profile of a given
absorption trough does not
necessarily correspond to the thermal width in equation (\ref{b13}). 
It should also be emphasized that the recent hydrodynamic simulations of the
Ly$\alpha$ forest, which have been so successful in accounting for a
lot of its observed properties, have similarly low temperatures (see
for instance Weinberg et al. 1996 \footnote{Their output is at redshift
of $2$ and so naturally they have a lower temperature. In general, the
temperature $T_0$ is dependent on the 
reionization history of the universe: crudely speaking, the closer the
epoch of reionization is to the epoch of observation ($\bar z = 3$ in
our case), the higher the temperature. Assuming reionization occurs
before a redshift of $5$, say, puts an upper bound on $T_0$ (Hui and
Gnedin 1996).}).

One might wonder if there exists an alternative spectrum reduction
method where the Voigt profile is not assumed to be the fundamental
shape of absorption 
troughs, and for such a method, how the column density is assigned to each
trough. The Voigt-profile-fitting-technique is nonetheless very important
because it is how all existing observational data on the column
density distribution are obtained. 

An alternative line identification algorithm was proposed by
Miralda-Escud\'{e} et al. (1996) and was also used by Hernquist et
al. (1995).  First a transmission ($e^{-\tau}$) threshold 
is chosen. Any 
part of the  
spectrum that is continuously below the threshold is identified as an
absorption line. The column density associated with it is defined by
\begin{equation}
N_{\rm HI} \equiv {1\over \sigma_{\alpha 0}} \int_{line} \tau(u_\circ) {du_\circ \over c} \, ,
\label{tcd}
\end{equation}
where $\sigma_{\alpha 0}$ is defined after equation (\ref{tau}). The limits of 
integration are taken to be over the absorption line, i.e. where the
transmission is continuously below the threshold. 
Note that if the {\rm narrow-maximum-limit} or the thin cloud assumption were
to hold, equation (\ref{cloud}) can be substituted into 
equation (\ref{tcd}) to show that $N_{\rm HI}$ does correspond to $\int
n_{\rm HI} dx/(1+z)$, assuming the threshold is high enough so that most
of the Voigt profile is included in the definition of the absorption
line.

Let us call the above procedure the Threshold-Algorithm.
We show in Fig.~\ref{cd7_21_50PCL1} the column density distribution
computed according to the algorithm (Crosses).
The symbol $d^2 N_{{\rm Ly}\alpha} /d N_{\rm HI}/dz$ denotes the number of
absorption lines per unit column density per unit redshift. 
The reason for the chosen range of column densities will be
given in Sec.~\ref{dpa}. The transmission is chosen to be at the mean
transmission of $0.89$.


The Threshold-Algorithm has the tendency
to underestimate the number of absorption lines compared to the
Voigt-profile-fitting-technique. One reason is that it does not deblend. In
other words, a given absorption line according to the
Threshold-Algorithm may contain more than one minimum in
transmission. Such an absorption line would be broken up to a few
lines if the Voigt-profile-fitting-technique is employed. To demonstrate this
effect, we modify the Threshold-Algorithm: for each
(parent) absorption line identified, we break it up into individual
components (children) where each component is bordered by local maxima
in the transmission within the confines of the parent. 
The column density for each child component is defined similarly as in
equation (\ref{tcd}) and the limits of integration are taken to be
the boundaries of each component. We will call it the
Threshold-Deblending-Algorithm. 

The resulting column density
distribution is denoted by the square symbols in
Fig.~\ref{cd7_21_50PCL1} for the transmission threshold of
$0.89$. One can see that indeed the number of lines of low column
densities go up. We should emphasize however that the
Threshold-Deblending-Algorithm cannot be used to analyze observational
data without modifications because in real life, noise creates local
transmission 
maxima within any parent absorption line.\footnote{In fact, numerical noise can
also have the same effect. We check it by defining local maxima in two
ways: local maxima over three cells and local maxima over five cells
with the slope on either side of the maxima not changing signs. It
turns out the resulting column density distributions are almost the
same.} 

For now, the Threshold-Algorithm is adopted as a simple way to
identify lines and assign column densities, which we will use to 
study the effects of the peculiar velocity on the column density
distribution.

\subsection{The Role of Peculiar Velocities}
\label{v}

\begin{figure}
\plotone{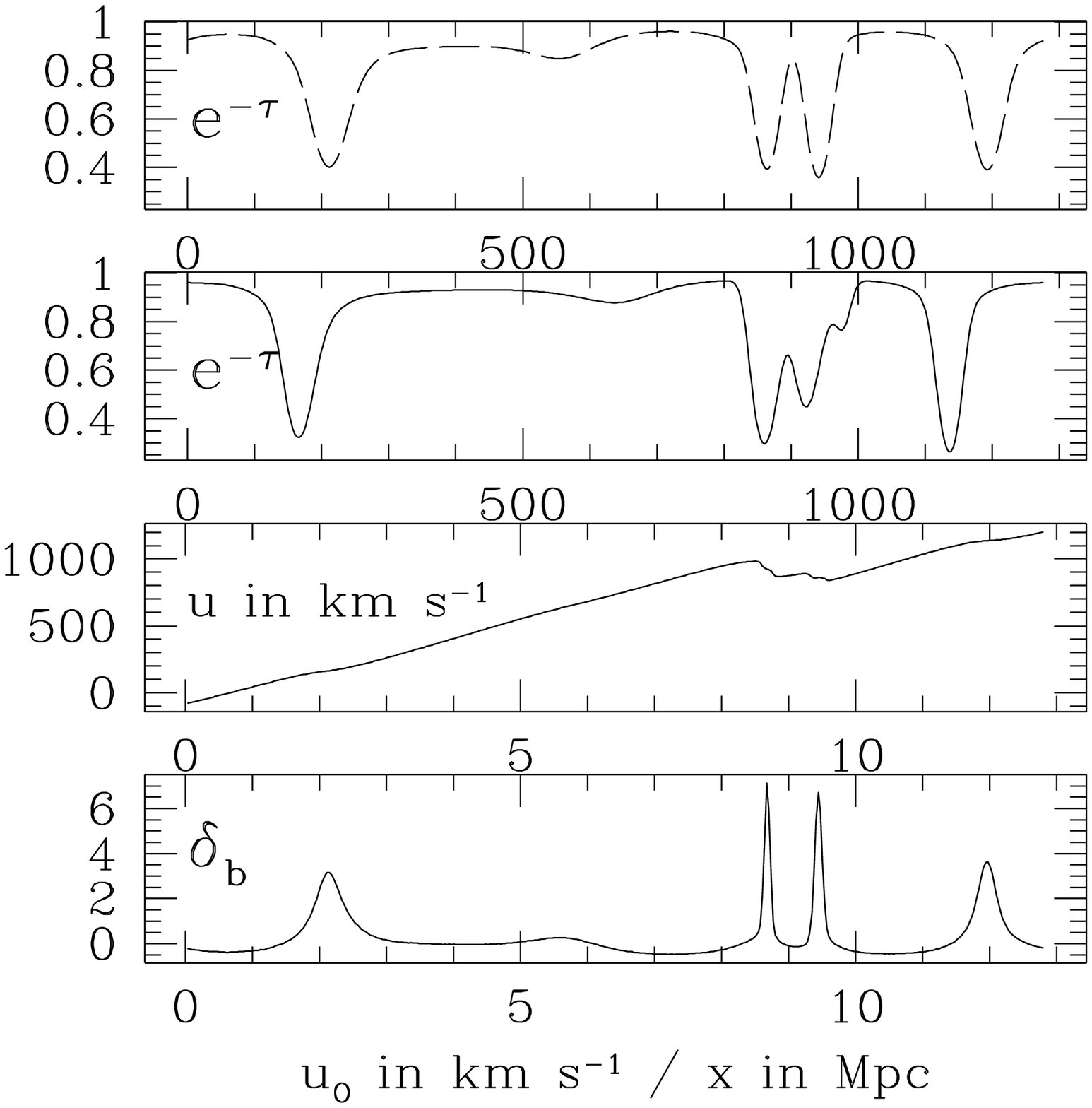}
\caption{\label{delvtrans7_21_50PCL1_52}\capdelvtrans} 
\end{figure}

\begin{figure}
\plotone{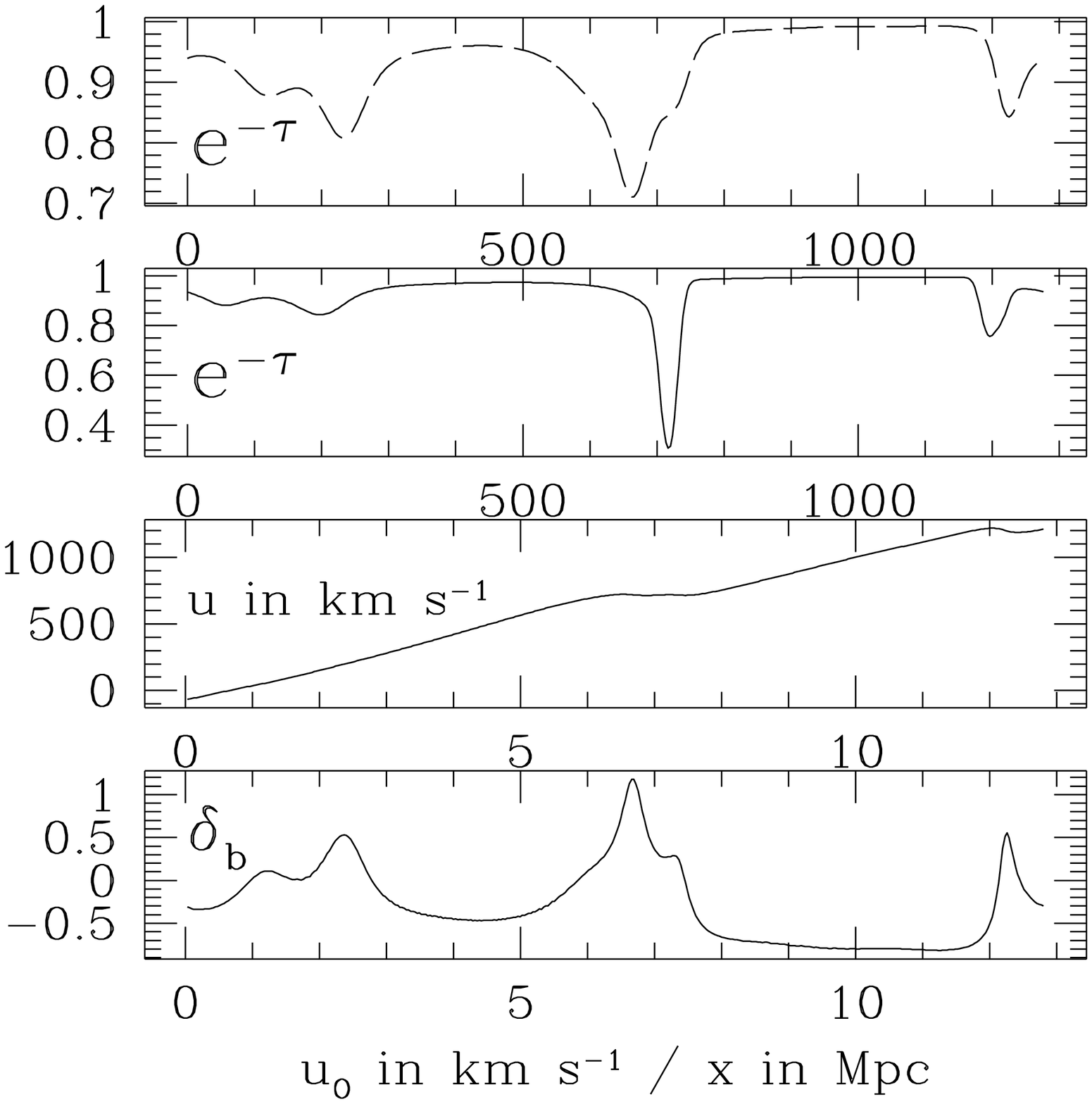}
\caption{\label{delvtrans7_21_50PCL1_6}\capdelvtransix}
\end{figure}

\begin{figure}
\plotone{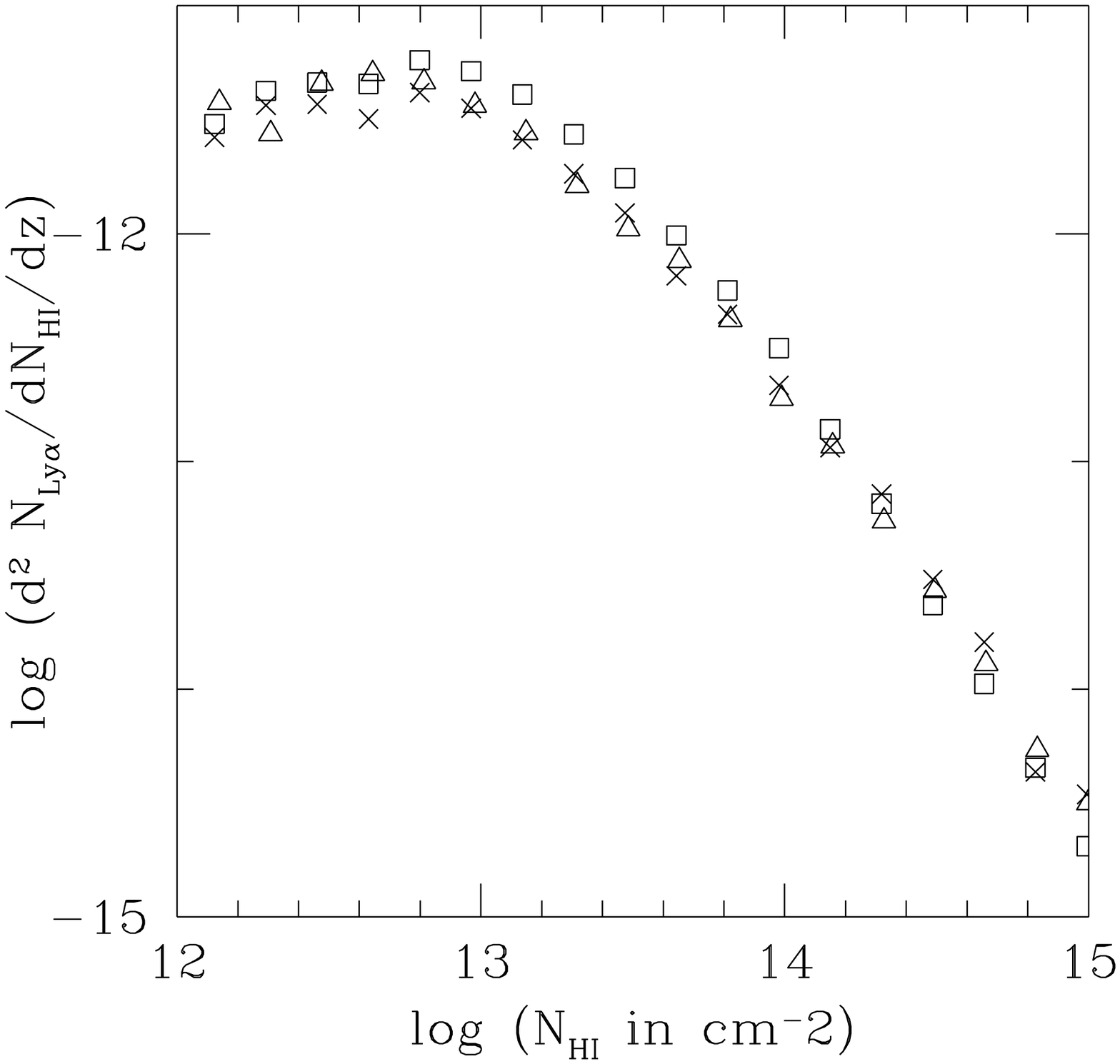}
\caption{\label{cd7_21_50PCL1}\capcda}
\end{figure}

The following experiment is performed to investigate the importance of
peculiar velocities. We generate absorption spectra and compute the
column density distribution using the same density field as that used
to produce the solid curves in Fig.~\ref{delvtrans7_21_50PCL1_52},
Fig.~\ref{delvtrans7_21_50PCL1_6} and the crosses in
Fig.~\ref{cd7_21_50PCL1} but we set all peculiar velocities to zero.

Let us first examine some examples of the absorption spectra.
The dashed curves in Fig.~\ref{delvtrans7_21_50PCL1_52} and 
Fig.~\ref{delvtrans7_21_50PCL1_6} are the resulting spectra after
putting all peculiar velocities to zero. 

A comparison of the dashed absorption spectrum with its solid
counterpart in each figure  shows that the
peculiar velocities play an important role in determining the shapes of
absorption lines.
Without peculiar velocities, the shapes of absorption troughs
mirror closely (perhaps a little thermally-broadened compared to)
those of the density peaks while with nonzero peculiar 
velocities, the absorption troughs can have quite different shapes 
from the underlying density field. Peculiar velocities can add or
erase structures.
An example of the former can be found in the pair of density peaks
around $x = 9 \, {\rm Mpc}$ and their corresponding  absorption
profiles in Fig.~\ref{delvtrans7_21_50PCL1_52}.
An example of the latter can be found in 
the density peak(s) around $x = 7 \, {\rm Mpc}$ and the corresponding
absorption 
trough(s) in Fig.~\ref{delvtrans7_21_50PCL1_6}. 

Broadly speaking, the effects of peculiar
velocities on absorption spectra fall into three
categories. They are distinguished by the value of $du/dx$ ($u$ is
defined in eq. [\ref{ux}]). 
First, there are regions in space where the peculiar velocity gradient
is small so that $du/dx$ is almost equal to its Hubble value $\bar
H/(1+\bar z)$ (eq. [\ref{ux}]). An example is the density peak
around $x = 2.2 \, {\rm Mpc}$ in Fig.~\ref{delvtrans7_21_50PCL1_6}. The
peculiar velocity shifts the position of the associated absorption
trough but does not affect its shape. 

Second, there are places where
the peculiar velocity gradient is opposite in sign and comparable in
magnitude to $\bar H/(1+\bar z)$, in which case $|du/dx|$ becomes very
small. Suppose also that $|d^2 u/d x^2|$ is small. The implication is
that  a small range in $u$ corresponds to a 
relatively large range in $x$. See for instance the density peak(s)
around $x = 7 \, {\rm Mpc}$ in Fig.~\ref{delvtrans7_21_50PCL1_6}, which is a
really broad structure in $x$ space but is relatively narrow in $u$
space if peculiar velocities are not put to zero. The small $|du/dx|$ or
the converging peculiar velocity flow around it 
helps to produce a narrow absorption trough
(second panel from the top in
Fig.~\ref{delvtrans7_21_50PCL1_6}). Contrast it with the 
corresponding absorption feature in the top panel of the same figure,
where peculiar velocities are set to zero.
The limiting case where $|du/dx|$ exactly vanishes is called a velocity
caustic (\cite{mcgill90}).

Third, there are regions where the peculiar velocity gradient
dominates in such a way that $du/dx$ is negative and $|du/dx|$ is not small.
An example can be found around the pair of density peaks at
$x = 9 \, {\rm Mpc}$ in Fig.~\ref{delvtrans7_21_50PCL1_52}. This is where
multiple-streaming occurs. A given range in $u$ corresponds to
disjoint pieces in 
$x$ space. As a result, the shapes of the associated absorption troughs
are significantly different from those of the underlying density peaks.

The three categories can be shown to correspond to the three evolutionary
stages of a pancake collapsing along the line of sight (McGill 1990).
Restricting equations (\ref{zad}) and (\ref{zav}) to one dimension and
putting it into 
equation (\ref{ux}), one obtains:
\begin{equation}
{du\over dx} = \dot a \left({1 + 2 a {d^2\psi\over dq^2}}\right) \left({1 + a
{d^2\psi\over dq^2}}\right)^{-1} \, ,
\label{dudx}
\end{equation}
where we have assumed a universe of critical matter density so that
$D_+ = a$. Restricting 
equation (\ref{zav}) to one dimension, it can also be shown that 
\begin{equation}
1 + \delta_{\rm b} = \left({1 + a{d^2\psi\over dq^2}}\right)^{-1} \, ,
\label{deltaone}
\end{equation}
where we have replaced $\delta_{\rm DM}$ by $\delta_{\rm b}$ assuming
the appropriate initial smoothing has been carried out as indicated in
Sec.~\ref{zasection}.
As $a$ grows, it can be seen that $du/dx$ goes through the three
different regimes outlined above for negative $d^2\psi/dq^2$. At the velocity caustic where $du/dx
= 0$, it can be shown that $\delta_{\rm b} = 1$ (\cite{mcgill90}). 
This conclusion does not hold in general of course because pancakes
can collapse in directions different from the line of sight. 
But it is true that velocity caustics are often found in regions of slight
overdensities.  

In principle, at a velocity caustic, an absorption line can arise
even without any variation in the density field at all (eq. [\ref{tau}]). 
In practice, one expects that converging peculiar velocity flows are
accompanied by density peaks.  This is consistent with the few
examples we have seen. 

Next, we consider how the column density distribution changes when the
same density field is used but all the peculiar velocities are put to
zero. This is shown in Fig.~\ref{cd7_21_50PCL1}.


The mean transmission of the spectra computed with zero peculiar velocities
differs from the mean transmission of the full spectra by less than a
percent. It is used as the transmission threshold in the line
identification procedure for both analyzes.
The resulting column density distributions are very similar. 

Hence, the peculiar velocity plays a relatively minor role in determining
the column density distribution. It changes the
shapes of absorption troughs without altering the overall 
number of lines and their column densities. This serves to motivate an
approximation we will introduce in the next section.

A final note on velocity caustics. The reader might worry that at a
velocity caustic, the optical 
depth may diverge while it is clear from equation (\ref{taux}) that
for a finite number density of neutral hydrogen, the optical depth
should always be a finite quantity. The resolution is that close to a
velocity caustic at $u = u_c$, $du/dx$ goes like $|u-u_c|^{1/2}$ (provided
the second derivative of $u$ with respect to $x$ is nonzero, otherwise
it will be $(u-u_c)^{2/3}$ if the third derivative does not vanish and so
on, by simple Taylor series expansion; see Shandarin and Zel'dovich
1989 for a similar analysis applied to real caustics as opposed to
velocity caustics; orbit-crossing occurs  in the former but not
the later). 
So under the integration in equation (\ref{tau}), the optical 
depth remains finite. We note also that because of the singular nature of 
${(du/dx)}^{-1}$ around $u = u_c$, the derivation leading to
equation (\ref{noncloud}) breaks down at a velocity caustic.

\section{The Statistics of Density Peaks}
\label{dpa}

In this section we explore a simple approximation in which each
density peak in $x$ space is identified with an absorption line. 
This is motivated by the fact that peculiar velocities do not play a
major role in determining the column density distribution and that
each maximum in density corresponds to a minimum in transmission or
maximum in absorption
if the peculiar velocities are set to zero and if the maximum in
density is separated from other maxima by a distance larger than that
given by the thermal broadening width.

To calculate $d^2 N_{{\rm Ly}\alpha}/d N_{\rm HI}/dz$, we relate $dz$ and $dx$
by ignoring peculiar velocities: $dz = c^{-1} \bar H dx$. Hence
\begin{equation}
{d^2 N_{{\rm Ly}\alpha}\over {dN_{\rm HI} dz}} = {d^2 N_{\rm pk}\over {dN_{\rm HI} dx}}
{c\over\bar H} \, ,
\label{NLyNpk}
\end{equation}
where $dN_{\rm pk}/dN_{\rm HI}/dx$ is the average comoving number
density of density 
peaks along a random line of sight per unit column density, $\bar H$
is the Hubble constant at the 
redshift of interest.  

For each density peak, we need a simple prescription for assigning a
column density.  
To that end, we perform the following expansion around each density maximum:
\begin{equation}
\ln [n_{\rm HI}(x)] = \ln [n_{\rm HI}(x_{\rm pk})] + {1\over 2}
\left.{d^2 \ln [n_{\rm HI}] \over 
dx^2}\right|_{x=x_{\rm pk}} (x-x_{\rm pk})^2 \, .
\label{npk1}
\end{equation}
It is a straightforward Taylor expansion around the position of the
peak $x_{\rm pk}$. The second derivative in the last term is negative.
The rationale behind expanding $\ln [n_{\rm HI}]$ rather than $n_{\rm
HI}$ itself is 
that $n_{\rm HI}$ is supposed to fall off quickly far away from the peak
(until, of course, another peak is encountered). In
other words, the above expansion implies that $n_{\rm HI}$ has a Gaussian
fall-off (instead of a power-law one if $n_{\rm HI}$ itself were Taylor
expanded). In a sense, this is close in spirit to the
Voigt-profile-fitting-technique. Suppose the broad-maximum-limit
(eq. [\ref{noncloud}]) holds
so that the local optical depth is simply proportional to the number
density of neutral hydrogen if one ignores peculiar velocities. Then,
fitting a minimum in optical depth with the Voigt or thermal
profile (eq. [\ref{tau}]) is equivalent to fitting the
corresponding neutral hydrogen density peak with a Gaussian. 

We then assign the following column density to the density peak:
\begin{equation}
N_{\rm HI} = \int_{\rm pk} {dx\over {1+\bar z}} n_{\rm HI}(x) =
\left.{n_{\rm HI}(x_{\rm pk}) 
(1+\bar z)^{-1} 
\sqrt{2\pi \left({- d^2 \ln [n_{\rm HI}] \over
dx^2}\right)^{-1}}}\right|_{x=x_{\rm pk}} \,
\label{npk2}
\end{equation}
where equation (\ref{npk1}) has been
used and where $\int_{\rm pk}$ denotes integration around the peak until it
decays sufficiently. All $x$ dependent terms on the right hand side
are evaluated at $x = x_{\rm pk}$. The above equation is also derived 
by the authors in a separate paper, using the Stationary Phase Method
(Gnedin and Hui 1996).

Using equations (\ref{nh1}) and (\ref{tgamma}), the above can be
rewritten as:
\begin{eqnarray}
\label{npk3}
N_{\rm HI} = && 1.63 \times 10^{13} {\rm cm^{-2}} \left({T_0\over 10^4
K}\right)^{-0.7} \left({\Omega_b 
h^2\over 0.0125}\right)^2 
\left({J_{\rm HI}\over 0.5}\right)^{-1} \left({1+\bar z\over 4}\right)^{5}
\\ \nonumber
&&
\left({{2-0.7(\gamma-1)}\over 1.65}\right)^{-0.5} A \,
\end{eqnarray}
where $A$ is defined by
\begin{equation}
A \equiv \left.{(1+\delta_{\rm b})^{2-0.7(\gamma-1)} \left({- d^2 \ln
[1+\delta_{\rm b}] \over 
{dx^2}}\right)^{-{1\over 2}} }\right|_{x = x_{\rm pk}} \, ,
\label{g}
\end{equation}
with $x$ being measured in ${\rm Mpc}$.

We will refer to our method as the Density-Peak-Ans\"{a}tz. It
consists of two parts: 1. associate each density peak in $x$
space with an absorption line; 2. assign a column density to each density
peak according to equation (\ref{npk3}).\footnote{Strictly speaking,
care should be taken not to count peaks that are separated in velocity
space by distance much less than the thermal width as contributing to
more than one absorption line. We will address
this later in the section.}

Making use of equations (\ref{NLyNpk}) and (\ref{npk3}), the column
density distribution can be written as 
\begin{eqnarray}
{d^2 N_{{\rm Ly}\alpha}\over {dN_{\rm HI} dz}} = && 6.25 \times 10^{-14}
\, {\rm cm}^{2}  
\left({T_0\over 10^4 K}\right)^{0.7} \left({\Omega_b 
h^2\over 0.0125}\right)^{-2}
\left({J_{\rm HI}\over 0.5}\right) \left({1+\bar z\over 4}\right)^{-5}
\\ \nonumber
&& \left[{{2-0.7(\gamma-1)}\over 
1.65}\right]^{0.5} {c\over \bar H} {d^2
N_{\rm pk} \over {dA
dx}}  \, ,
\label{goodie}
\end{eqnarray}
where $x$ and $c \bar H^{-1}$ are in the same unit of ${\rm Mpc}$.
Most of the factors above come from the scaling between $A$ and $N_{\rm HI}$,
and $c/\bar H$ provides the conversion from comoving 
coordinate $x$ to redshift $z$. The last factor $d^2 N_{\rm pk}/dA/dx$
involves is the number density of peaks in $x$ space having the quantity $A$
within the range $dA$. 

Let us define $\xi \equiv \ln [1+\delta_{\rm b}]$. Suppose one is given 
$P(\xi,\xi',\xi'') d\xi d\xi' d\xi''$ which is the 
probability that $\xi$ and its first and second derivatives with
respect to $x$ fall in the specified ranges at a point. Then,
\begin{equation}
{dN_{\rm pk} \over dx} = \int_{-\infty}^{\infty} d\xi \int_{-\infty}^{0}
d\xi'' |\xi''| P(\xi,\xi'=0,\xi'') \, ,
\label{totalnumber}
\end{equation}
where $dN_{\rm pk}\over dx$ is the integral of ${d^2 N_{\rm pk} \over {dA
dx}}$ over all $A$ (Bardeen et al. 1986).

By a change of variable and a differentiation and making use of
equation (\ref{g}), one can obtain
\begin{equation}
{d^2 N_{\rm pk} \over {dA dx}} = {1\over [2-0.7(\gamma-1)]A} \int_{-\infty}^{0}
d\xi'' |\xi''| P(\xi,\xi'=0,\xi'') \,
\label{differential}
\end{equation}
where $\xi$ should be expressed in terms of $\xi''$ and $A$ using
equation (\ref{g}).

Note that the above two equations are completely general and 
no assumption about the Gaussianity of the underlying fields has been made.
The hard part is of course to come up with the probability function
$P$. The one point probability distribution of just $\xi$ or density
has been calculated for the Zel'dovich approximation
(\cite{kbgnd94}). We find the one point joint-probability distribution
of density and its first and second derivatives along a  line of
sight difficult to calculate 
analytically for the Zel'dovich approximation. 
A numerical approach is adopted in this paper and the number of peaks
is counted along random lines of sight in actual three-dimensional
realizations. 
In a separate paper, we discuss an analytical calculation based
upon not the Zel'dovich approximation but the lognormal approximation,
where $\xi$ 
is assumed to be a Gaussian random field (Gnedin and Hui 1996).
A comparison between the two will be made here.


\subsection{Testing the Density-Peak-Ans\"{a}tz}
\label{testing}

\begin{figure}
\plotone{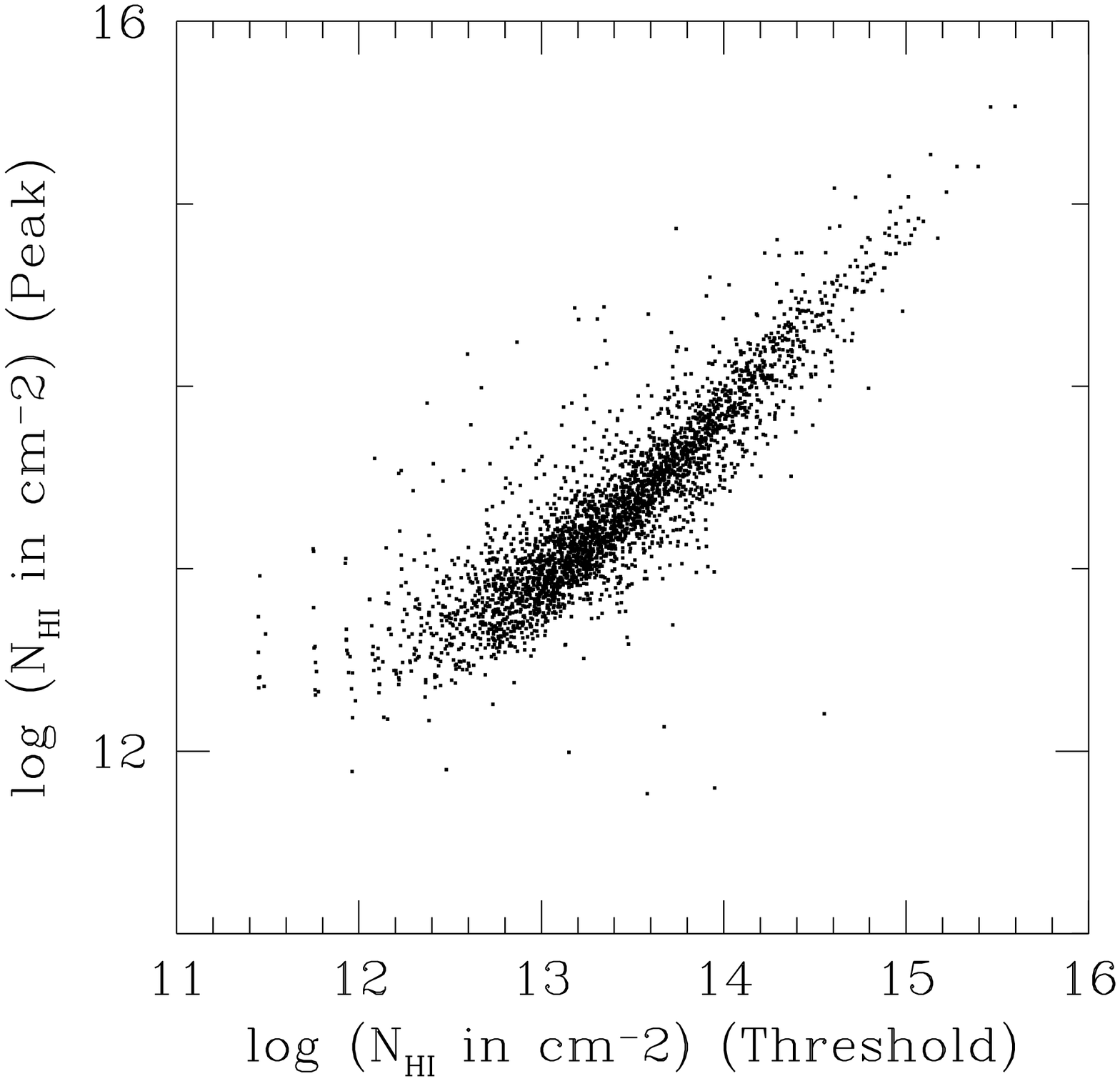}
\caption[Comparison of density-peak-ans\"{a}tz together with the
Zel'dovich approximation with a hydrodynamic
simulation.]{\label{modcdeblend7_21_50PCL1}\capmodcd} 
\end{figure}

We test the Density-Peak-Ans\"{a}tz in two different ways.
First, we make a scatter plot of the column density
obtained using the Threshold-Deblending-Algorithm versus the column
density obtained by searching for the maximum density peak that
contributes to each absorption line identified using the threshold
method and then applying equation
(\ref{npk3}). The result is shown in Fig.~\ref{modcdeblend7_21_50PCL1}.
It shows that while the agreement is far from perfect, the column
densities assigned using the Density-Peak-Ans\"{a}tz and using the
Threshold-Algorithm are broadly consistent.

\begin{figure}
\plotone{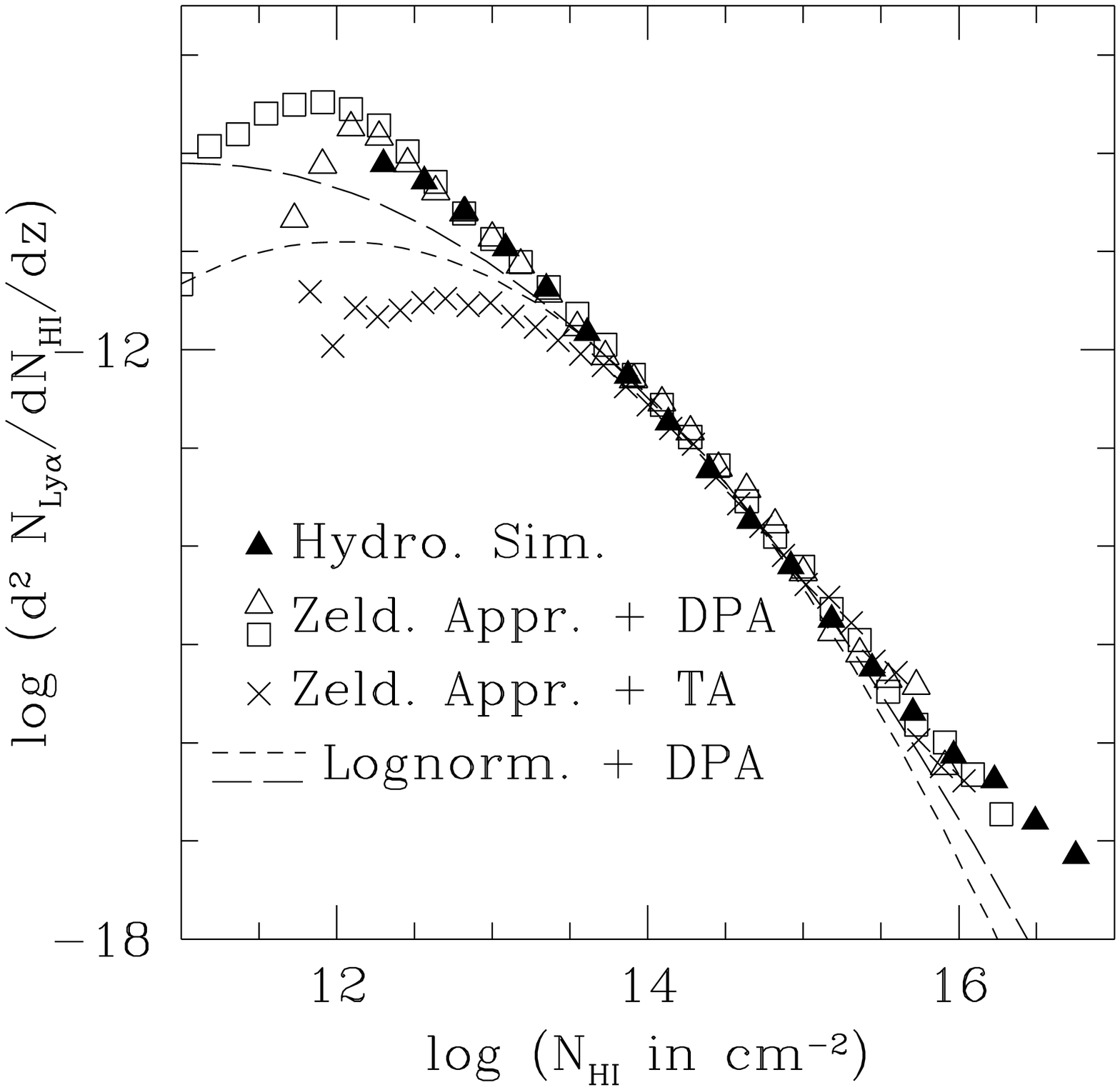}
\caption{\label{zhang}\capzhang}
\end{figure}

The important question, however, is whether the
Density-Peak-Ans\"{a}tz, coupled with the Zel'dovich
approximation with appropriate initial smoothing, gives the correct
number of absorption lines as a function of column density.  
We compare the column density distribution obtained using our
approximate methods against that obtained by applying the 
Voigt-profile-fitting-technique to synthetic spectra from a full
hydrodynamic simulation (see discussion below and Zhang et al. 1996 for
details). This is done in Fig.~\ref{zhang}. Note that the 
values of $\gamma$ and $T_0$ given in the caption of Fig.~\ref{zhang} is
obtained directly from the hydrodynamic simulation. The temperature-density
relation is not an exact power law but is well approximated by one for
the relevant range of densities. We also put in the same figure the column
density distribution obtained using the Threshold-Algorithm, coupled with the
Zel'dovich approximation. The predictions of the lognormal approximation are
shown as well for comparison.


The level of agreement between the exact hydrodynamic
computation and our calculation based on the Density-Peak-Ans\"{a}tz
coupled with the Zel'dovich approximation is encouraging. 
Two sets of points are shown for our approximate calculation 
using the Density-Peak-Ans\"{a}tz, one (open triangles) with exactly
the same box size and grid spacing as the hydrodynamic simulation and
the other (open squares) with larger box size and smaller
grid spacing. They both agree very well with the exact computation.
We will explore the effects of changing the resolution in the next
sub-section. A third set of points (crosses) shows that the Threshold-Algorithm
described in Sec.~\ref{v} underestimates the number of lines at low
column densities. 

The agreement between the results of the hydrodynamic simulation
and our approximation method is perhaps telling us something
interesting about the low column density systems. Note that
for the Zel'dovich approximation computations presented in Fig.~\ref{zhang},
an initial smoothing scale ($k_{\rm S} = 2.3 \, {\rm Mpc^{-1}}$) larger than
the 
Jeans length ($k_J = 8.4 \, {\rm Mpc^{-1}}$) was chosen to 
deal with the problem of orbit-crossing. The agreement can then be
understood as a result of relatively little contribution to the low column
density systems ($10^{12.5} - 10^{14.5} \, {\rm cm^{-2}}$) from
structure on small length scales suppressed by our 
$k_{\rm S}^{-1}$ smoothing even though those scales could be
larger than the Jeans length. 
The low column density systems mostly consist of relatively broad density
peaks. 

Some explanation is also in order regarding the spectral analysis method used
by Zhang et al. (1996) in obtaining the distribution shown in Fig.~\ref{zhang}.
It consists of identifying absorption features above a specified opacity
and deblending them into individual lines centered at local
maxima in optical depth and fitting each with a suitable Voigt profile. 
This procedure is designed to be similar to most observers' analysis
techniques but is not exactly the same. In particular most observers' method
consists of seeking a ``minimal'' set of Voigt profiles, the superposition of
which reproduces the input absorption spectrum to within some specified error
consistent 
with noise. It should be noted, however, that the definition of 
``minimal'' tends to vary from one observer to another.
The superposition of Voigt profiles identified using the
method of Zhang et al. seems to reproduce the input spectrum quite well (see
Fig. 2 
of Zhang et al. 1996), although the agreement has not been carefully
quantified.
One should keep in mind possible differences
in the column density distributions obtained using different methods (see
Dav\'e et al. 1997 for related discussions).

We also show in Fig.~\ref{zhang} two sets of curves based on the
lognormal approximation but using the same Density-Peak-Ans\"{a}tz
(see Gnedin and Hui 1996). One of them has the same amount of
initial smoothing as that of the truncated Zel'dovich approximation
and the other has less smoothing so as to match the 
final (not linear) rms density fluctuation of the Zel'dovich
computation. In both cases, the lognormal approximation tends to predict too
much flattening of 
the column density distribution at low column densities. In general
the lognormal approximation tends to give column density distributions
that deviate quite significantly from power-law unless a very small
smoothing length $k_{\rm S}^{-1}$ is chosen.
(From the figure, it might appear that the lognormal
approximation gives more lines than the Zel'dovich
approximation at the very low column densities but
it is really a resolution effect: see the next sub-section.)

The reader might have noticed that we have included in
Fig.~\ref{zhang} a wider range of column densities 
than is warranted by the nature of our approximations. For instance,
objects with column densities higher than $10^{16} \,{\rm cm^{-2}}$ are
almost certainly highly nonlinear and we do not expect the truncated
Zel'dovich approximation to work well in this regime. For the low
column densities, the finite resolution should cause us to
underestimate the number of absorption lines. In the next sub-section,
we give quantitative estimates of the range of column densities within
which the Density-Peak-Ans\"{a}tz, used in conjunction with the truncated
Zel'dovich approximation, can be counted upon to give reliable 
column density distributions. 

We will also discuss two different ways of defining a density peak in
the next sub-section: local maxima over three cells or local maxima
over five cells with the slope on either side of the maxima not
changing signs. The three-cell criterion is used in Fig.~\ref{zhang}. 
One expects however that some of the three-cell peaks are not real but
merely artifacts of numerical noise, especially those with low column
densities. The five-cell criterion, on the other hand, probably fails
to include some narrow peaks which are real.
We will see that the two different criteria give almost identical results
above a certain column density.

One aspect of the Density-Peak-Ans\"{a}tz we have glossed over is that
two density peaks that are separated by a distance in velocity space
much less than the thermal width should be counted as contributing
to not two but one absorption line.
A more sophisticated approach would be to group together such
density peaks and use the sum of their column densities as the
column density of one single absorption line. We find that for the 
the range of validity discussed in the following sub-section, it makes
little difference. It is conceivable, however, that this effect
cannot be ignored for simulations with higher resolution than we have, or at
higher redshifts, where line blending is more important.

\subsection{The Range of Validity}
\label{rov}

For the computation presented in Fig.~\ref{zhang}, 
the column density (given by the Density-Peak-Ans\"{a}tz) above
which the mean $\delta_{\rm b}$ exceeds $5$ is about 
$10^{14.1} \,{\rm cm^{-2}}$. For the parameters listed in the caption of
Fig.~\ref{zhang}, $N_{\rm HI} = 3.6 \times 10^{13} \,  A \,\,{\rm
cm^{-2}}$ (eq. 
[\ref{npk3}]). We therefore take $A = 3.5$ as an upper
limit beyond which we cannot expect our approximations to be
reliable. Shock-heating should be relatively unimportant for $\delta_{\rm b}$
less than about $5$ (see Hui \& Gnedin 1996).

Note that according to Fig.~\ref{zhang}, comparing with the
hydrodynamic simulation data, the Density-Peak-Ans\"{a}tz
coupled with the truncated Zel'dovich approximation, seems to give a
reliable number density of absorption lines even for column densities 
higher than $10^{14.1} {\rm cm^{-2}}$. The level of agreement at such high
column densities (and by extension, such high $\delta_{\rm b}$) is 
surprising. We will adopt the conservative upper limit of $A = 3.5$. 

\begin{figure}
\plotone{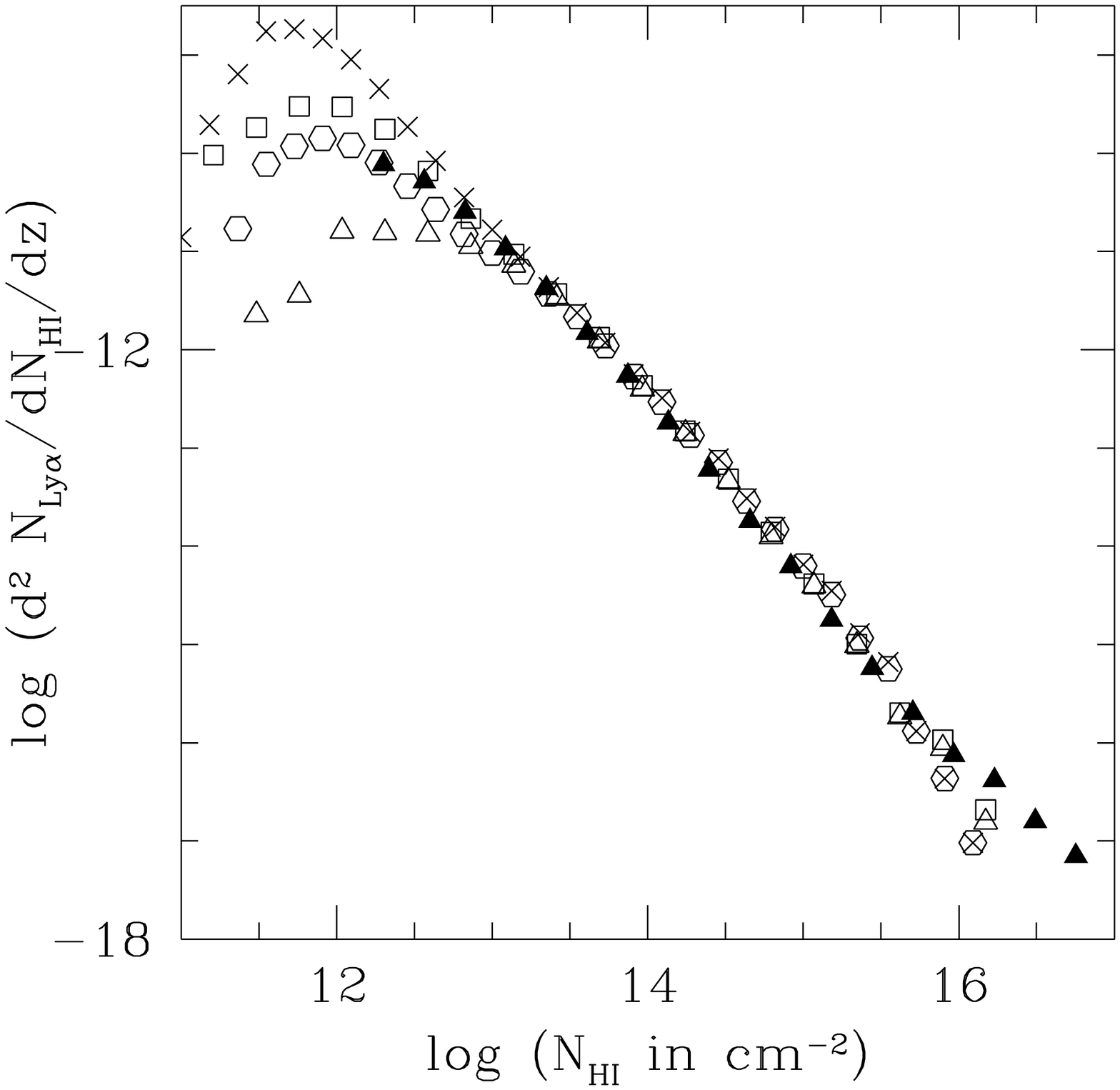}
\caption{\label{resolution}\capresolution}
\end{figure}

To determine the column density below which finite resolution results
in an underestimate of the number of absorption lines, we perform a
simulation using the truncated Zel'dovich approximation with the same
parameters as the open squares in Fig.~\ref{zhang} but higher
resolution: same box size of $12.8 \, \,{\rm Mpc}$ but smaller grid
spacing of $0.0284 \, \,{\rm Mpc}$. A comparison of the 
resulting column density distributions is made in Fig.~\ref{resolution}.


Note that we have included two definitions of density peaks (three-cell
and five-cell). For each simulation, the true column density
distribution is probably somewhere between the two in the places they
differ. 

We take the low column density cut-off to be $10^{12.8} \,{\rm cm^{-2}}$
for the lower resolution simulation (box size of $12.8 \, {\rm Mpc}$ ,
with grid spacing of 
$0.05 \, \,{\rm Mpc}$) using the three-cell definition of peaks.
It can be seen that the higher
resolution simulation differs from the lower one only at column
densities less than roughly this cut-off value.
Moreover, above this column density, the three-cell and
five-cell criteria give almost identical results. 

The parameters in the simulations in Fig.~\ref{resolution} are such
that $N_{\rm HI} = 3.6 
\times 10^{13} \, A \,{\rm cm^{-2}}$ (eq. [\ref{npk3}]). Hence the above
column density cut-off implies a lower limit of $0.18$ for $A$. 
From now on, we will use the three-cell definition of
density peaks.

For readers interested in applying our formalism at different redshifts,
we recommend that they choose the appropriate range of $A$ 
using the same methods above: upper limit set by non-linearity
and lower limit set by running simulations of varying resolutions.
The range is not expected to change significantly with redshift.

In the following section, we systematically investigate how the column
density distribution depends on the cosmological parameters and
properties of the intergalactic medium. All the simulations presented in the
next two sections have the same resolution and box size, $256^3$ grid
points with grid spacing of $0.05 \, \,{\rm Mpc}$. For each of them, we will
only plot the part of the column density distribution that falls
within the limits of $0.18 < A < 3.5$. The column densities these
limits correspond to depend on the properties of the intergalactic
medium and the redshift (eq. [\ref{npk3}]). Note that our conservative
limits for $A$ greatly 
reduce the range of column densities we can examine, but within these
limits we can be reasonably confident that the truncated Zel'dovich
approximation together with the Density-Peak-Ans\"{a}tz should yield accurate
predictions for the column density distribution.


\section{The Column Density Distribution: Dependence
on the Ionization Flux, Temperature, Equation of State 
and the Mean Baryon Density} 
\label{cdm}

We use the Cold Dark Matter (CDM) model presented in Fig.~\ref{zhang} 
to study systematically how the column density distribution
depends on properties of the intergalactic medium: the 
level of radiation background, the mean baryon density and its
thermal properties. The tools we use to calculate the column density
distributions are the truncated Zel'dovich approximation and the
Density-Peak-Ans\"{a}tz.

The transfer functions for this model and all the other models considered 
in this paper are taken from Ma (1996). A summary of parameters of all models
can be found in Tables~\ref{cdmmodels} and ~\ref{mdmmodels}.


\subsection{Dependence on Overall Temperature, Ionization Flux and
Baryon Density}


Let us first consider the case in which the
equation of state is fixed at $T\propto (1+\delta_{\rm b})^{0.5}$.
As is shown in equation (\ref{npk3}), the column density of a density
peak with a 
given $\delta_{\rm b}$ (overdensity) is proportional to the following
combination of parameters:
\begin{equation}
F\equiv \left({T_0\over 10^4 K}\right)^{-0.7} \left({\Omega_b h^2\over
0.0125}\right)^2 
\left({J_{\rm HI}\over 0.5}\right)^{-1} \, .
\label{Fdef}
\end{equation}
Hence, by equation (\ref{goodie}), if $F$ is rescaled by a certain factor (by
changing $T_0$, $\Omega_b$ or $J_{\rm HI}$ or their combinations), the number
of absorption lines is also changed by the same factor at an
appropriately rescaled column density. 

Suppose $F$ is rescaled to $F'$ such that $F'=r F$, then
\begin{equation}
\left.{{d^2N'_{{\rm Ly}\alpha}\over {dN'_{\rm HI} dz}}}\right|_{N'_{\rm HI}= r
N_{\rm HI}} = {1\over r}\left.{{d^2N_{{\rm Ly}\alpha}\over {dN_{\rm HI}
dz}}}\right|_{N_{\rm HI}} \, .
\label{Fr}
\end{equation}
It implies that if the column density distribution is a pure power
law, then in a log-log plot of the number of absorption lines per unit
column density per unit redshift versus column density, the straight line would
simply be shifted to the right or left (or up/down) by rescaling.
In reality, the column density distribution only approximately
obeys a power law and so there should be a slight change of
slope at any given column density as a result of rescaling. 

\begin{figure}
\plotone{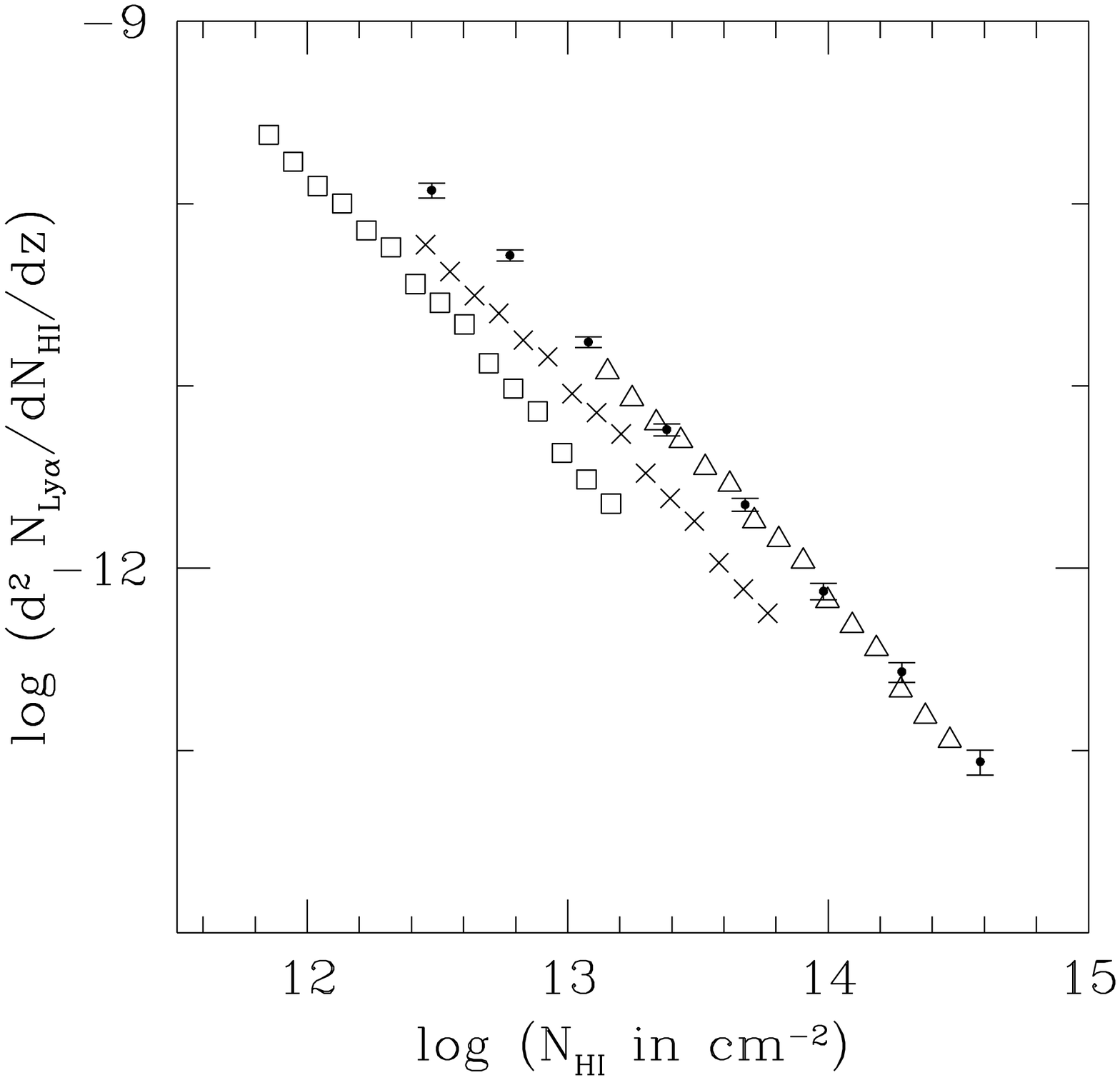}
\caption{\label{Fcd}\capFcd}
\end{figure}

The effects of rescaling can be seen clearly in Fig.~\ref{Fcd}, where
$F$ is allowed to 
take the values $0.25$, $1$ and 
$5$. Keeping $\Omega_b h^2 = 0.0125$ and $J_{\rm HI} = 0.5$, it
corresponds to changing $T_0$ from about $72000 \, {\rm K}$ to $1000\,
{\rm K}$. Alternatively, 
keeping $T_0$ and $\Omega_b h^2$ fixed at their canonical values (as
shown in eq. [\ref{Fdef}]), it corresponds to allowing $J_{\rm HI}$
to vary between $2$ and $0.1$. See Hui and Gnedin (1996) for a
discussion of the 
dependence of $T_0$ on reionization history. $T_0$ is expected to fall
within the range quoted above. 

The conventional value of $\Omega_b h^2 = 0.0125$ has been challenged
by recent measurements of light element abundance in high redshift
absorption systems. Tytler and Burles (1996) obtain a value of
$0.024$, which for $T_0 = 10^4\,  {\rm K}$ and $J_{\rm HI}=0.5$, implies $F=3.686$, well
within the range of $F$ plotted in Fig.~\ref{Fcd}. The analysis of
Rugers and Hogan
(1996), on the other hand, favors the  value $0.006$, which means
$F=0.23$ for the same values of $T$ and $J_{\rm HI}$. The lowest set of
points in Fig.~\ref{Fcd} has to be lowered further to accommodate this value of
the baryon density.

The observational data are taken from Hu et al. (1995), measured at
about redshift of 
$3$ and corrected for incompleteness.
The incompleteness correction was obtained 
by applying the same spectral analysis technique for the observed data to
simulations of randomly distributed Voigt-profiles with a known column density
distribution and measuring how much the analysis method underestimates the
number of low column density lines. The amount ranges from about no
correction necessary at $N_{\rm HI}\, > \, 2 \times 10^{13} \, {\rm cm^{-2}}$
to a factor of $5$ increase in the number of lines at $N_{\rm HI} \, \sim \, 3
\times 10^{12} \, {\rm cm^{-2}}$.

We note in passing that strictly speaking, altering $\Omega_b$, in addition to 
rescaling the number of absorption lines as discussed above, also
changes the transfer function in a non-trivial way. But the effect is very
small for models in which the dark matter (non-baryons) dominate. 
In fact, using the BBKS (\cite{bbks86}) transfer function, which does
not take into 
account the effect of baryons at all, instead of the Ma (1996)
transfer function, which does take it into account for $\Omega_b h^2 =
0.0125$ with $h=0.5$, gives essentially the
same column density distribution for the range of column densities
discussed here. For models where the baryon content is proportionally
higher, Low-density Cold Dark Matter Models for instance, changing
$\Omega_b h^2$ has a more pronounced effect on the transfer function.

\subsection{Dependence on the Equation of State or the
Temperature-density Relation}

\begin{figure}
\plotone{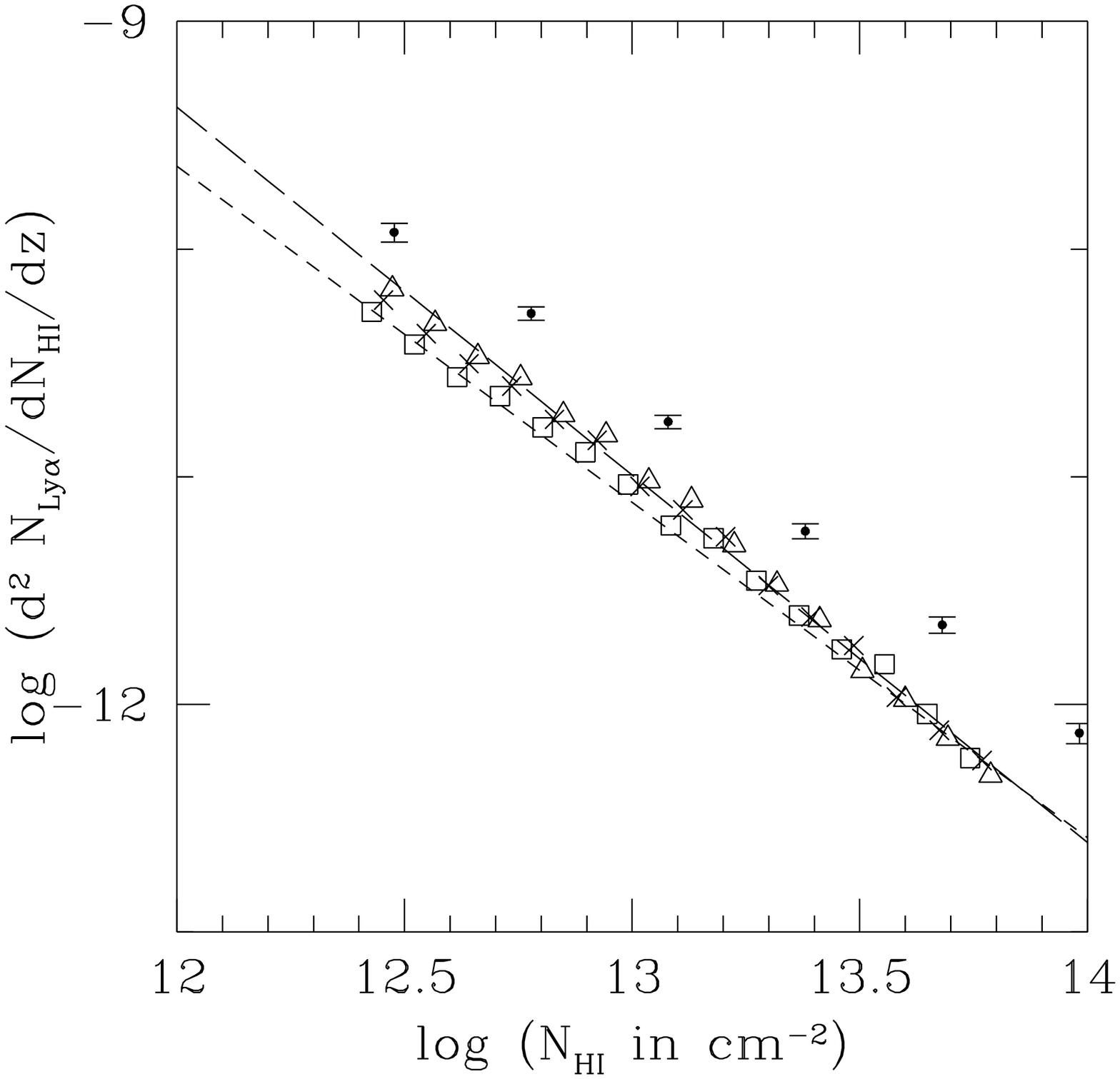}
\caption{\label{eqs}\capeqs}
\end{figure}

Let us hold fixed $T_0$, $\Omega_b h^2$ and $J_{\rm HI}$ at their
canonical values as shown in equation (\ref{Fdef}) but allow the equation of
state to change, for the same CDM model as above. As is pointed out in 
Sec.~\ref{tXrho}, the temperature-density relation for low enough
overdensity is well-approximated by a power law where the power index is around
$0.5$, but can change slightly depending on the reionization
history. We plot in Fig.~\ref{eqs} the column density distributions for 
$\gamma=1.2,1.5,1.7$ where $\gamma$ is defined by $T\propto
(1+\delta_{\rm b})^{\gamma-1}$. It should adequately cover the possible range of
$\gamma$ (\cite{hg96}).

The first thing to notice is that the column density distribution
remains almost the same for the three different values of
$\gamma$. This is because $\gamma$ affects column density through the
power index of $(1+\delta_{\rm b})$, which is $2-0.7(\gamma-1)$ (eq. [\ref{g}]).
The index does not change significantly for the range of $\gamma$
considered. A larger index (smaller $\gamma$) means for a density peak
with a given 
$1+\delta_{\rm b}$ (and its second derivative), the column density is larger
or smaller depending on whether $1+\delta_{\rm b}$ is bigger or smaller than
one. The net effect is to decrease the slope of the column density
distribution. The effect, though very small for the values of $\gamma$
plotted, can still be seen in Fig.~\ref{eqs}.
We also show the approximate slopes given by an analytical formula
(eq. \ref{slope}) which will be discussed later. Note how the column
density distribution does not exactly follow a power law but can be
approximated by one.

Hence as a crude approximation, we conclude that the mean temperature,
radiation intensity and baryon density mainly determine the overall
normalization of the column density distribution. The equation of
state, on the other hand, mostly affects the slope of the column
density distribution, but its effect is small for a reasonable
range of $\gamma$.

\section{The Slope of the Column Density Distribution}
\label{slopecd}

It has been shown that the normalization of the column density
distribution is influenced by the thermal and ionization states  of
the intergalactic medium, which are not well-constrained observationally.
The slope of the distribution, on the other hand, is only weakly 
affected by the equation of state or temperature-density relation.

The slope of the column density distribution is therefore relatively
free of uncertainties due to our ignorance of the thermal and ionization
properties of the intergalactic medium. We now turn our attention
to the effect of the power spectrum on the slope of the distribution.

From equations (\ref{npk3}) and (\ref{g}), it can be seen that the
column density $N_{\rm HI}$ is proportional 
to $(1+\delta_b)^{2-0.7(\gamma-1)}$ times $1/\sqrt{\xi''}$, which
basically defines a length scale. Taking into account the correlation
between this length scale and the overdensity, we find from our
simulations (which use the Zel'dovich approximation) a useful approximate
relation for column densities between about $10^{12.5}$ and $10^{14.5} \, {\rm
cm^{-2}}$ (or more accurately, for the range of validity discussed in
Sec.~\ref{rov}):   
\begin{equation}
N_{\rm HI} \propto (1+\delta_b)^{1.68-0.7(\gamma-1)} \, ,
\label{nh1delta}
\end{equation}
which roughly means the length scale $1/\sqrt{\xi''}$ is approximately
proportional \footnote{A log-log plot of
$1/\sqrt{\xi''}$ versus $(1+\delta_b)$ actually shows a lot of scatter
but equation (\ref{nh1delta}) appears to capture the overall
dependence of $N_{\rm HI}$ on $1+\delta_b$.}
to  $(1+\delta_b)^{-0.32}$.

Since we are interested in the slope of the column density
distribution, the relevant quantity to consider is:
\begin{equation}
-\beta \equiv {{d \ln {d^2 N_{{\rm Ly}\alpha}\over {dN_{\rm HI}
dz}}}\over {d \ln 
N_{\rm HI}}} = -1 + {m\over {1.68-0.7(\gamma-1)}} \, \, , \, \, 
m\equiv{ {d \ln \int_{-\infty}^{0} 
d\xi'' |\xi''| P(\xi,\xi'=0,\xi'') }\over {d \xi}} \, . 
\label{slope0}
\end{equation}
The equality follows from equations (\ref{nh1delta}), (\ref{npk3}),
(\ref{goodie}) and (\ref{differential}) and noting that $\xi \equiv
1+\delta_b$. The column density distribution  can be approximated
by the simple power law $N_{\rm HI}^{-\beta}$ if $\beta$ defined above
is only weakly dependent on $\xi$ or $N_{\rm HI}$.

Lacking an analytical expression for $P$ under the Zel'dovich
approximation, we can nonetheless guess what the general properties of the
quantity $m$ are.
First of all, $m$ depends on $\xi$ in general because the integral
$\int_{-\infty}^{0} 
d\xi'' |\xi''| P(\xi,\xi'=0,\xi'')$ cannot
be a simple power law in $\xi$. This is because 
we expect the integral to vanish for very large and 
very small $\xi$ 
and peak at some 
intermediate $\xi$. This implies one should not expect an exact
power-law for the column density distribution, although pieces of it
might be approximated by power-law.
Suppose $\xi_{\rm pk}$ is the value of $\xi$
where the integral $\int_{-\infty}^{0}
d\xi'' |\xi''| P(\xi,\xi'=0,\xi'')$ reaches its maximum value, then
the column densities 
we are interested in must correspond to $\xi > \xi_{\rm pk}$ 
where ${ {d \ln \int_{-\infty}^{0}
d\xi'' |\xi''| P(\xi,\xi'=0,\xi'') }/ {d \xi}}$ is negative.
This is based on the knowledge that the computed (as well as observed)
slope in equation (\ref{slope0}) is less than $-1$
(the factor ${1.68-0.7[\gamma-1]}$ is always positive for reasonable values of
$\gamma$).

Under the Zel'dovich approximation, the quantities $\xi$ and its derivatives
can be related to the displacement potential $\psi({\bf q})$ using
equations (\ref{zad}) and (\ref{zav}) and so one can express $P(\xi,
\xi'. \xi'')$ in terms of the probability density for derivatives of $\psi$:
\begin{eqnarray}
\int_{-\infty}^{0} d\xi'' |\xi''| P(\xi,\xi'=0,\xi'') =&& \int_{\-\infty}^{0}
d\xi'' d\psi_{ij} d\psi_{ijk} d\psi_{ijkl} |\xi''| e^{-\xi}
P(\psi_{ij}, \psi_{ijk}, \psi_{ijkl}) \\ \nonumber
&&\delta_D [\xi - \xi(\psi_{ij})] \delta_D [\xi'(e^i, \psi_{ij},\psi_{ijk})]
\delta_D [\xi'' - \xi''(e^i, \psi_{ij}, \psi_{ijk}, \psi_{ijkl})] \, ,
\label{mza}
\end{eqnarray}
where $\delta_D$ denotes the Dirac delta function, $\psi_{ij}$ is
the derivative of the displacement potential $\psi$ with respect to
$q_i$ and $q_j$ (eq. [\ref{zad}]) and similarly for $\psi_{ijk}$ and
$\psi_{ijkl}$ and $e^i$ is the unit vector pointing along the line sight. 
The quantity $P(\psi_{ij}, \psi_{ijk}, \psi_{ijkl})$ is the probability
density of $\psi_{ij}$ and its derivatives in Lagrangian space and the factor
$e^{-\xi}$ converts it into its counterpart in Eulerian space (see Kofman et
al. 1994). The
quantities $\xi$ and its derivatives along the  
line of sight can be expressed as functions of $\psi_{ij}$, $\psi_{ijk}$,
$\psi_{ijkl}$ and $e^i$. Isotropy of the universe implies that one
can average the above expression over all possible orientations of $e^i$.

For the cosmological models we consider in this paper, 
the probability density $P(\psi_{ij}, \psi_{ijk}, \psi_{ijkl})$ is a
multivariate Gaussian function that depends on $3$ parameters: 
$\sigma_0$, $\sigma_1$ and $\sigma_2$
defined as follows (see Bardeen et al. 1986):
\begin{equation}
\sigma_j = D_+(t) \sqrt{\int_0^{\infty} 4 \pi k^{2+2j} P(k)
e^{-({k/k_{\rm S}})^2} dk} \, , \quad \, \,
j=0,1,2
\label{sigma0}
\end{equation}
where $D_+$ is the linear growth factor which is equal to $(1+\bar
z)^{-1}$ for a universe at critical matter density and $k_{\rm S}$
is an appropriate smoothing scale. This follows from the structure of
the various expectation values: $<\psi_{ij}\psi_{mn}> \propto \sigma_0^2$,
$<\psi_{ij}\psi_{mnlp}> \propto \sigma_1^2$, $<\psi_{ijkl}\psi_{mnop}> \propto
\sigma_2^2$ and $<\psi_{ijk} \psi_{mnl}> \propto \sigma_1^2$.
The fact that $<\psi_{ij}\psi_{mnl}>$ and $<\psi_{ijkl}\psi_{mno}>$ vanish
by isotropy means that the probability density factors into two separate
multivariate Gaussian functions: $P(\psi_{ij}, \psi_{ijk}, \psi_{ijkl})
= P(\psi_{ij},\psi_{ijkl}) P(\psi_{ijk})$, with the first factor
depending on $\sigma_0$, $\sigma_1$ and $\sigma_2$ and the second factor
depending on $\sigma_1$.

One can replace $\sigma_1$ and $\sigma_2$ by combinations of $\sigma_0$ and the
following two new parameters, which Bardeen et al. (1986) defined:
\begin{equation}
R_{*} \equiv \sqrt{3} {\sigma_1 \over \sigma_2} \, , \quad \gamma_{B}
\equiv {\sigma_1^2\over {\sigma_2 \sigma_0}} \, ,
\label{rstargamma}
\end{equation}
where we have renamed $\gamma_{B}$ to distinguish it from $\gamma$ we
use in this paper. 

While $R_*$ defines a length scale, the quantity $\gamma_B$ is a measure
of the slope of the power spectrum. We find it convenient to use the
following quantity $n_{\rm eff}$ in place of $\gamma_B$ (Bardeen et al. 1986):
\begin{equation}
n_{\rm eff} \equiv {{5\gamma_B^2 - 3} \over {1 - \gamma_B^2}} \, .
\label{neff}
\end{equation}
It is easy to show that the above quantity coincides exactly with the slope of
the power spectrum if it obeys a pure power-law. For $\gamma_B = 0.5$, $n_{\rm
eff} = -2.33$. The $n_{\rm eff}$ defined here should be distinguished
from $n$ in Tables~\ref{cdmmodels} and \ref{mdmmodels}: $n$ is the
slope of the power spectrum at large scales whereas $n_{\rm eff}$ is the slope
at the (small) smoothing scale. 

The advantage of the new notation is that it is possible to show, by
changing variables $\bar\psi_{ijk} \equiv \psi_{ijk} R_*$, 
$\bar\psi_{ijkl} \equiv \psi_{ijkl} R_*^2$ and $\bar\xi'' = \xi'' R_*^2$, 
the integrals in equation (\ref{mza}) are independent of $R_*$ except for a
normalization factor. This implies, by equations (\ref{slope0}) and
(\ref{mza}), 
that 
the slope $\beta$ depends on the equation of state through $\gamma$ and on
the power spectrum only through $\sigma_0$ and $n_{\rm eff}$.

It is hard to make analytic progress from this point on, because
$\xi$ and its derivatives are complicated functions of the derivatives
of $\psi$. We resort to our numerical 
simulations to extract the parameter dependence of $\beta$.
Using the above arguments and observing that for all models we
consider in this paper, $\sigma_0 \sim 1$ and $n_{\rm eff} \sim -2.33$
($\gamma_B \sim 0.5$), we assume the following form for $\beta$:
\begin{equation}
\beta = 1 + {{B_1 +B_2 (\sigma_0 - 1) + B_3 (n_{\rm eff}
+2.33) }\over {1.68-0.7(\gamma-1)}} \, ,
\label{BBB}
\end{equation}
where $B_1$, $B_2$ and $B_3$ are constant coefficients of a Taylor
series expansion of $m$ (eq. [\ref{slope0}]).

We determine these constants by computing $\beta$ for a series of
CDM models of varying $\sigma_0$ and $\gamma_B$. The column density
distributions are obtained by using the Zel'dovich
approximation together with the Density-Peak-Ans\"{a}tz, as described in
previous sections. We pick out particularly models with similar $n_{\rm eff}$'s
but very different $\sigma_0$'s and vice versa. 
It is found that the following expression for $\beta$ fits reasonably
well the slope of the column density distributions for the CDM models,
as well as the Cold + Hot Dark Matter (CHDM) models that we will also study
later: 
\begin{equation}
\beta = 1 + {{0.96 - 0.8 (\sigma_0 - 1) - 0.4
(n_{\rm eff}+2.33)}\over{1.68-0.7(\gamma-1)}} \, ,
\label{slope}
\end{equation}
where $\beta$, $\gamma$, $\sigma_0$ and $n_{\rm eff}$ are defined in 
equations (\ref{slope0}), 
(\ref{tgamma}), (\ref{sigma0}) and (\ref{neff}) respectively.

\begin{figure}
\plotone{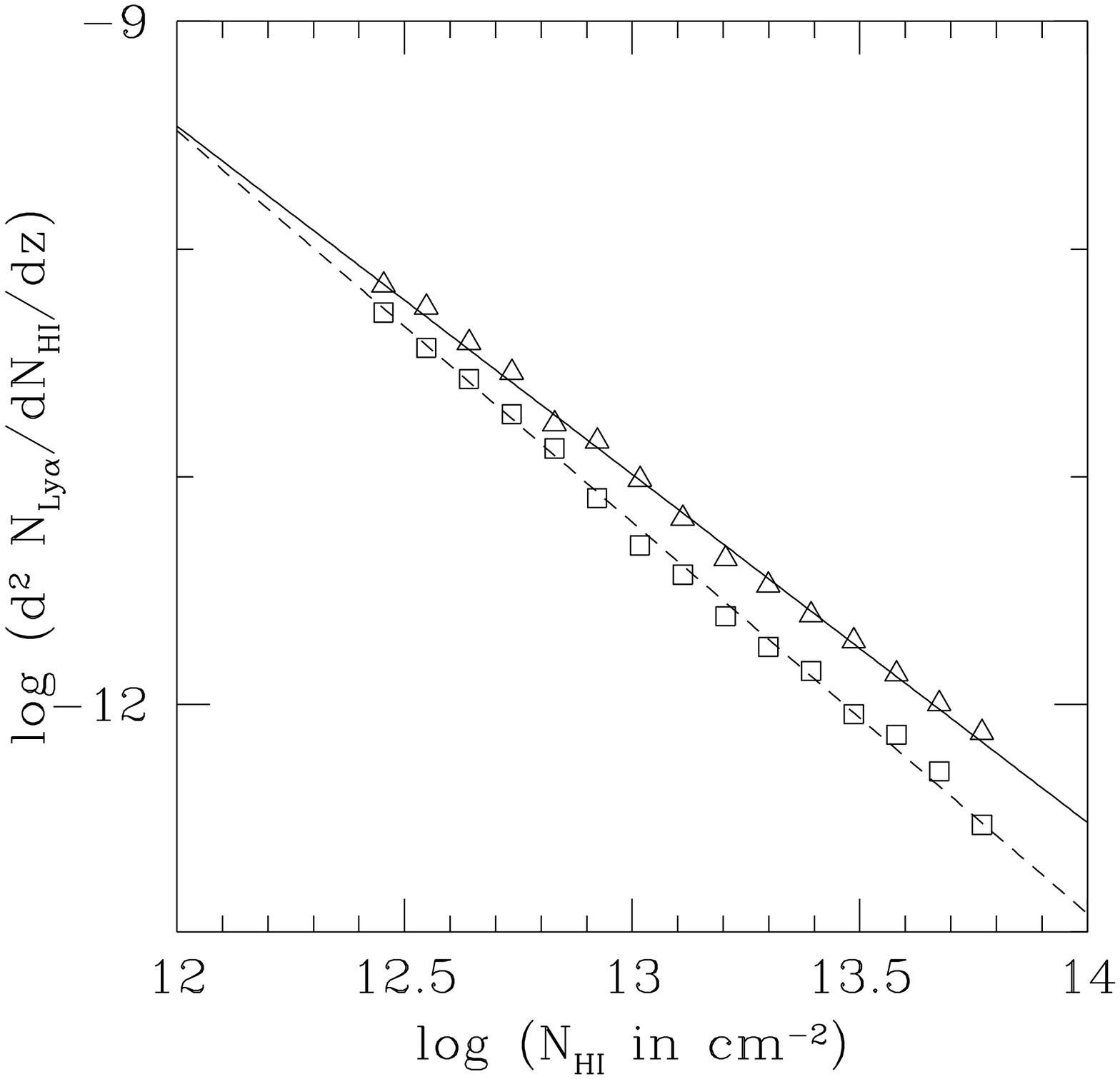}
\caption{\label{norm}\capnorm}
\end{figure}

\begin{figure}
\plotone{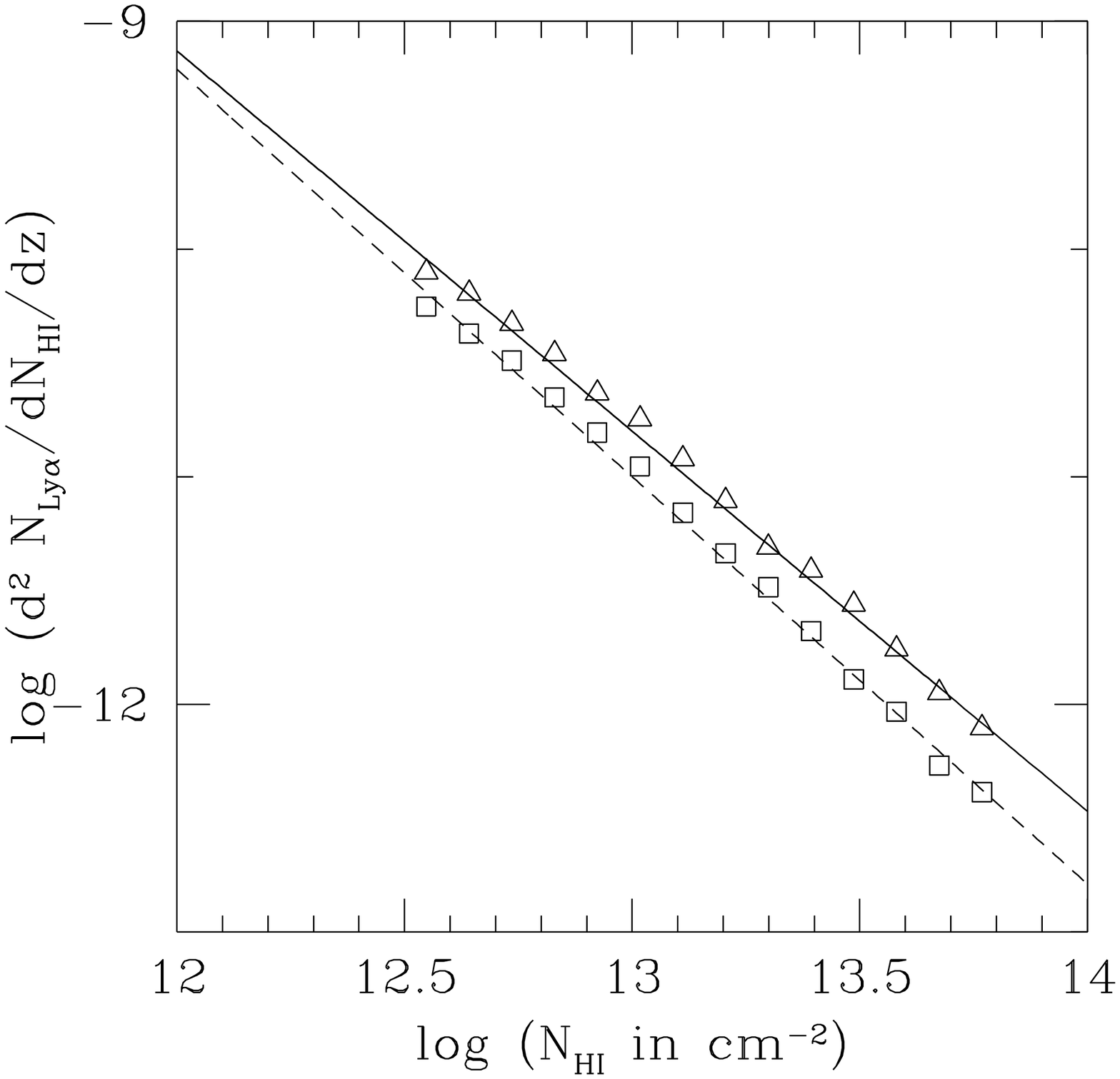}
\caption{\label{an}\capan}
\end{figure}

We show two sets of examples in Fig.~\ref{norm} and Fig.~\ref{an}.
The former has two CDM models with the same $n_{\rm eff}$ but
different $\sigma_0$'s while the opposite is true for the latter. 
To find models with such properties, we find it necessary sometimes to smooth
on scales larger than the 
orbit-crossing scale $k_{\rm S}^{-1}$ defined by equation (\ref{ks})
(examples are models CDM3 and CDM4 shown in the Figures; see also
Table~\ref{cdmmodels}). 
For this reason, the column density distributions computed in this
section should be viewed as predictions of the corresponding CDM models
at the specified smoothing scales only: they do not necessarily coincide
with the predictions of these models if more appropriate smoothing scales are
chosen. Nor should the CDM models shown in the two figures be considered
realistic models of the universe.

Examples of how well equation (\ref{slope})  
describes the variation with the equation of state for a CDM model
can be found in Fig.~\ref{eqs}. More examples for CHDM models
are also shown in Figures ~\ref{Fnu2_42}, ~\ref{Fnu2_33b}, ~\ref{Fnu2_44} and
~\ref{eqsnu2_42}.  

Before we go on to discuss physical interpretations of the above expression for
$\beta$, let us make some general remarks and state a few caveats.
First, as we discussed before, $\beta$ is
expected to vary with column density 
but can be approximated as a constant over a restricted range.
That a power-law column density distribution is only an approximation
is clear even in some of the figures mentioned above. 
We find it appropriate to allow for a maximum error-bar of $\pm 0.1$
for $\beta$ in  
equation (\ref{slope}), given $\gamma$, $\sigma_0$ and 
$\gamma_B$. The coefficients $B_1$, $B_2$ and $B_3$ (eq. [\ref{BBB}])
should be viewed as correspondingly uncertain.
Furthermore it should be kept in mind that  equation
(\ref{slope}) holds only for $\sigma_0 \sim 1$ and $n_{\rm eff} \sim -2.33$.
as the constants $B_1$, $B_2$ and $B_3$ are only meant to be 
coefficients of a Taylor series expansion.

\begin{figure}
\plotone{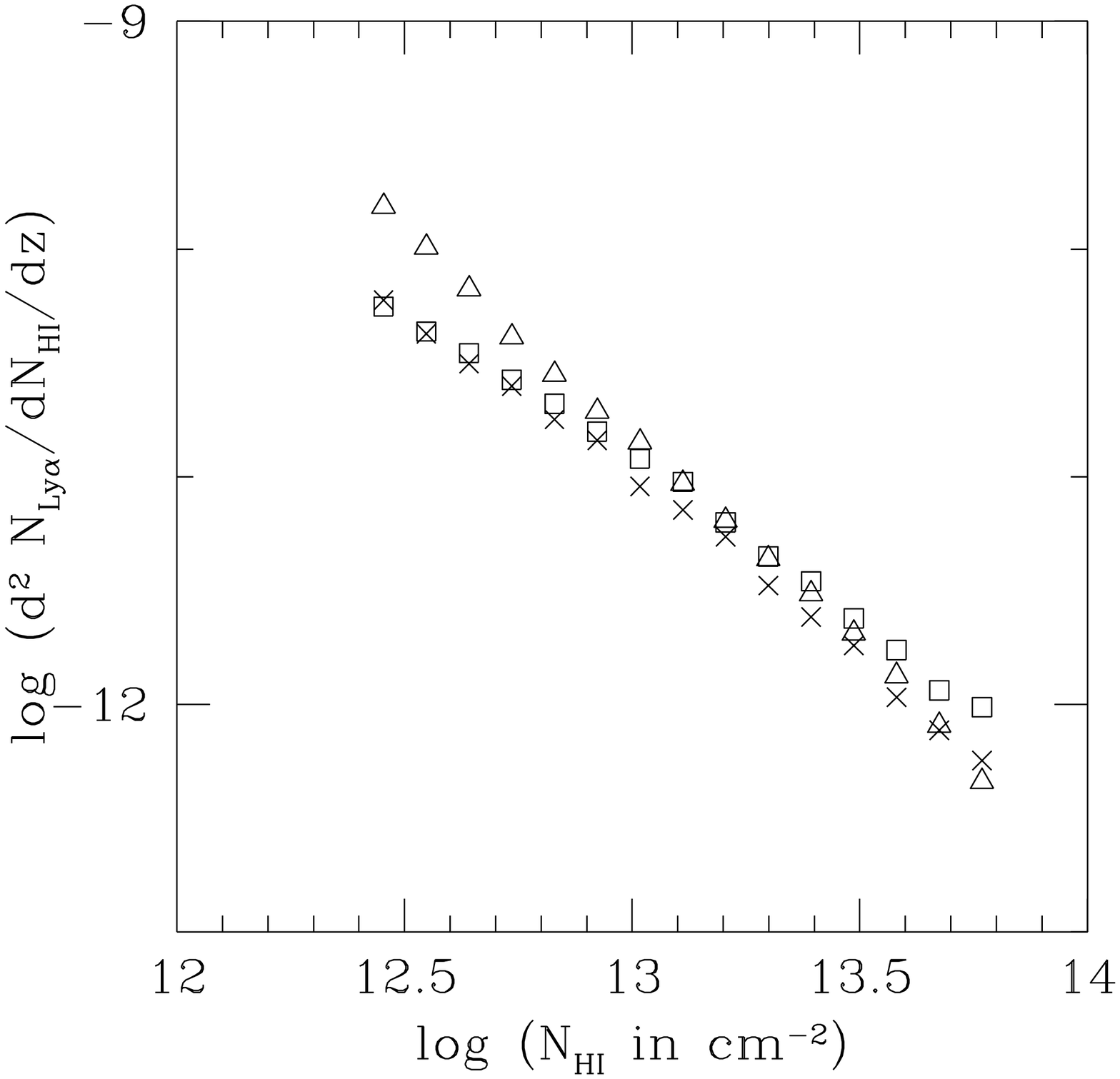}
\caption{\label{smooth}\capsmooth}
\end{figure}

Finally, one obvious problem of using equation (\ref{slope}) to make prediction
for a given cosmological model is that both $\sigma_0$ and $n_{\rm eff}$ varies
with the smoothing scale. An example of how the choice of smoothing scale
can affect the column density distribution is shown in Fig.~\ref{smooth}.
The comparison between results of a hydrodynamic
simulation and our Zel'dovich computation in Fig.~\ref{zhang} seems to 
support the choice of smoothing in equation
(\ref{ks}). However, one could reasonably argue that the Jeans scale
(eq. [\ref{jk}]) is a more physically motivated smoothing scale while the
choice 
in equation (\ref{ks}) 
was made only as a device to counter the effect of orbit-crossing and
hence one cannot be sure that one is not thereby erasing
real structures on scales smaller than the orbit-crossing scale but larger
than the Jeans length, which can contribute significantly to the low
column density Lyman-alpha forest.
One way to settle this question is to make more detailed comparisons with
more hydrodynamic simulations, which is outside the scope of the
present work and we will leave for a future paper.
It suffices to say that, for models with sufficiently
little small scale power such as the CHDM models considered in
Sec.~\ref{mdm}, which have orbit-crossing scales close to
or smaller than the Jeans length, it is probably safe to smooth
according to the prescription laid down in Sec.~\ref{zasection} and that the
precise smoothing scale might not matter very much 
(see Fig.~\ref{smoothMDMnu02}). For more nonlinear models, we can only offer
the comparison in Fig.~\ref{zhang} as one piece of evidence that smoothing on
the orbit-crossing scale works reasonably well.
However, for whatever smoothing scale one decides upon, we expect equation
(\ref{slope}) to hold approximately, provided that $\sigma_0 \sim 1$ and
$n_{\rm eff} \sim -2.33$. 

With these being said, let us try to understand qualitatively
the parameter dependence of $\beta$ as expressed in equation (\ref{slope}).

First, the dependence on $\gamma$. Recall that
$N_{\rm HI} \propto (1+\delta_{\rm b})^{1.68-0.7(\gamma-1)}$ (taking into
account the correlation between $1+\delta_{\rm b}$ and its second derivative).
This means that higher density peaks
translate into higher $N_{\rm HI}$ and vice versa. Lowering $\gamma$
increases the effect of this translation i.e. a given density peak with $1 +
\delta_{\rm b} > 1$ is associated with a higher column density if $\gamma$ is
reduced (the opposite is true for $1 + \delta_{\rm b} < 1$). The net effect is
to stretch a power-law-like column density distribution and make it
flatter. The effect is small for a reasonable range of $\gamma$. It should be
between $1.3$ and $1.62$ at $\bar z =3$ if the universe reionizes before $\bar
z \sim 5$ (see Hui and Gnedin 1996). A somewhat larger range is shown in
Fig.~\ref{eqs}. 

Second, the dependence on $\sigma_0$: a cosmological model
with lower $\sigma_0$ is in a more ``linear'' state of evolution compared to
higher $\sigma_0$ models. In other words, a lower $\sigma_0$ model has
proportionally more intermediate density peaks compared to high density ones,
hence the steeper column density distribution (associating once again high
density peaks with high column densities using the $N_{\rm HI}
\propto (1+\delta_{\rm b})^{1.68-0.7(\gamma-1)}$ scaling). For sufficiently low
column densities, however, the absorption lines arise from very underdense
regions which should be more common in higher $\sigma_0$ models. Hence at very
low column densities, the high $\sigma_0$ model should win: it has more very
low density peaks. Where this might occur we cannot tell from our simulations
because of the limited resolution. For the range of column densities we can
measure reliably, the slope of the column density distribution simply steepens
as $\sigma_0$ is lowered.

The dependence of $\beta$ on $n_{\rm eff}$ is more subtle. 
To understand it, we resort to a Press-Schechter-type argument (Press \&
Schechter 1974). According to the Press-Schechter theory, 
the low mass slope of $n_C (M)$ (the number density of clumps of mass $M$
per  
unit range $dM$) is $-1.5 + n_{\rm eff}/6$ which means that smaller $n_{\rm
eff}$ ($n_{\rm eff} < -1$)implies a steeper slope of the clump number density
distribution,  
exhibiting the same trend as in the case of equation (\ref{slope}). The $n_{\rm
eff}$ dependence 
arises from the fact that $\sigma_0(k) \propto k^{(n_{\rm eff}+3)/2}$
($\sigma_0$ with smoothing scale $k$) and assuming $M \propto k^{-3}$, together
with $M n_C (M) \propto k^3 / 
\sigma_0(k)$ (for high $k$). The
factor of $k^3$  appears because $n_C (M)$ is a number density. 
The dependence of $n_C (M)$ on $\sigma_0(k)$ follows from 
a simple conjecture: the probability that a given point belongs to a clump of
mass 
$M$  or higher is equal to $2\int_{\delta_{\rm th}}^\infty P(\delta) d\delta$
where $\delta_{\rm th}$ is a fixed threshold and $P(\delta)$ is the Gaussian 
probability distribution for overdensity $\delta$ with dispersion $\sigma_0
(k)$. 

One can carry the above reasoning over to absoprtion lines, keeping
in mind something like the Threshold Algorithm for identifying lines
as the analogue of the threshold-criterion of the Press-Schechter
theory for a collapsed clump.
Assume $N_{\rm HI} \propto k^{-1}$
(because in the linear regime, it is the peak width, not the peak height
that provides the lowest order contribution to column density),
it ``follows'' that
$N_{\rm HI} (d^2 N_{Ly\alpha}/dN_{\rm HI} dz) \propto k / \sigma_0(k)$
, again in the high $k$ limit. The factor of $k$ is to account for
the quantity of interest being a one-dimensional number density. 
It is easy to see this implies $\beta = (1-n_{\rm eff})/2$ i.e.
smaller $n_{\rm eff}$ (larger $|n_{\rm eff}|$) implies steeper column
density distribution. This admittedly
crude argument actually gives a value for $\beta$ that agrees surprisingly well
with what we have found using the Zel'dovich approximation and 
Density-Peak-Ans\"{a}tz. For example, for the CDM model in Fig.~\ref{zhang},
$n_{\rm eff} = -2$ (see Table~\ref{cdmmodels}), $\beta = 1.5$ according
to the adapted-Press-Schechter argument. This of course brings up the
question why there does not seem to be any $\sigma_0$ dependence of $\beta$ in
the above argument. It is in fact possible to develop this line of reasoning
further to include the $\sigma_0$ dependence, but we will leave this to a later
paper. 

Finally, it is interesting to point out that because most realistic
cosmological models have similar power on large scales (COBE or cluster scale),
small $\sigma_0$ on small scales almost necessarily means a steeper fall-off
of the power spectrum, namely smaller $n_{\rm eff}$. So, in practice, both the
amplitude $\sigma_0$ and the slope $n_{\rm eff}$ of the power spectrum work in
the same direction in decreasing/increasing the slope of the column density
distribution. Hence, as a rule of thumb, less small scale power implies
a steeper column density distribution, for the column densities of interest
in this paper.

\section{The Column Density Distribution for CHDM models}
\label{mdm}

In this section, we turn our attention to CHDM models, which have relatively
little small scale power and for which the method of truncated Zel'dovich
approximation is ideally suited. The orbit-crossing scales for them
are either smaller than or only slightly larger than the Jeans length
and so one has reasons to believe any structure erased by our smoothing
procedure does not contribute significantly to the number of absorption lines
in the range of column densities we are interested in. Predictions for
the CHDM models are also of current interest because no hydrodynamic
simulations of the Lyman-alpha Forest have been performed for these models.

We list in Table \ref{mdmmodels} the CHDM models 
considered in this section.
They are all $\Omega_0 = 1$ models with $\Omega_b = 0.05$. Both the
$\Omega_\nu = 0.2$ and $\Omega_\nu = 0.1$ versions are considered.
They have been shown to give good agreement
with the observational data on large scales ($k$ around $0.02 - 0.4 \,
\,{\rm Mpc^{-1}}$) (see Fig. 6 and 7 of Ma 1996).  The $\Omega_\nu = 0.3$
models seem to conflict with the observed abundance of damped Ly$\alpha$
systems, which correspond to roughly $k$ around $0.1 - 1.0
\,{\rm Mpc^{-1}}$ comoving in the linear power spectrum
(\cite{mm94}; \cite{kc94}; \cite{mb94}). 
We include one $\Omega_\nu = 0.3$ CHDM model in Table \ref{mdmmodels}
for the sake of comparison. As is shown convincingly by Ma (1996), all
models need some amount of tilt to match observations.

We compute as before the column density distribution for each model using the
Density-Peak-Ans\"{a}tz and the Zel'dovich approximation with
appropriate smoothing. The (density weighted) power spectrum for each
CHDM model is taken 
from Ma (1996). 

\begin{figure}
\plotone{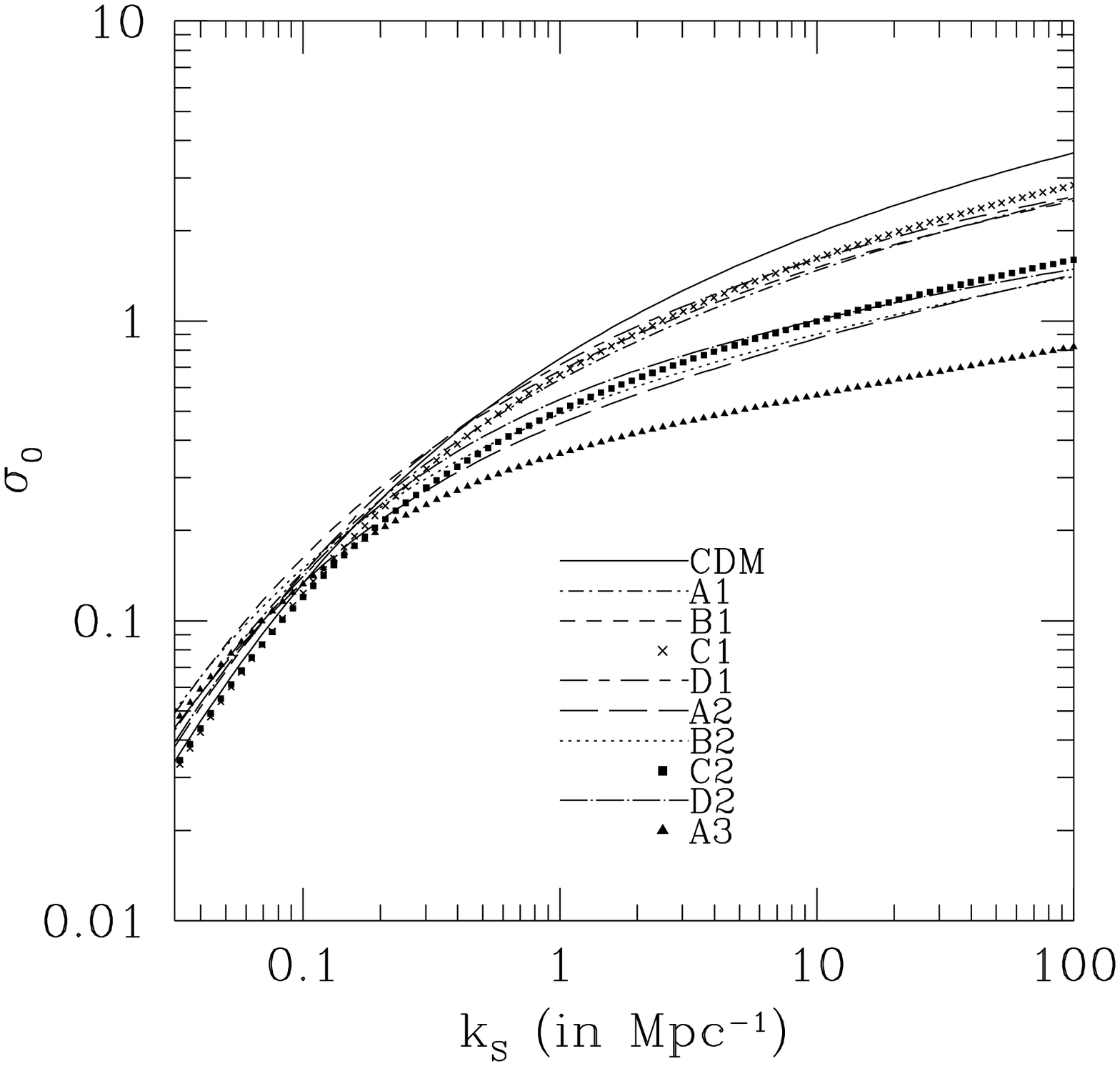}
\caption{\label{psmMDM}\cappsmMDM}
\end{figure}

As we have discussed in the last section, a plot of
rms smoothed linear density fluctuation 
$\sigma_0$ versus smoothing scale $k_{\rm S}$ (eq. [\ref{sigma0}]) is a
very good indicator of what column density distribution to
expect. This is done in Fig.~\ref{psmMDM} for the CHDM models
tabulated in Table \ref{mdmmodels}. The no-tilt $\sigma_8 = 0.7$ CDM
model is also plotted for comparison. 

Because of neutrino free streaming, all CHDM models have less power (smaller
$\sigma_0$) and steeper
spectral slope (smaller $n_{\rm eff}$) than the CDM model on small
scales. Those with more neutrino 
content ($\Omega_\nu = 0.2$) have even less power than the others.
In fact, the $\Omega_\nu = 0.2$ models have $\sigma_0 < 1$ on all
scales larger than the Jeans scale ($k_{\rm S} < k_{\rm J} \sim 10 \,
{\rm Mpc}^{-1}$). One expects the
Zel'dovich approximation to work particularly well for these models
because the amount of orbit-crossing will not be significant,
even without initial truncation.

\begin{figure}
\plotone{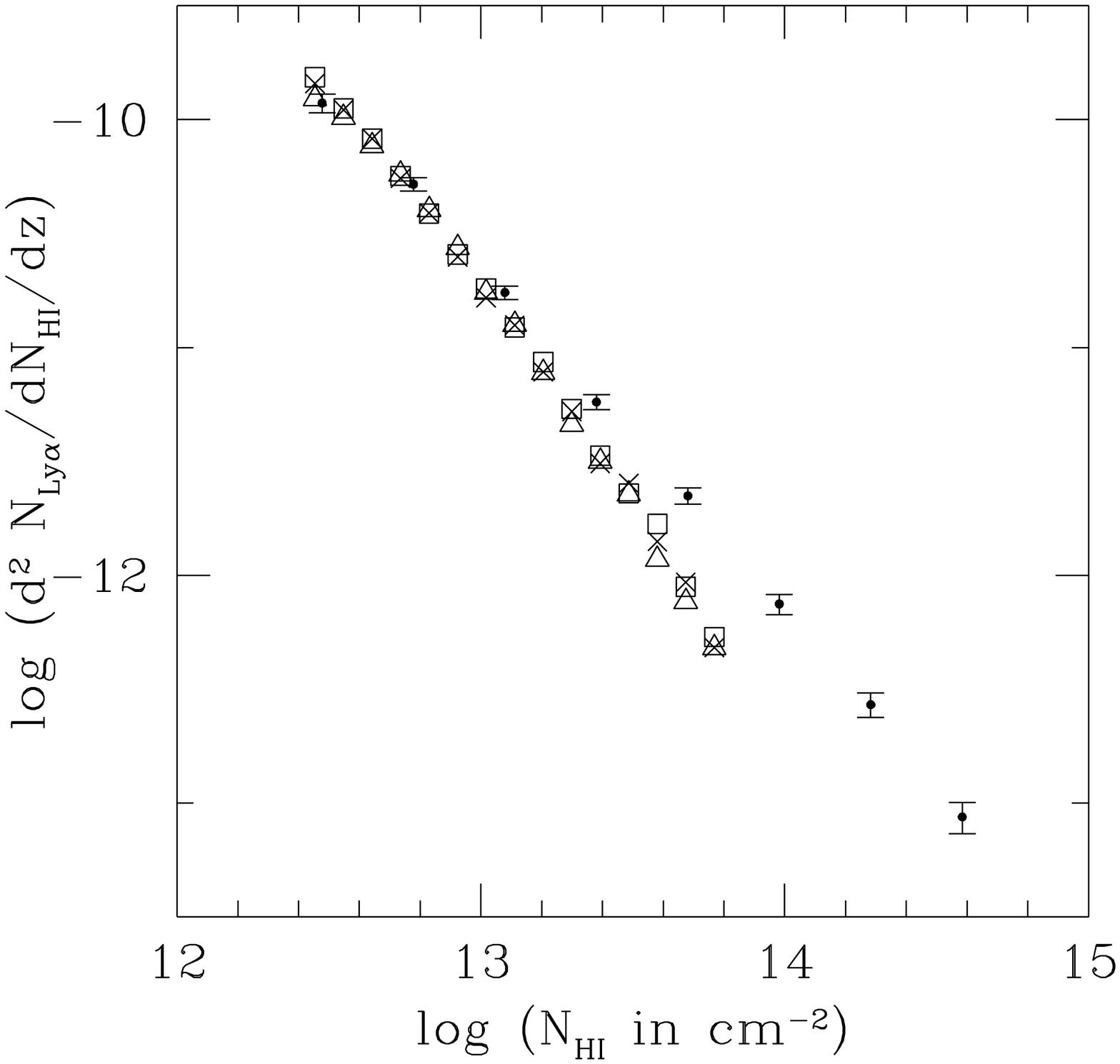}
\caption{\label{smoothMDMnu02}\capsmoothMDM}
\end{figure}

This is borne out by the next test: we compute the column
density distribution for one CHDM model B2 and examine the effect of
choosing different smoothing scales. The result is plotted in
Fig.~\ref{smoothMDMnu02}. 
The column density distribution in the range plotted does not change
much at all for the three different smoothing scales plotted. 
Contrast this
with the case of $\sigma_8 = 0.7$ CDM (Fig.~\ref{smooth}) where the column
density distribution is more sensitive to changes in the smoothing scale.
That's why the truncation scale has to be chosen with some care for
the CDM model: not too 
small ($k_{\rm S}$ too big) so that too much orbit crossing has occurred and
not too large so 
that too much small scale structure is erased. The
standard prescription (Sec.~\ref{zasection}) seems to work well
according to Fig.~\ref{zhang}.

For the CHDM model considered (in fact, it holds true for all
other $\Omega_\nu = 0.2$ models here), the amount of small scale power
is so insignificant that excluding them by smoothing does not affect the
overall column density distribution at all (except possibly that one
loses the small scale fluctuations that can give rise to very low
column density absorption i.e. lower than our resolution limit).
We have also done similar tests for the
$\Omega_\nu = 0.1$ models, their response to changes in the truncation
scale is somewhere between the $\sigma_8 =
0.7$ CDM model and the $\Omega_\nu = 0.2$ CHDM models, as can be
expected based on their difference in Fig.~\ref{psmMDM}. 

We adopt the following truncation scales for the CHDM models.
For the $\Omega_\nu = 0.1$ models, the standard prescription described
in Sec.~\ref{zasection} is used (i.e. $k_{\rm S} = 1.5 \, k_{\rm NL}$). 
The $\Omega_\nu = 0.2$ models, according to the above prescription,
would have truncation scales less than the Jeans length ($k_{\rm S} >
k_{\rm J}$) and so by the arguments presented in Sec.~\ref{zasection}, $k_{\rm S} = k_{\rm J}$
is adopted. Again, we emphasize that for this class of models that
have relatively little power on small scales, the precise truncation
scale is not important. A summary of the truncation scales for all
models can be found in Table \ref{mdmmodels}.

\begin{figure}
\plotone{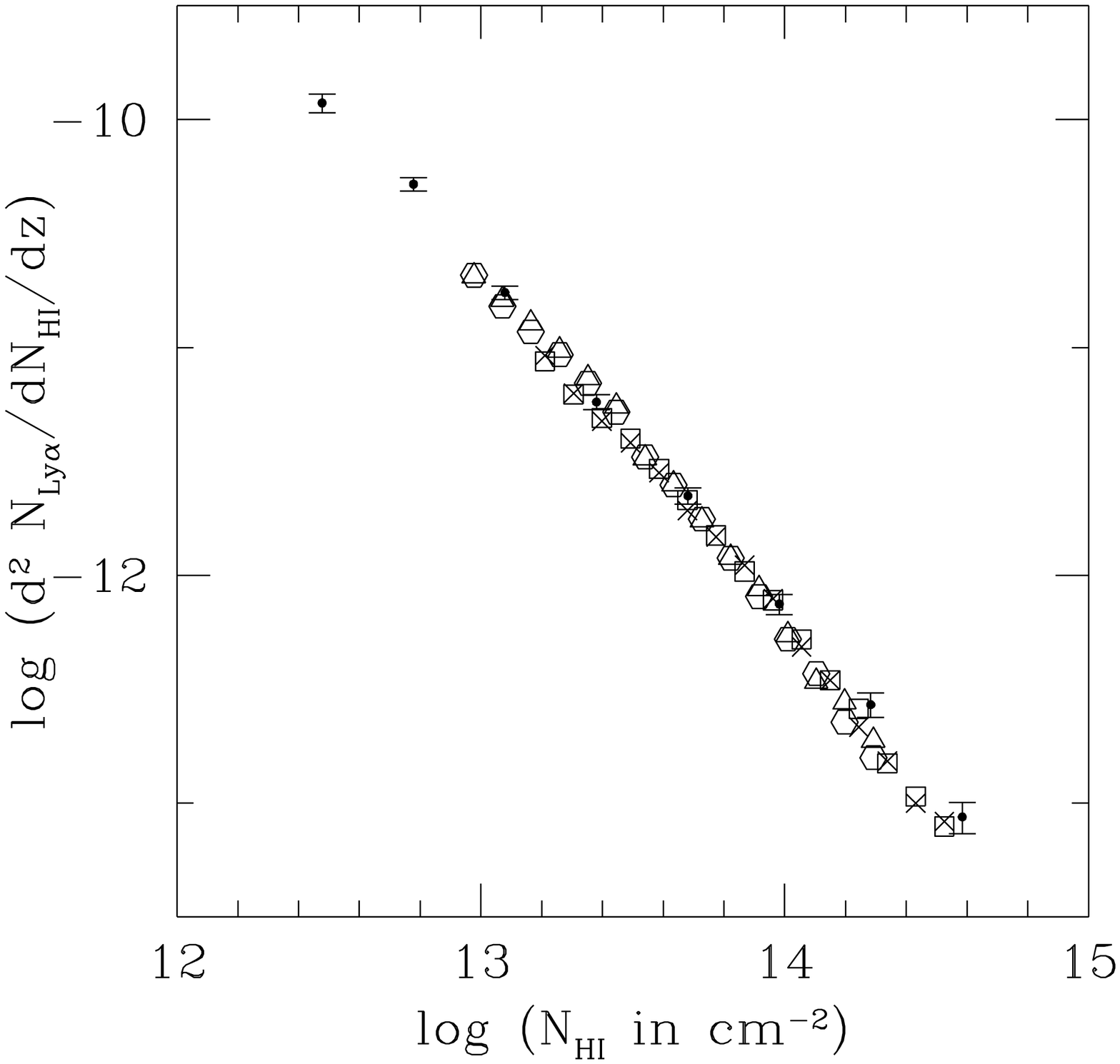}
\caption{\label{mdmcdnu1}\capmdmcdnuone}
\end{figure}
\begin{figure}
\plotone{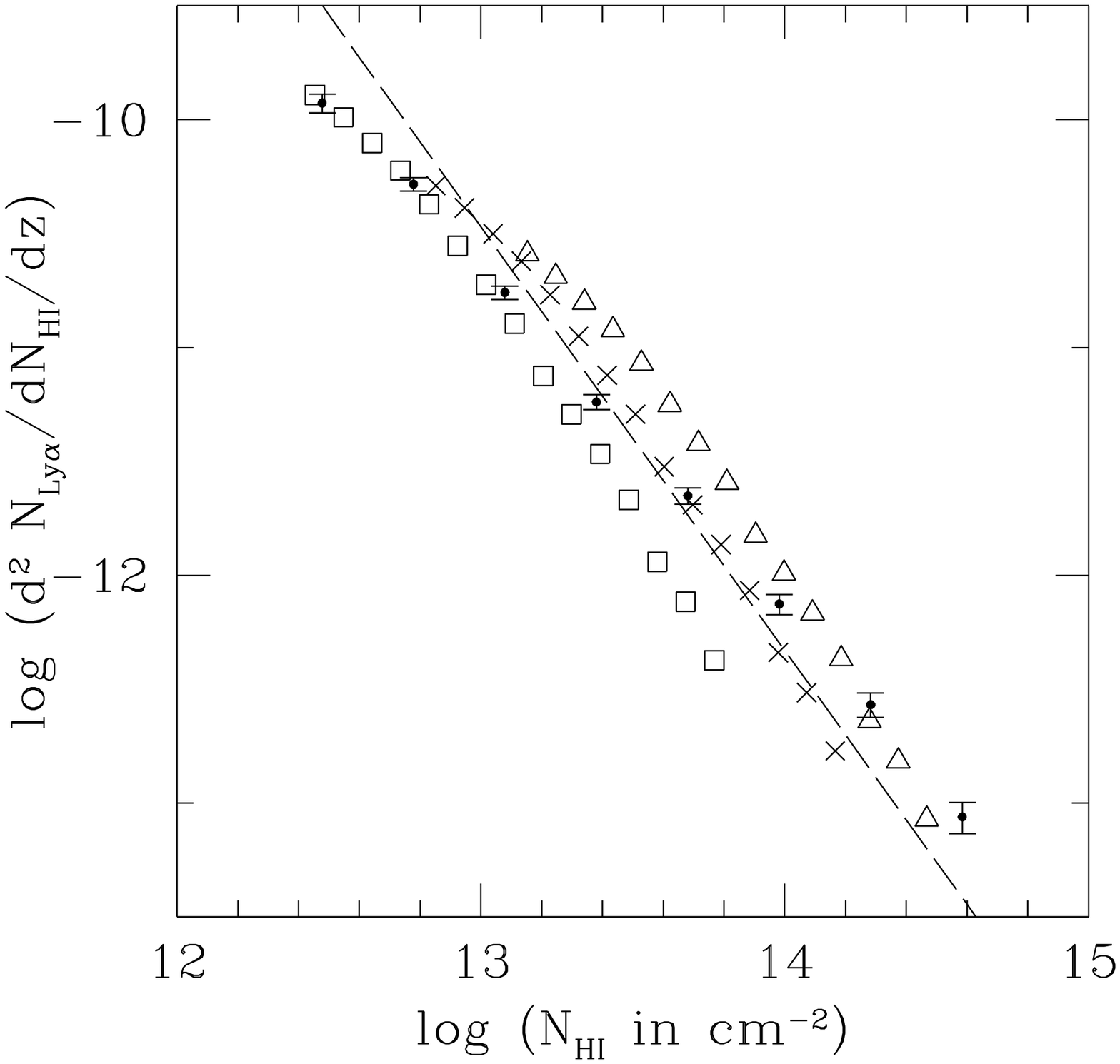}
\caption{\label{Fnu2_42}\capFnutwofourtwo}
\end{figure}
\begin{figure}
\plotone{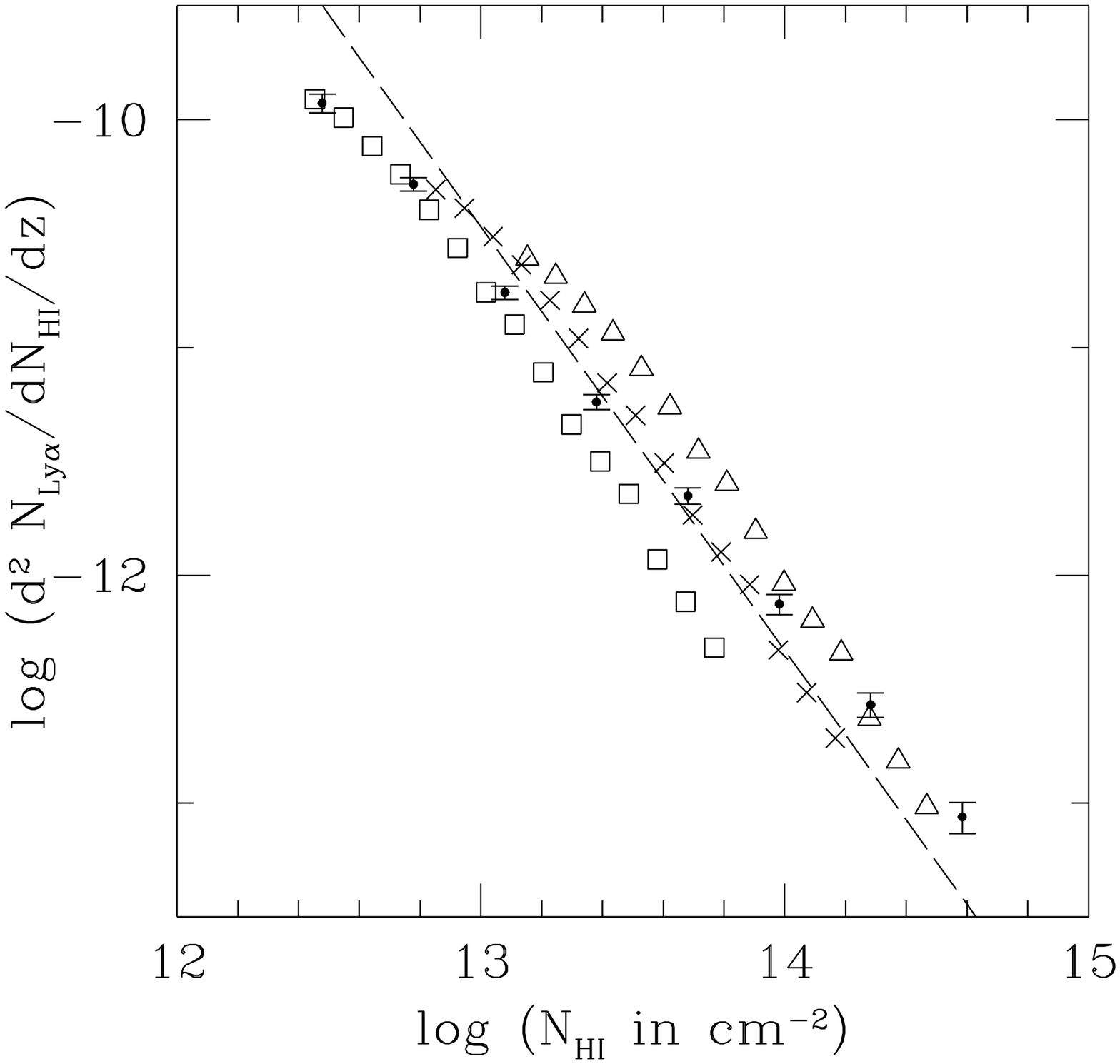}
\caption {\label{Fnu2_33b}\capFnutwothreethreeb}
\end{figure}

The CHDM models with $\Omega_\nu = 0.1$ are plotted in
Fig.~\ref{mdmcdnu1}. Values of $F$ that give reasonable match to
the observational data are chosen for each model. Note how the
low-Hubble-constant-models ($h = 0.5$) require a 
slightly lower $F$ (eq. [\ref{Fdef}]) than the
higher-Hubble-constant-models. The equation of state is chosen to be
the same for all models ($\gamma = 1.5$, see equation [\ref{tgamma}]). 
The level of agreement with the observational data, for the given choices
of parameters, is satisfactory. Notice how the low Hubble-constant
($h = 0.5$) models tend to have steeper column density distributions,
because they have less power on the relevant scales (see
Fig.~\ref{psmMDM}). Their slopes can be brought into better agreement
with that of the observational data if a smaller $\gamma$ is used.





For the $\Omega_\nu = 0.2$ models, we cannot find values of $F$ that
give the
same level of agreement with observations for $\gamma = 1.5$. Two
examples are shown in 
Fig.~\ref{Fnu2_33b} and Fig.~\ref{Fnu2_42}. Both have $h=0.5$ and small
amounts of tilt. For each, three sets of theoretical predictions are
plotted, one for each value of $F$: $1$, $2.5$ or $5$. For $\Omega_b
h^2 = 
0.0125$, the conventional Big-bang nucleosynthesis value, and $T_0 =
10^4 {\rm K}$, they correspond to radiation intensity $J_{\rm HI}$ of 
$0.5$, $0.2$ and $0.1$ (eq. [\ref{Fdef}]). As we have shown
before, changing $F$ mainly shifts the sets of points without altering
the slope significantly.
For the column density between about $10^{12.5}$ and $10^{14.5} \,
{\rm cm^{-2}}$, the slope  
of the predicted distribution seems to be too steep compared to the
observational 
data. $\beta$ is about $1.86$, with some flattening at the lower
column densities, compared with the observed value of about $1.5$.

\begin{figure}
\plotone{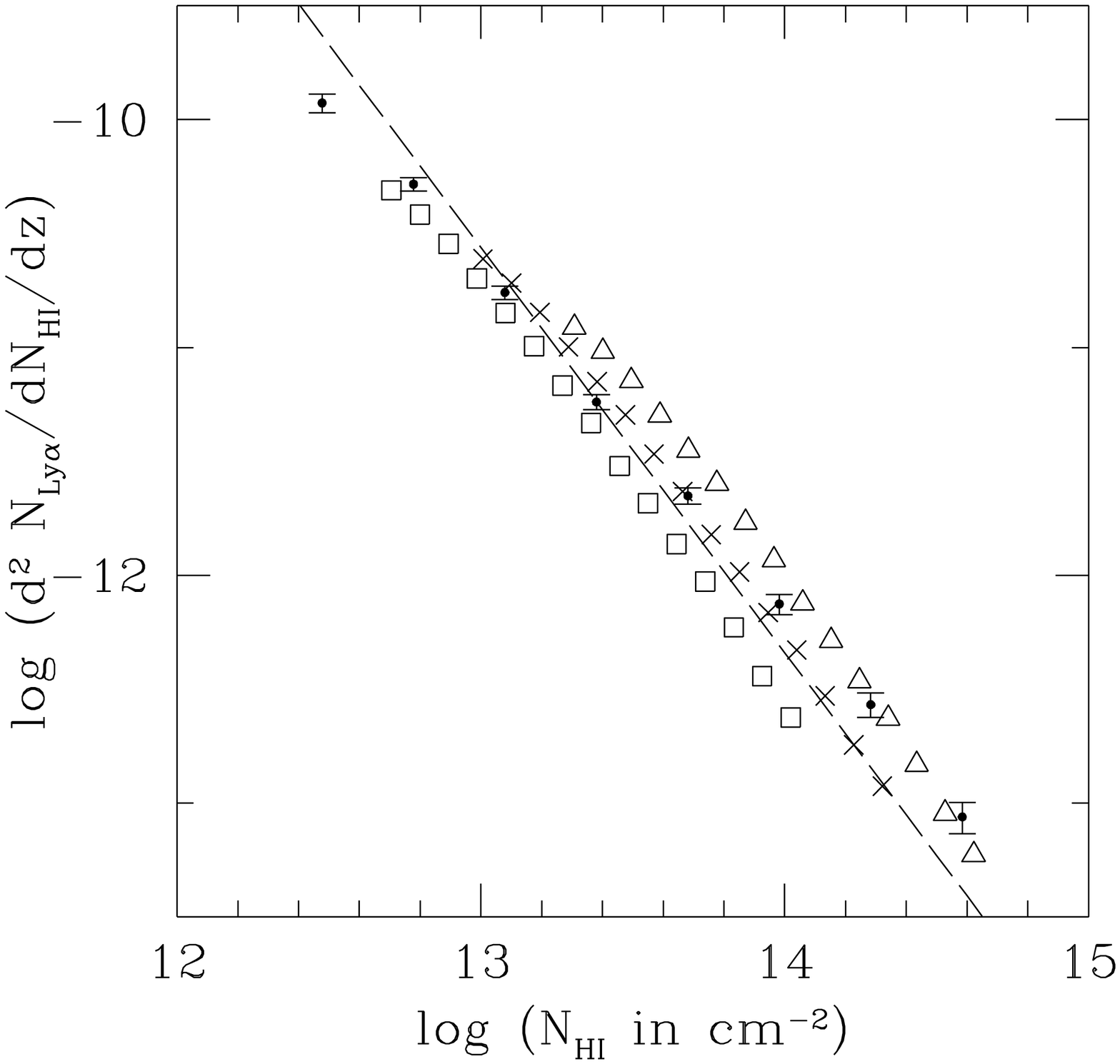}
\caption {\label{Fnu2_44}\capFnutwofourfour}
\end{figure}

\begin{figure}
\plotone{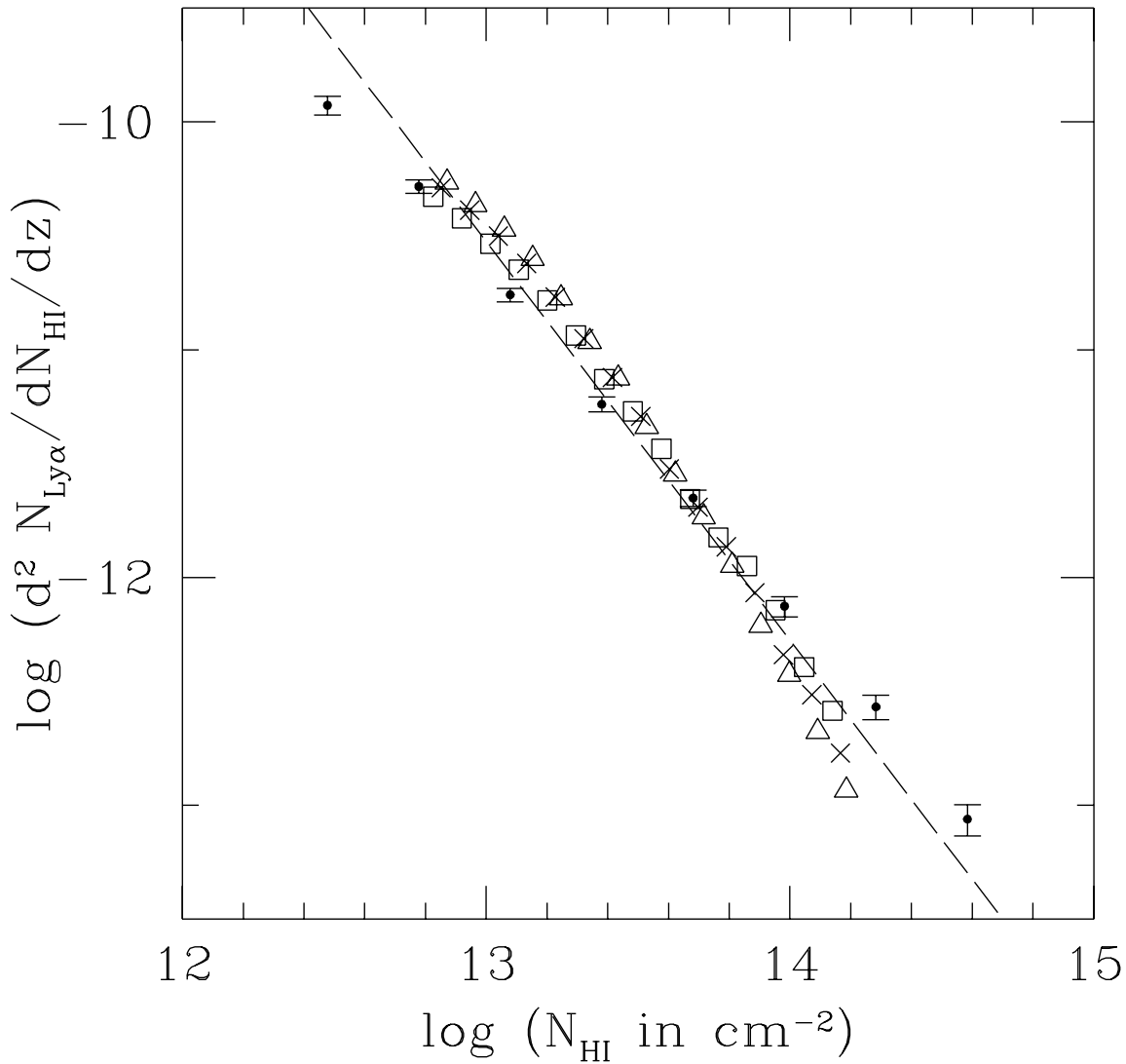}
\caption{\label{eqsnu2_42}
The column density distribution of the $A2$ CHDM model
(Table~\protect{\ref{mdmmodels}}) for three different values of
$\gamma$ (eq. [\protect{\ref{tgamma}}]). $F = 2.5$ (eq.
[\protect{\ref{Fdef}}])
is adopted. Three values of $\gamma$ are 
shown: $\gamma=1.2$ 
(open squares), $\gamma=1.5$ (crosses) and $\gamma=1.7$ (open
triangles). Points with error-bars are the
observational data as in Fig.~\protect{\ref{Fcd}}. The dashed line has
a slope of $\beta = 1.75$, which is the value given in equation
(\protect{\ref{slope}}) for $\gamma=1.2$ and $\sigma_0$ as given in
Table~\protect{\ref{mdmmodels}}.}
\end{figure}

Another $\Omega_\nu = 0.2$ CHDM model ($D2$ in Table \ref{mdmmodels}),
which has a higher Hubble 
constant ($h = 0.65$), is  shown in Fig.~\ref{Fnu2_44}. (The $C2$ CHDM
model which also has $h = 0.65$, gives very similar column density
distribution.)
The slope of its column density distribution is not as steep as the
previous ones. This is expected because the higher Hubble constant
models have slightly more power
on relevant scales, as is evident in Fig.~\ref{psmMDM}. In fact, one
might argue that the middle set of points 
in Fig.~\ref{Fnu2_44}, the one having $F=3.57$, matches the observational data
reasonably well if both observational and theoretically errors are
carefully 
taken into account. However, it is still true these two models predict a 
steeper column density distribution for $N_{\rm HI}$ between about $10^{12.5}$
and $10^{14.5} {\rm cm^{-2}}$, compared to the $\Omega_\nu = 0.1$ CHDM models
(Fig.~\ref{mdmcdnu1}).

It is not hard to understand the column density distributions of the
CHDM models presented if one refers back to Fig.~\ref{psmMDM} or Table
\ref{mdmmodels}. The $\Omega_\nu = 0.2$ models have less power (smaller
$\sigma_0$) and 
steeper spectral slope (larger $|n_{\rm eff}|$) than those with $\Omega_\nu
= 0.1$ on scales $1 {\rm Mpc^{-1}} \approxlt  k_{\rm S} \approxlt 10 {\rm
Mpc^{-1}}$, which are 
relevant for the range of column densities we are interested in.
As we have explained before, the column density distributions are
therefore steeper for the $\Omega_\nu = 0.2$ models at this range of
column densities.
Among the $\Omega_\nu = 0.2$ models, those with a lower Hubble
constant produce comparatively steeper column density distributions
because they have even less small scale power and steeper spectral slope than
the ones with a higher Hubble constant.

In discussing the predictions of the slope of the column density distribution
for these models, one should bear in mind uncertainties due to the choice
of smoothing scale, as we have presented in the last section.
However, the fact that these CHDM models have relatively little power
on small scales works in our favor: the exact choice of the smoothing scale
does not affect the slope of the distribution very much (see
Fig. \ref{smoothMDMnu02}). This is especially true for the $\Omega_\nu=0.2$
models which have orbit-crossing scales (eq. [\ref{ks}]) smaller than the
Jeans length. So, the general conclusion of steeper column density distribution
for the tilted CHDM models considered
here compared to, say, the CDM1a model (see Table \ref{cdmmodels} and
Fig. \ref{zhang}; the results of which compare favorably with a hydrodynamic
simulation) should be robust. 


In Sec.~\ref{cdm}, we have discussed how the equation of state or
temperature-density relation
can also change the slope of the column density distribution, 
although the effect is small for a reasonable range of $\gamma$. 
We
show in Fig.~\ref{eqsnu2_42} the effects of altering the equation
of state on the column density distribution for one particular
CHDM model ($A2$). $F$ is fixed at $2.5$, the value that seems to give a
column density distribution closest to the observational data.
Smaller $\gamma$, as we have noted before, helps flatten the column
density distribution but the flattening seems to be not quite enough even
for $\gamma = 1.2$. We show in the same figure a dashed line with a slope
of $-1.75$ (which follows from equation (\ref{slope}) by putting
$\gamma = 1.2$ and substituting
the appropriate values of $\sigma_0$ and $n_{\rm eff}$ as given in
Table~\protect{\ref{mdmmodels}}).
It should be kept in mind that we can always
shift the column density distribution up and down by rescaling $F$
(Sec.~\ref{cdm}), so the normalization is not important. 
It seems $\gamma < 1.2$ is
needed for this model to give the right slope of the distribution, at
least the right slope to within the $95 \%$ confidence limits of the
observed $\beta$ (1.37,1.51). The same conclusion holds for the
other low Hubble constant $\Omega_\nu = 0.2$ model ($B2$). 
We should emphasize, however, that a more detailed comparison between
the predictions of the models and observations, taking into account
noise and biases of the line identification techniques, is necessary
before any model can be considered ruled out. 

The high Hubble constant ($h = 0.65$) $\Omega_\nu = 0.2$ models
$C2$ and $D2$, on the other hand, have intrinsically flatter
distributions and a reasonable match between theory and observations
can be made by choosing $\gamma$ in the range $1.2 - 1.7$. 


\section{Conclusion}
\label{gpconc}

We have systematically developed a set of tools to compute in an
efficient manner the column density distribution given a
cosmological model. One fundamental assumption of the approximations
involved is that most of the Ly$\alpha$ forest with column densities
in the range $10^{12.5} - 10^{14.5} \,{\rm cm^{-2}}$ originates from
regions of low overdensities 
or even underdensities which have not undergone orbit-crossing. 
The result of a comparison with a hydrodynamic simulation lends
support to it. 

One major conclusion we reach, in the process of developing the tools,
is that the peculiar velocities play a minor role in determining
the column density distribution at our column densities of interest,
even though they are very important in determining the shapes of
individual absorption line profiles. We take advantage of this fact
and develop a method we call the Density-Peak-Ans\"{a}tz in which each
density peak is identified as an absorption line and assigned a column
density based on its local properties. The column density 
distribution then becomes a statistic of density peaks.

In Sec.~\ref{cdm} and \ref{slopecd}, we investigate the factors controlling the
column 
density distribution, which  
can be divided into two categories. One mostly affects the normalization, while
the other mostly influences the  slope. Those that fall into the former
category include the ionizing radiation intensity, the mean
temperature of the intergalactic medium and the mean baryon
density. Uncertainties in their values are such that almost any
viable cosmological model which has the correct slope of the column
density distribution can be made to match observations by a
judicious choice of parameters.

The factors that mostly affect the slope of the 
distribution include the equation of state and more strongly so, the
amplitude and slope of the (linear) power spectrum on scales $1 \,{\rm
Mpc^{-1}} \, \, \approxlt k 
\approxlt \, \,   10
\,{\rm Mpc^{-1}}$. Models that have less power on these scales tend to
have comparatively more intermediate density peaks than high density ones and
hence have relatively steeper column density distributions.
Models with a steeper power spectrum on these scales have $k^2 /\sigma_0(k)$ 
that falls off more quickly with increasing $k^{-1}$ ($\sigma_0 (k) \propto
k^{(n_{\rm eff}+3)/2}$) and therefore
a steeper column density distribution, assuming 
high column density corresponds to high $k^{-1}$ (Press-Schechter argument; see
Sec. \ref{slopecd}).
Equations of state which are closer to isothermal (smaller $\gamma$
where $\gamma$ satisfies $T \propto (1+\delta_{\rm b})^{\gamma-1}$) tend to
produce flatter column density distributions. 
However, within the
reasonable range of $\gamma$ (see Hui and Gnedin 1996), its precise
value depending upon the reionization history, the effect of changing
the equation of state is small. 
We put forward an approximate
expression relating the slope of the column density distribution to 
$\gamma$ and the amplitude ($\sigma_0$) and slope ($n_{\rm eff}$) of the power
spectrum on small scales 
(eq. [\ref{slope}]), which describes reasonably well all the models we study in
this 
paper. 

Hence, the slope of the column density distribution provides a
measure of the amplitude and slope of the power spectrum on small scales for a
given cosmological model and given temperature-density relation. We
apply our techniques to study a class of 
CHDM models which are known to have less power on small scales
compared to other popular CDM models. We conclude that the CHDM models
indeed produce steeper column density distributions compared to the
CDM models. In particular,
the low Hubble constant ($h = 0.5$) $\Omega_\nu = 0.2$ CHDM
models, which have the least amount of power on small scales among the
models we study, have column density distributions which can be made
consistent with observations only for $\gamma$ less than the values we
consider reasonable. We emphasize however
that only after a more detailed comparison between theories and observations,
including all the effects of noise and biases of the
line-identification methods, can any model be considered ruled out by the
observed column density distribution.  

We therefore conclude that a lot of work still needs to be done both
on the observational and theoretical fronts. The biases of the
line-identification techniques used for data reduction deserve close
study so that the error bars in the observed column density
distributions can be better understood and perhaps reduced. 
Numerical simulations on the CHDM models should be carried out to
further test
the accuracy of the approximations made in the present work.
Detailed comparisons with hydrodynamic simulations will shed light
on the appropriate choice of smoothing scales for approximate methods
such as the one we present here.
The effect of a fluctuating radiation field, instead of a uniform one
as is assumed here, has to be investigated.
Moreover, in terms of constraining models, it is also important to
examine other possible statistics. We have shown, for instance, that
the column density distribution is relatively independent of peculiar
velocities. Are there other statistics that can take advantage of the
different peculiar velocity structures predicted by different
cosmological models?

In short, the study of the Ly$\alpha$ forest has
entered an exciting stage. There is a gold mine of information
contained in the quasar absorption spectra waiting to
be discovered.

\acknowledgements
We would like to thank Edmund Bertschinger for his encouragement and
many insightful suggestions which greatly improved the paper. 
We are also grateful to Jeremiah Ostriker for his helpful comments.
It is a pleasure to thank Thomas Buchert, Lars Hernquist, Fred Rasio, Ur\u{o}s
Seljak and Matias Zaldarriaga for useful discussions and Esther Hu for
kindly sending us her data on the column density distribution. We thank
HongGuang Bi and Arthur Davidsen for correspondence regarding their preprint
and we gratefully acknowledge useful comments from Edward 
Wright. We are indebted to our referee, David Weinberg, for many thoughtful
comments and interesting discussions, particularly relating to
Sec.~\ref{slopecd}.
This work is supported in part by funds provided by the U.S. Department
of Energy (D.O.E.) under cooperative research agreement DE-FC02-94ER40818 and
in part by NSF grant AST-9318185 and NASA grant NAG5-2816.

\appendix
\section{Smoothing at Jeans Scale}
\label{jsm}

The effect of gas pressure is to smooth the baryon density field 
compared to its dark matter counterpart. The length scale below which
this becomes important is the Jeans scale. In linear theory, the
baryon overdensity obeys the following equation in a dark matter
dominated universe (Bi et al. 1992; \cite{peebles1}):
\begin{equation}
{\partial^2 \tilde\delta_{\rm b}\over {\partial t^2}} +
2H{\partial\tilde\delta_{\rm b}\over\partial t} = 4\pi G
\bar\rho_{DM}\tilde\delta_{\rm DM} - {\gamma k_B \bar T\over \mu a^2} k^2 
\tilde\delta_{\rm b} \, ,
\label{jeans}
\end{equation}
where the tilde denotes functions in Fourier space as before, 
$H$ is the Hubble constant, $k_B$ is the Boltzmann constant, $G$ is the
Newton constant, $\bar\rho_{DM}$ is the average dark matter
mass density, $\bar T$ is the
average  temperature of the gas and $\mu$ is the mean mass of each gas
particle (for a fully ionized gas composed of hydrogen and helium with
primordial abundances, it is about $0.6$ times the proton mass).
The relation between the temperature (not its
average but its actual value) and $1+\delta_{\rm b}$ is described by
$\gamma$, the  
temperature being proportional to $(1+\delta_{\rm b})^{\gamma -1}$. 

The Jeans scale is defined in equation (\ref{jk}).
For a dark matter dominated universe, one can replace $\bar\rho_{DM}$
by the total mean density of the universe. 

For the special case of $\bar T \propto a^{-1}$, making use of an 
equation for $\tilde\delta_{\rm DM}$ which is the same as equation
(\ref{jeans}) except for the absence of the temperature term,
 it can be shown that
\begin{equation}
\tilde\delta_{\rm b}({\bf k}) = {\tilde\delta_{\rm DM}({\bf k}) \over 
{(1+{k^2/ k_{\rm J}^2})}} \, , 
\label{jeansm}
\end{equation}
if one ignores decaying modes. It expresses in a quantitative way the
expectation that the overdensity in baryons is the same as that of
dark matter on large 
scales (low $k$) but is lower on small scales (high $k$). 
For $\bar T$ with some other power-law dependence on $a$, solutions for
equation 
(\ref{jeans}) are more complicated but the low and high $k$ limits are
the same: $\tilde\delta_{\rm b}=\tilde\delta_{\rm DM}$ for small $k$ and 
$\tilde\delta_{\rm b}=\tilde\delta_{\rm DM}k_{\rm J}^2/k^2$ for large
$k$ (Bi et al. 1992).  We note, however, that in practice,
the gas temperature is expected to rise from almost zero before reionization to
around $10^4 - 10^5 \, {\rm K}$ after, and so the power-law time-dependence of
$T$ is probably not realized. The actual linear smoothing scale of
the gas should be a little different from the one given above
and one has to be careful about boundary conditions for $\delta_b$ and
its derivative at the onset of reionization (see Gnedin \& Hui 1997).

\section{Thermal and Ionization Evolution}

The evolution of temperature is governed by:
\begin{equation}
{dT\over dt} = -2HT + {2T\over{3(1+\delta_{\rm b})}} {d\delta_{\rm b}\over dt}
- 
{T\over \sum_i \tilde X_i} {d\sum_i \tilde X_i\over dt} + {2\over {3k_B n_b}}
{dQ\over dt} \, , 
\label{T}
\end{equation}
where $d/dt$ is the Lagrangian derivative following each fluid
element, $n_b$ is
the proper number density of all gas particles and $T$ is the
temperature which depends on both space and time. The symbol $\tilde X_i$ is
defined by 
$n_i \equiv (1+\delta_{\rm b}) \tilde X_i\, {\bar\rho_b/m_p}$,
where $n_i$ is the proper number density of the species $i$, $\bar\rho_b$ 
is the mean mass density of baryons at the time of interest, $m_p$
is the mass of the proton and $\delta_{\rm b}$
is the overdensity as in equation (\ref{deltadef}). 
The neutral
fraction of hydrogen, $X_{\rm HI}$ (distinct from $\tilde X_{\rm HI}$) as in
equation (\ref{Xdef}), is then 
$\tilde X_{\rm HI}/(\tilde X_{\rm HI}+\tilde X_{HII})$. Note that $\tilde X_i$
is a 
function of space and 
time in general.  

The first two terms on the right hand side take care of adiabatic cooling
or heating. The third accounts for the change of internal energy due
to the change in the number of particles. The last term $dQ/dt$ is the
heat gain (or negative heat loss) per unit volume by the gas particles
from the surrounding radiation field. At a redshift of $2$ to $4$ and for
densities of our interest, the main
source of heat gain is photoionization and the main source of heat
loss is through the recombination of ionized hydrogen and the free
electron. At higher redshifts, other processes become important, such
as Compton cooling. More discussion on these
processes will be presented in Hui and Gnedin (1996). We note that one
particularly simple solution of equation (\ref{T}) is $T \propto a^{-2}
(1+\delta_{\rm b})^{2/3}$, which holds when the last two terms on
the right hand side can be ignored i.e. pure adiabatic expansion or
compression. 

The above equation has to be supplemented by one that determines the
abundance of each particle type, which takes the form:
\begin{equation}
{d\tilde X_i \over dt} = -\tilde X_i P_h + \sum_{j,k} \tilde X_j \tilde X_k 
R \left[{{\bar\rho_b (1+\delta_b)}\over m_p}\right]\, .  
\label{n}
\end{equation}
For instance, if $\tilde X_i = \tilde X_{\rm HI}$, $P_h$ is the photoionization
rate. It is given by:
\begin{equation}
P_h = \int_{\nu_{\rm HI}}^{\infty} 4 \pi J_\nu \sigma_{\rm HI}
{d\nu\over {\rm h}\nu} \, ,
\label{P}
\end{equation}
where ${\rm h}$ is the Planck constant, ${\rm h} \nu_{\rm HI} = 13.6 \,
\,{\rm eV}$, $\sigma_{\rm HI}$ is the cross-section for 
photoionization as a function of the frequency $\nu$ and $J_\nu$ is
the specific 
intensity. The photoionization rate $P_h$ depends on the normalization
as well as spectrum 
of $J_\nu$. The specific intensity $J_\nu$ is generally taken to have
a power law spectrum, $\nu^{-1}$ 
to $\nu^{-1.5}$, for frequencies just above $\nu_{\rm HI}$. The
spectrum at higher frequencies is less important for the
photoionization rate of hydrogen. A convenient way to hide our
ignorance of the spectrum is to define $J_{\rm HI}$ as in equation
(\ref{JHI}). 

For $\tilde X_i = \tilde X_{\rm HI}$, $R$ is the recombination rate of ionized
hydrogen 
and the free electron ($\tilde X_j = \tilde X_e$ and $\tilde X_k = \tilde
X_{\rm HII}$ in equation [\ref{n}]): 
\begin{equation}
R \sim 4 \times 10^{-13} ({T\over {10^4 K}})^{-0.7} \,{\rm cm^{3} s^{-1}} \, .
\label{R}
\end{equation}

For $J_{\rm HI}$ with the values noted in Sec.~\ref{tXrho}, the photoionization
time-scale is much shorter than the Hubble time. This means that
hydrogen is highly ionized and is essentially in ionization
equilibrium. The two terms on the right 
hand side of equation (\ref{n}) almost
balance each other, which implies equation (\ref{nh1}).

We now have all the equations in place to compute the thermal
and ionization evolution. 
The overdensity $\delta_{\rm b}$ is evolved using the Zel'dovich
approximation.
Its rate of growth is substituted into equation (\ref{T}), which 
is solved together with equation (\ref{n}). The initial
conditions are as follows.  The gas temperature $T$ is
equal to the cosmic microwave background temperature at $z = 100$
(maintained by Compton scattering) and evolves adiabatically
after that until the universe is reionized by the UV
background. Abundances are assumed to be primordial, which is consistent with
observations so far for column densities less than about $10^{14.5} \,
{\rm cm^{-2}}$
(See Songaila and Cowie 1996. Cooling processes due to metals
are not important for our densities 
of interest in any case). All species are neutral until reionization
occurs. One can integrate equations (\ref{T}) and (\ref{n}) forward starting
from any time between $z = 100$ and the beginning of reionization to
obtain $T$ and $\tilde X_{\rm HI}$ or $X_{\rm HI}$.

\clearpage

\begin{table}
\begin{center}
\begin{tabular}{|ccccccc|}\hline
Label & $\sigma_8$ & $n$ & $k_{\rm S}/{\rm Mpc^{-1}}$ & $\Omega_b$ & $\sigma_0$
& $n_{\rm eff}$ \\ \hline
\hline
CDM1a & $0.7$ & $1.0$ & $2.3$ & $0.05$ & $1.12$ & $-2$ \\
CDM1b & $0.7$ & $1.0$ & $2.3$ & $0.06$ & $1.12$ & $-2$ \\
CDM2 & $0.8$ & $1.0$ & $1.7$ & $0.05$ & $1.13$ & $-1.94$ \\
CDM3 & $0.57$ & $1.0$ & $1.7$ & $0.05$ & $0.81$ & $-1.94$ \\ 
CDM4 & $0.46$ & $1.0$ & $4.9$ & $0.05$ & $1.01$ & $-2.17$ \\
CDM5 & $0.8$ & $0.5$ & $8.4$ & $0.05$ & $1.01$ & $-2.59$ \\ \hline
\end{tabular}
\end{center}
\caption{\label{cdmmodels} A list of all the CDM models discussed in
this paper. All have $h = 0.5$. The larger scale power-spectral index is
$n$. Every model has $\Omega_b = 0.05$ 
except for CDM1b, which has a higher baryon content and is shown in
Fig.~\protect{\ref{zhang}}. The truncation scales $k_{\rm S}$ for CDM1a, CDM1b
and CDM2 are defined by
$k_{\rm S} = 1.5 \, k_{\rm NL}$ (eq. [\protect{\ref{ks}}]).
The smoothing scales for the rest of the models are chosen for the purpose
of testing (see Sec.~\protect{\ref{slopecd}}) but
always larger than the length scale defined above.
The rms density fluctuation $\sigma_0$ and the small scale effective slope of
the power 
spectrum $n_{\rm eff}$ are defined in equations
(\protect{\ref{sigma0}}) and (\protect{\ref{neff}}), for the value of $k_{\rm
S}$ given here. We 
assume $\bar z = 3$.  The rms linear density fluctuation in a sphere of radius
$8 h^{-1} \, 
{\rm Mpc}$ is equal to $\sigma_8$.}
\end{table}

\begin{table}
\begin{center}
\begin{tabular}{|cccccccccc|}\hline
Label & $\Omega_\nu$ & $h$ & $n$ & $Q_{\rm rms}/\,{\rm \mu K}$ & $T/S$ &
$k_{\rm S}/\,{\rm Mpc^{-1}}$ & $\sigma_8$ & $\sigma_0$ & $n_{\rm eff}$ \\
\hline \hline  
A1 & $0.1$ & $0.5$ & $0.95$ & $18.5$ & $7(1-n)$ & $3.8$ & $0.73$ & $1.08$ &
$-2.23$ \\
A2 & $0.2$ & $0.5$ & $0.95$ & $18.5$ & $7(1-n)$ & $8.4$ & $0.66$ & $0.84$ &
$-2.48$ \\
A3 & $0.3$ & $0.5$ & $0.95$ & $18.5$ & $7(1-n)$ & not apply & $0.64$ & not
apply & not apply \\
B1 & $0.1$ & $0.5$ & $0.9$ & $19.2$ & $0$ & $3.3$ & $0.82$ & $1.08$ & $-2.25$
\\ 
B2 & $0.2$ & $0.5$ & $0.9$ & $19.2$ & $0$ & $8.4$ & $0.73$ & $0.87$ & $-2.52$
\\ 
C1 & $0.1$ & $0.65$ & $0.9$ & $19.2$ & $7(1-n)$ & $3.1$ & $0.82$ & $1.09$ &
$-2.14$ \\
C2 & $0.2$ & $0.65$ & $0.9$ & $19.2$ & $7(1-n)$ & $10.9$ & $0.76$ & $1.02$ &
$-2.5$ \\
D1 & $0.1$ & $0.65$ & $0.8$ & $20.5$ & $0$ & $2.8$ & $0.94$ & $1.08$ & $-2.19$
\\ 
D2 & $0.2$ & $0.65$ & $0.8$ & $20.5$ & $0$ & $10.9$ & $0.87$ & $1.02$ & $-2.57$
\\ \hline 
\end{tabular}
\end{center}
\caption{\label{mdmmodels} A list of all the CHDM models discussed in
this paper. All have $\Omega_b = 0.05$. The parameters are defined as
follows: $\Omega_\nu$ is the density
parameter in neutrino, $n$ is the large scale spectral index of
the power spectrum, $Q_{\rm rms}$ is the COBE quadrupole in $\mu K$ and $T/S$
is the tensor to scalar ratio. The smoothing wavenumber $k_{\rm S}$ for
each $\Omega_\nu = 0.1$ model 
is $1.5 \, k_{\rm NL}$ (eq. [\protect{\ref{ks}}]) and $k_{\rm S}$ for each
$\Omega_\nu 
= 0.2$ model is the Jeans scale for $\gamma = 1.5$, $T_0 = 10^4 \, K$
and the corresponding $h$ (see Sec.~\protect{\ref{tXrho}}). No
simulation is run for $A3$, so no 
$k_{\rm S}$ is listed. The rms linear density fluctuation $\sigma_0$ and the
small scale effective spectral slope $n_{\rm eff}$ are
defined in equations (\protect{\ref{sigma0}}) and (\protect{\ref{neff}}), for
each value of $k_{\rm S}$ given in the seventh column. The models and their
power spectra are 
taken from Ma (1996).}
\end{table}

\clearpage

\clearpage

\end{document}